\documentclass[12pt]{article}
\pdfoutput=1
\usepackage{sectsty}
\sectionfont{\large\bfseries}
\subsectionfont{\normalsize\sc}
\subsubsectionfont{\normalsize\it}
\usepackage{amsmath, amsthm,comment}
\usepackage{amsfonts}
\usepackage{amssymb,graphics,psfrag}
\usepackage{array,epsfig,stmaryrd,graphicx}
\usepackage{cite}
\usepackage{graphicx}
\usepackage{makeidx}
\usepackage{multicol}
\usepackage{geometry}
\usepackage{amsfonts}
\usepackage{dsfont}
\usepackage{mathrsfs}
\usepackage{slashed}
\usepackage{heck}
\providecommand{\U}[1]{\protect\rule{.1in}{.1in}}
\pdfoutput=1

\newcommand{\beq}{\begin{equation}}
\newcommand{\eeq}{\end{equation}}
\newcommand{\be}{\begin{equation}}
\newcommand{\ee}{\end{equation}}
\newcommand{\bea}{\begin{eqnarray}}
\newcommand{\eea}{\end{eqnarray}}
\newcommand{\ben}{\begin{eqnarray*}}
\newcommand{\een}{\end{eqnarray*}}
\newcommand{\ba}{\begin{aligned}}
\newcommand{\ea}{\end{aligned}}
\newcommand{\bt}{\begin{tabular}}
\newcommand{\et}{\end{tabular}}
\newcommand{\bc}{\begin{center}}
\newcommand{\ec}{\end{center}}

\newcommand{\cref}{{\bf [check ref]}}

\newcommand{\bs}{\begin{subarray}{c}}
\newcommand{\es}{\end{subarray}}

\renewcommand{\overrightarrow}{\vec}

\def\spc{\hspace{.8pt}}
\def\bee{\begin{equation}}
\def\eee{\end{equation}}
\def\is{\! & \! = \! & \!}
\renewcommand{\footnotesize}{\small}

\def\ddd{\overline\partial}

\begin{document}

\renewcommand{\footnotesize}{\small}

\addtolength{\abovedisplayskip}{0mm}
\addtolength{\belowdisplayskip}{0mm}

\date{May, 2010}

\preprint{PUPT-2338}

\institution{IAS}{\centerline{${}^{1}$School of Natural Sciences, Institute for Advanced Study, Princeton, NJ 08540, USA}}

\institution{PU}{\centerline{${}^{2}$Department of Physics, Princeton University, Princeton, NJ 08544, USA}}

\title{Evidence for F(uzz) Theory}

\authors{Jonathan J. Heckman\worksat{\IAS}\footnote{e-mail: \texttt{jheckman@sns.ias.edu}} and Herman Verlinde\worksat{\PU}\footnote{e-mail: \texttt{verlinde@princeton.edu}}}

\abstract{We show that in the decoupling limit of an F-theory compactification,
the internal directions of the seven-branes must wrap a non-commutative four-cycle ${\cal S}$.
We introduce a general method for obtaining fuzzy geometric spaces via toric geometry, and
develop tools for engineering four-dimensional GUT models from this non-commutative setup.  We
obtain the chiral matter content and Yukawa couplings, and show that the theory has a finite
Kaluza-Klein spectrum. The value of $1/\alpha_{GUT}$ is predicted to be equal to the number of
fuzzy points on the internal four-cycle ${\cal S}$. This relation puts a non-trivial restriction
on the space of gauge theories that can arise as a limit of F-theory. By viewing the seven-brane
as tiled by D3-branes sitting at the $N$ fuzzy points of the geometry, we argue that this theory
admits a holographic dual description in the large $N$ limit. We also entertain the
possibility of constructing string models with large fuzzy extra dimensions,
but with a high scale for quantum gravity.}

\maketitle

\tableofcontents

\pagebreak

\section{Introduction}

In local string model building,
 the degrees of freedom that comprise the Standard Model are assumed to be localized
within a small neighborhood of the compactification manifold, arranged such that one can {\it in principle} take a decoupling limit,
in which the four-dimensional Planck scale is sent off to infinity
\cite{VerlindeWijnholtBottomUp,KiritsisBottomUp,UrangaBottomUp,BerensteinJejjalaLeigh}.
The gauge and matter fields are engineered via branes wrapping small cycles of a  geometric singularity, and decoupling is achieved by a combined
decompactification and zero slope limit. The decoupled theory reduces to a pure four-dimensional quantum field theory (QFT), with
{\it a priori} freely tunable coupling constants. In recent years, there has been some notable progress
in local model building of supersymmetric grand unified field theories (GUTs) via F-theory
\cite{DWI,BHVI,WatariTATARHETF,BHVII,DWII}. See \cite{Heckman:2010bq} for a recent review of F-theory GUTs.
In this setting, the existence of a decoupling limit turns out to be quite restrictive:
it requires that the gauge fields must live on spacetime filling seven-branes that wrap
a del Pezzo surface. As argued in \cite{BHVII}, decoupling gravity from an F-theory GUT
imposes significant restrictions, and provides a way to narrow the search for potentially
promising vacua. The local approach in M-theory is less well developed, but presumably gives rise to
a similarly restrictive framework.

Local string model building is only the first stage of a modular bottom-up approach to string phenomenology \cite{VerlindeWijnholtBottomUp,KiritsisBottomUp,UrangaBottomUp,BerensteinJejjalaLeigh} in which the task of
embedding particle physics within string theory is made more tractable, by decomposing it
into a sequence of four steps:\\[1mm]\addtolength{\baselineskip}{1mm}
(i)~Geometrically engineer a (semi-)realistic GUT via a local M-theory or F-theory model.
(ii)~Investigate its low energy phenomenology, while assuming that couplings can be tuned.
(iii)~Formulate the topological and geometric compatibility conditions that
follow from the requirement that the local model can be embedded inside a global string compactification.
(iv)~Establish the existence of stabilized compactifications that fulfill these requirements.
\smallskip

\addtolength{\baselineskip}{-1mm}
\noindent
Each stage has its own set of challenges and potential for learning new lessons.
In the first two steps one develops a link between the effective field theory language of GUT phenomenology and the geometric language of extra dimensions, strings and branes. This local geometric perspective is flexible yet restrictive, and when combined with minimal naturalness requirements,
produces a concrete and predictive framework. Via the last two steps one imposes the requirement that  the local model can be extended into a UV complete  theory with gravity. This second half of the program is harder, as it involves dealing with the full complexity
and apparent arbitrariness of the closed string landscape, but is equally essential, since one should expect to uncover additional non-trivial restrictions and predictions.  Central model building components,
such as supersymmetry breaking, may involve all four stages.

Here we concern ourselves with the first step, the map between the geometric
ingredients of F-theory and four-dimensional quantum field theory. In a
strict decoupling limit, the two systems should be equivalent: all higher dimensional
and stringy degrees of freedom should either be explicitly encoded in terms of four-dimensional
quantum fields, or integrated out and absorbed via computable threshold corrections.
This  raises the obvious question: \\[1mm]
{\it What is the subspace of all four-dimensional QFTs that
can be obtained via a decoupling limit of F-theory?}

\smallskip

\noindent
We will find that the landscape of decoupling limits of F-theory is distinctly smaller than the space of all four-dimensional QFTs.
Specifically, for any F-theory GUT with a given matter content, we will establish a concrete upper
bound on the gauge coupling. Conversely, for a given gauge coupling, we will find restrictions on the possible matter content of the
gauge theory.

\subsection{F(uzz) Theory as the Decoupling Limit of F-theory}

\noindent
Suppose we set out to engineer a grand unified theory by means of a stack of seven-branes, wrapped on a
contractible del Pezzo four-cycle ${\cal S}$ of an F-theory compactification. To gain a better
understanding and control over the physical content of the model, it is helpful to isolate the four-dimensional QFT data from
the string and extra dimensional degrees of freedom. However, as long as the four-cycle ${\cal S}$
has some finite size, the seven-brane theory still has an elaborate spectrum of Kaluza-Klein (KK) and string excitations,
and interacts with the closed strings in its neighborhood. To get a decoupled four-dimensional
gauge theory, we must let the four-cycle ${\cal S}$ shrink to zero
size\footnote{We use units with $2\pi \alpha' =1$. The zero slope limit then corresponds to working at energy scales
below the scale of massive string excitations.}
 \bea
 \label{decoupling}
 {\rm Vol}_{\rm closed}({\cal S}) 
 \;
  \to\; 0,
\eea
where ${\rm Vol}_{\rm closed}({\cal S})$ denotes the volume of the four-cycle
${\cal S}$ evaluated in the {\it closed string} metric. The relevant RG scale, which typically
would be equal to $M_{GUT}$, is assumed to be far below the string scale. This assumption
amounts to taking a zero slope limit. The small volume limit pushes up the KK mass scale,
and the zero slope limit decouples the massive string excitations. The combined
limit decouples the closed string dynamics, and sends the four-dimensional
Planck mass to infinity.\footnote{The assumption that gravity can be
decoupled from  a local F-theory model imposes conditions on possible
global completions \cite{DWIII,Cordova:2009fg,Grimm:2009yu}.}

Local geometric engineering of four-dimensional gauge theories thus typically requires
that parts of the internal geometry become highly curved. Besides the decoupling argument,
there is also a more dynamical motivation for considering the small volume limit. A general
important issue is how the moduli that determine the local size and shape of the seven-brane worldvolume
in fact get stabilized. One common way that volume moduli are stabilized is through instanton
effects. This is potentially problematic for the GUT volume modulus, because
if we are to treat the gauge coupling as a fixed parameter up to high scales
(in accord with the paradigm of unification), the corresponding modulus must
have a very large mass. On the other hand, the naive mass-squared induced by
an instanton of the GUT\ theory is suppressed by $e^{-2\pi/\alpha_{GUT}}$. In particular, to achieve an appropriate
GUT\ scale mass for the volume modulus, it would seem necessary to consider a
regime of parameters where some volume has collapsed to zero size, so as to enhance the associated instanton induced
mass scale. Related to this is the problem noted in \cite{Dine:1985he} that one expects at least some
of the parameters of the model to be stabilized at strong coupling or high curvature.
Our approach is to turn this problem on its head. Rather than trying to avoid high curvature or strong coupling,
we instead set out to make use of the fact that closed string moduli may naturally be stabilized in a small volume regime.

\def\BB{{\cal B}}

The commonly used approach to study string and brane dynamics in a highly curved setting is to make use of the fact that
the geometric methods remain accurate  for characterizing topological and holomorphic gauge theory data, which
can be computed at large volume and extrapolated
to the small volume limit. However, as we will now argue, there is another physical mechanism and
independent reason for why the geometric perspective remains valid and useful even in the zero volume limit.

\smallskip

To explain our argument,
let us consider the value of the four-dimensional gauge coupling in the decoupling limit.
In the context of compactifications of perturbative type IIB\ vacua
with D7-branes, the four-dimensional gauge coupling is given by
\bea
\frac{4\pi}{g_{GUT}^2} \spc =\, \frac{{\rm Vol}_{\rm open}({\cal S})}{(2\pi)^2g_s } \, 
\eea
where ${\rm Vol}_{\rm open}({\cal S})$ denotes the internal volume of the GUT\ seven-brane as measured by the {\it open string} metric.
Since we are interested in producing four-dimensional gauge theories with a perturbative coupling,  we must ensure that the open string volume remains finite,  while taking the small closed string volume limit (\ref{decoupling}). From the DBI action, we read off that at large closed string volume the
open string volume can be expressed as
\bea
\label{gcoupling}
{\rm Vol}_{\rm open}({\cal S}) 
 = \; 
{\rm Vol}_{\rm closed}({\cal S})
\, + \, \, \int_{{\cal S}} \BB \wedge \BB + \cdots . \; 
\eea
Here $\BB$ is the background two-form flux on the D7-branes, given by the sum
\bea
\BB = F + \hat{B}
\eea
of the worldvolume $U(1)$ flux $F = dA$ and the pull-back ${\hat B}$ of the NS two-form field $B_{\rm\spc NS}$ to the seven-brane worldvolume. In addition,
the ``$\cdots$'' refers to various correction terms associated with the K\"ahler form. In the
$\rm{Vol}_{\rm closed}(S) \rightarrow 0$ limit we can keep ${\rm Vol}_{\rm open}({\cal S})$
finite, provided we switch on a non-zero background $\BB$-flux
along the internal directions. The decoupling limit (\ref{decoupling}) then
produces a four-dimensional gauge theory with finite gauge coupling
\bea
\label{symvolume}
\frac{4\pi}{g_{GUT}^2} \spc =\, \frac{1}{(2\pi)^2g_s } \spc  \int_{{\cal S}}  \BB \wedge \BB\, .
\eea
Note that because $\BB \wedge \BB$ is a topological quantity, we expect this
formula to be correct in the decoupling limit we are considering.

In flat space, taking the zero slope limit at zero closed string volume, in tandem with holding the
two-form field $\BB$ and the open string metric $G_{ij}^{\rm open}$
fixed coincides with the Seiberg-Witten limit. This
produces a non-commutative gauge theory along the internal directions of
the seven-brane \cite{SeibergWitten}.\footnote{In most discussions of D-brane engineering
of non-commutative gauge theories, one only considers the NS 2-form $B_{\rm \spc NS}$. The
$U(1)$ flux $F=dA$ can be viewed as the integral part of $B_{\rm NS}$. Indeed, it is clear that in
the presence of a $U(1)$ magnetic field and in a suitable low energy limit, the charged endpoints of the
open strings will relax into the states of the lowest Landau level, which can be thought of
as occupying unit Planck cells of a non-commutative geometry.} Non-commutative theories on
tori \cite{Connes:1997cr} and in flat space \cite{SeibergWitten} have been studied extensively in the literature.
Some generalizations of the flat space case have been studied in the string theory literature
for example in \cite{Alekseev:1999bs,Alekseev:2000fd}. See for example
\cite{ConnesBook,Douglas:2001ba,Balachandran:2005ew}
for reviews of some aspects of non-commutative geometry and its
possible applications to physics. Though somewhat orthogonal to the limit we
consider here, related fluxes are often considered in theories of magnetized
branes (see \cite{AntoniadisReview} for a review).

Our present application requires a non-trivial generalization of the flat space
situation considered in \cite{SeibergWitten}, but it seems reasonable to assume that the same conclusion
should carry over to our context. In other words, we learn that
from the perspective of the seven-brane, the internal four-cycle should be treated as a non-commutative or ``fuzzy'' space
${\cal S}_{\rm NC}$ with a finite volume and a non-trivial symplectic form equal to $\BB$.
At a schematic level, the commutative coordinates $(z,\bar{z})$ on ${\cal S}$ should be
replaced by operators $(Z,Z^\dagger)$ which satisfy a commutation
relation of the form\footnote{Here, and in the following, we assume that $\BB$ is a (1,1)-form.}
\bea
\label{ncoordinates}\quad
[\, Z_i,Z^\dag_{j}\, ] = \theta_{i  j}\, , \qquad \quad
\theta_{i j} = \BB^{-1}_{ij}\, .
\eea
The four uncompactified space-time dimensions remain as ordinary commuting directions.

Motivated by this observation, we will set out to develop some basic algebraic tools that are needed to build local F-theory models with  a non-commutative internal geometry~${\cal S}$.
In this paper, we will show that all the necessary ingredients can be realized on an internal fuzzy geometry with only a finite number of points.
The number of points on ${\cal S}$ is equal to the dimension of the associated
state space on which the non-commuting coordinates $Z_i$ and $Z_{j}^{\dag}$ act as linear operators. The space of
functions on ${\cal S}$ is identified with the space of all  linear maps from this state space to itself, and is therefore finite-dimensional. 
Hence by design, the decoupled seven-brane theory has a finite Kaluza-Klein expansion, and reduces
to a pure four-dimensional quantum field theory with a finite number of fields. Our goal in the following sections is to develop
the basic elements of the non-commutative geometry probed by the seven-brane and
give a precise algebraic prescription for computing the spectrum of fields, the Lagrangian and the gauge and Yukawa
interactions of the four-dimensional QFT.

At a practical level, making the internal space non-commutative also provides us with a regulator for the seven-brane, which softens
the usually problematic high energy behavior of gauge theory in extra dimensions. Ordinarily, one would appeal to a lattice formulation,
which comes at the cost of destroying some of the geometric structure of the internal space. By contrast, the fuzzy prescription
provides us a way to retain the holomorphic geometry used to define the compactification in the first place. Schematically,
the higher-dimensional fields become operators via the replacement:
\bea
\phi(x_{\mu},z,\overline{z}) \rightarrow \phi(x_{\mu},Z^{\dag},Z) = \sum_{n,m} \phi_{n,m}(x^{\mu}) Z^{\dag n} Z^{m}
\eea
where each $\phi_{n,m}(x_{\mu})$ corresponds to a four-dimensional field. As we show, when the internal space ${\cal S}$ is compact,
the oscillators $Z$ and $Z^{\dag}$ act on a finite-dimensional Hilbert space. As a consequence, our power series truncates to a
polynomial, and only a finite number of the $\phi_{n,m}$ wind up being dynamical four-dimensional fields. Similar conclusions about
models with fuzzy extra dimensions have been drawn from a somewhat different viewpoint for example in \cite{Aschieri:2006uw,Steinacker:2007ay,Chatzistavrakidis:2009ix}.

\begin{figure}[t]
\begin{flushright}
 \epsfig{figure=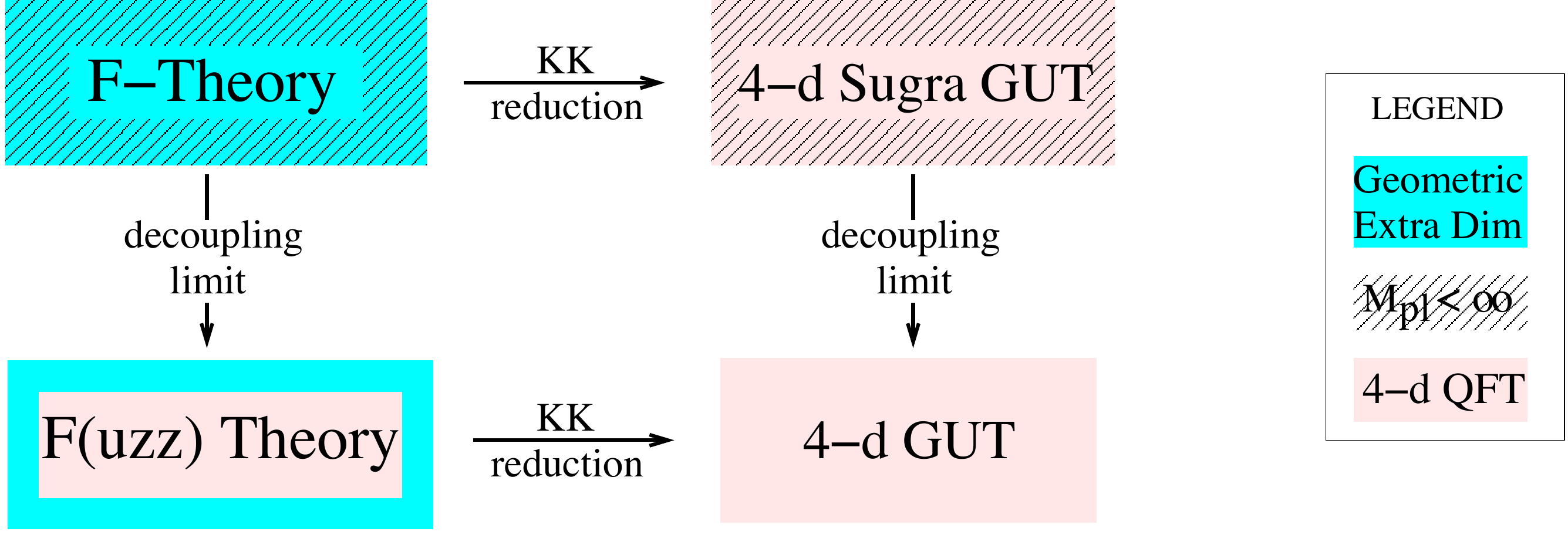,scale=.6}
\end{flushright}
  \caption{F(uzz) theory arises from F-theory via a decoupling limit in which $M_{pl}$ is sent to infinity.
  It represents a four-dimensional QFT, but  preserves the  geometric higher dimensional perspective of the local F-theory construction.
  The extra dimensions are non-commutative, and give rise to a finite KK spectrum. The arrows in the above diagram are not surjective
  and the diagram does not commute: each reverse arrow represents an embedding into a more complete theory, which give rise to
  non-trivial consistency requirements on the possible GUT models that can arise from F-theory.}
  \label{fuzzfig}
\end{figure}

Each point on the non-commutative space ${\cal S}$ can be viewed as a localized quantum state, that occupies one unit Planck cell. The size of the Planck cells is governed by the ${\cal B}$-field, and the total number of  points is given by the symplectic volume
\bea
\label{number}
 \# \ \text{  points in}\;\, {\cal S} =\frac{1}{(2\pi)^2} \int_{\cal S}  \BB \wedge \BB 
 .
\eea
For small $\BB$, there are only a few fuzzy points. To recover the commutative continuum
gauge theory, one needs to take the large $\BB$-field limit. In order to quantize ${\cal S}$,
the periods of $\BB$ around any two-cycle $\Sigma_a$ within ${\cal S}$ must be an integer
multiple of $2\pi$. This in particular ensures that the right-hand side of (\ref{number}) is an
integer.

At a general point in moduli space, when ${\cal S}$ has finite size,
the periods of $\BB$ do not automatically take quantized values.
The quantization constraint is a natural consequence of the small volume limit,
both from a dynamical and a topological perspective. Like any stringy axion, the $B_{NS}$-periods acquire a periodic
potential, generated via string instantons wrapping the associated 2-cycles. It is reasonable to expect that
in the small volume limit, when the string instantons are unsuppressed,  the induced potential enforces
the constraint that  $\BB$ has integer periods. Topologically, shrinking a two-cycle $\Sigma$ can be
thought of as  going towards a geometric transition point, at which the three-chain bounded
by  $\Sigma$ turns into a three-cycle $B_3$. Cutting off the non-compact geometry, $B_{3}$ becomes compact.
It is then consistent to find that $\int_{\Sigma} \BB \simeq \int_{B_{3}}H_{3}$
attains a quantized value at this geometric transition point.
Thus, after taking the zero volume limit,
the periods of $\BB$ do not represent freely tunable moduli, but are
more properly viewed as discrete parameters.

An illustrative alternative perspective on the situation is as follows. The
integral part of the periods of ${\cal B}$ and of the symplectic volume
$\int_S {\cal B} \wedge {\cal B}$ represent, respectively, wrapped
five-branes and three-branes, that are dissolved inside of the seven-brane.
In the decoupling limit, the internal open string volume of the wrapped
seven-brane is fully supported by the induced D3-brane charge. This
suggests that one can view the seven-brane as completely built up
from constituent D3-branes, each occupying one Planck cell. The fuzzy
points on ${\cal S}$ are in one-to-one correspondence with the
constituents D3-branes. We will return to this perspective later,
in section \ref{sec:Holo}.

\subsubsection*{$({\alpha_{GUT}})^{-1} \simeq\, \text{Number of Fuzzy Points}$}

Our considerations uncover some useful new insights into the structure of the
subspace of four-dimensional QFTs that can arise as a decoupling limit of F-theory. As a particular example, we can
combine equations (\ref{symvolume}) and (\ref{number}) to obtain the following intriguing relation between the
four-dimensional gauge coupling and the number of fuzzy points on the internal space
\bea
\label{couplingrel}
\qquad \frac{4\pi}{g_{GUT}^2 } \; =   \frac{1}{g_s} \; \Bigl(
\; \# \ \text{ points in ${\cal S}$}  \; \Bigr)
\eea
A priori, it would seem that one could relax the proportionality between the GUT\ gauge coupling
and the number of  points on ${\cal S}$ by varying the value of the string coupling $g_s$  over its full range.
However, $g_s$ is not an arbitrarily adjustable coupling constant.

In an F-theory compactification, the axio-dilaton $\tau = C_0 + i/g_s$  represents
the shape of the elliptic fiber of the Calabi-Yau fourfold. Like any complex structure modulus, this shape can be
stabilized by turning on appropriate fluxes. In general $\tau$ does not stabilize at a constant value, but
varies non-trivially over the compactification manifold.  In the neighborhood of seven-branes, $\tau$ behaves in a well
prescribed but possibly singular way. (For example, near a D7-brane it behaves as $\tau \sim \frac{1}{2 \pi i} \log z$.)
The string coupling at the seven-brane thus depends on the value of the short-distance cut-off near the brane.
In this paper we shall assume that $\tau$ varies adiabatically over the holomorphic divisor defined by the GUT seven-brane, 
and so we shall specify a cut-off value for $\tau$ compatible with its profile near an E-type Yukawa point. In principle 
such position dependence can also be taken into account \cite{HVinprog}, though throughout this paper we shall neglect this subtlety.

A closely related point is that building an F-theory GUT necessarily involves non-perturbative
$(p,q)$ seven-branes, around which $\tau$ has a monodromy given by some non-trivial element of the $SL(2,{\mathbb Z})$ S-duality group.
One could argue that a self-contained discussion of non-commutative seven-branes must be formulated in an $SL(2,{\mathbb Z})$
covariant way. The non-commutativity parameter of a $(p,q)$ seven-brane gauge theory is given by the overall $U(1)$ flux
\bea
\BB = {F} \! +  p \hat{B}_{NS}\! - q \hat{B}_{RR}.
\eea
Similarly, we need to generalize the formula (\ref{symvolume}) for the four-dimensional gauge coupling by replacing the prefactor $1/g_s$ by the
$(p,q)$ seven-brane tension (in string frame). Upon switching to Einstein frame, this provides an S-duality invariant characterization of the
parameters of the seven-brane gauge theory. Note in particular, that the covariance under S-duality allows one to impose the restriction that
$g_s$ is always less than one (or more accurately, less than $2/\sqrt{3}$). See \cite{Bergshoeff:2006gs} for a discussion of the
DBI action for $(p,q)$ seven-branes.\footnote{For bound states of seven-branes of different $(p,q)$-type, turning on
a $\BB$-field would appear differently to the constituent branes. For example, the exceptional seven-branes with gauge group
$E_6$, $E_7$ and $E_8$ can be realized as bound states of the form $E_{n} = A^{n-1} B C^{2}$, where $A$, $B$ and $C$ correspond
to seven-branes of different $(p,q)$ type. As the $\BB$-flux experienced by each type of seven-brane factor is different, it is as if we had switched on a
GUT breaking flux, breaking the corresponding group to $SU(n-1) \times SU(2) \times U(1)$.}

Now, in a local F-theory model, the requirement that the theory incorporates all the necessary matter and
Yukawa couplings dictates that the seven-brane worldvolume must contain a locus where the singularity type is enhanced to $E_6$.
At the $E_6$ singularity, the value of $\tau$ is fixed to be equal to the cubic root of unity.
Combining this with our argument above, we learn that in an F-theory GUT, the string coupling is naturally stabilized
at the special value $g_s = {2}/{\sqrt{3}}\, .$ Hence it is reasonable to fix $g_s$ to be close to this maximal value.
Setting $g_s \simeq 1$, equation (\ref{couplingrel}) turns into a
direct equality between $1/\alpha_{GUT}$  and the number of  points on ${\cal S}$
\be
\frac{1}{\alpha_{GUT}} \;  \simeq \;
\; \# \ \text{ points in ${\cal S}$} \;
\ee
Given a typical value $\alpha_{GUT} \simeq 1/25$, this prescribes that the number of  points in ${\cal S}$ must be equal
to an integer on the order of 20 to 30. Given that ${\cal S}$ is four-dimensional, this is a rather small number -- which means that
non-commutative effects  can be sizable.\footnote{Note that even if one relaxes the assumption that the local string coupling is close to one,
the requirement that $g_s \lesssim 1$ (since one can always go to an S-duality frame in which this is the case) gives the same strong
upper bound on the number of  points.}

\subsection{Outline}

The rest of this paper is organized as follows. In section \ref{sec:FREVIEW} we review the main ingredients of commutative F-theory. One of
our goals will be to translate the holomorphic data of the commutative theory to the non-commutative setting. In section \ref{Fuzztor}
we develop some aspects of fuzzy toric geometry. A general non-commutative K\"ahler manifold can be written
as a subspace of a toric space and we define forms and differential calculus on such subspaces.
Using this construction, in section \ref{sec:FUZZTHEORY} we
formulate the main ingredients of Fuzz theory, and develop the theory of seven-branes that wrap a non-commutative four-cycle. Sections \ref{sec:FORMAL} and \ref{sec:Physical} contain some applications of Fuzz theory. In section \ref{sec:Holo} we comment on the limit where the number of fuzzy points $N$ becomes large,
and on the possible closed string duals. Section \ref{sec:CONCLUDE} contains
our conclusions. Additional details of fuzzy $\mathbb{P}^{1}$ and $\mathbb{P}^{1} \times \mathbb{P}^{1}$
are given in Appendices A and B.

\section{Review of Commutative F-theory}\label{sec:FREVIEW}

In this section we briefly review the main geometric elements of F-theory
constructions \cite{VafaFTHEORY,MorrisonVafaI,MorrisonVafaII}.

To achieve four-dimensional $\mathcal{N} = 1$
supersymmetry, we compactify F-theory on an elliptically fibered Calabi-Yau fourfold,
fibered over a complex threefold $B_{3}$.  The elliptic fiber is infinitesimally small, and the base $B_3$ represents the actual compactification manifold of the IIB string theory. A minimal presentation of the
Calabi-Yau fourfold is in terms of the Weierstrass equation%
\begin{equation}
\label{weierstrass}
y^{2}=x^{3}+f(z_{i})x+g(z_{i}),
\end{equation}
where the $f$'s and $g$'s are given by sections of suitable line-bundles over the base $B_3$. We can represent $f$ and $g$ as polynomials of
the complex coordinates $z_i$ that parameterize $B_{3}$.
At any point $z_j$ on the base, equation (\ref{weierstrass}) defines a torus with shape modulus $\tau$,
which specifies the value of the IIB axion and dilaton field via $\tau = C_0 + i e^{-\phi}$. In F-theory compactications, this axio-dilaton field
varies non-trivially over the internal directions.

The location of
seven-branes in the compactification is specified by the discriminant locus
\be
\label{discr}
\Delta=4f^{3}+27g^{2} \, = \, 0\text{,}
\ee
where the elliptic curve degenerates. The discriminant locus (\ref{discr})
in general factorizes into several irreducible components:%
\bea
\Delta \spc =\, \underset{\alpha}{%
{\displaystyle\prod}
}\, \Delta_{\alpha}(z_i)\, = \, 0 \text{.}%
\eea
Each vanishing locus $(\Delta_{\alpha}=0)$ defines a hypersurface ${\cal S}_\alpha$ in $B_{3}$ where
a seven-brane is located.
In the direct neighborhood of a seven-brane worldvolume ${\cal S}_\alpha$,
the Calabi-Yau fourfold can be represented as a local $K3$ fibration over ${\cal S}_\alpha$.
The ADE degeneration of the local K3 then dictates the gauge group of
the worldvolume gauge theory on the seven-brane.

Our main interest in the present context will be the study of the worldvolume theory of the
\textquotedblleft GUT seven-brane\textquotedblright\ wrapping the hypersurface
\be
{\cal S}\, =\, \bigl(\spc \Delta_{GUT} =0\, \bigr).
\ee
This seven-brane hosts the gauge sector of the F-theory GUT, which for concreteness we will take
to have gauge group $G=SU(5)$. In accord with the decoupling principle, we will restrict our attention to the case
 that the four-cycle ${\cal S}$ can shrink to zero size within the base $B_3$.  This condition requires that ${\cal S}$ describes
 a del Pezzo surface. The detailed geometric aspects of del Pezzo geometries will not be important to us in what follows,
 except that it is always possible to represent them as intersections of hypersurfaces inside of a toric space.

In the vicinity of ${\cal S}$, we can
switch to a local coordinate $z$ such that $z=0$ indicates the location of the
GUT\ seven-brane. In this case, the presence of an $SU(5)$ gauge theory on the
GUT\ brane indicates the discriminant locally looks as:
\bea
\Delta=z^{5}\underset{\Delta_{\alpha}\neq\Delta_{GUT}}{%
{\displaystyle\prod}
}\Delta_{\alpha}(z_{j},z)\text{.}%
\eea
Here we have treated the local coordinate $z$ somewhat differently from
the other coordinates $z_{j}$ parameterizing the local region of the
threefold. At a typical point at $z=0$, the equation $\Delta=0$ defines an $A_4$ singularity.
Further enhancement in the singularity type indicates
where chiral matter localizes. Each vanishing locus
\be
\Sigma_\alpha \spc = \, \bigl(\spc \Delta_{\alpha}(z_{j},z) = 0\spc \bigr) \cap (z = 0)
\ee
defines a complex curve inside of ${\cal S}$, at which
the GUT\ seven-brane intersects with one or more other  seven-branes.
We will refer to these intersections $\Sigma_\alpha$ as ``matter curves'',
since they give rise to six-dimensional matter. This matter arises
from the ground states of open strings that connect the two stacks of
seven-branes, and correspondingly, are charged as ``bifundamentals''
under the two gauge groups.

The existence of chiral four-dimensional matter can be assured by activating appropriate background fluxes on the worldvolume of
the flavor seven-branes. These fluxes specify that the higher-dimensional matter
transform as sections of vector bundles defined on the matter curves,
and  the  localized chiral zero modes are then  guaranteed to exist and are counted
via the appropriate index theorem. The intersections between  the matter curves indicate additional enhancements of the singularity type,
and give rise to Yukawa couplings of the compactification. We will now briefly describe in a bit more detail
how the matter curves and their intersections can be described from the point of view of the worldvolume gauge theory of the GUT seven-brane.

\subsection{Seven-Brane Gauge Theory}

The worldvolume theory of the GUT seven-brane is given by a partially twisted version of the maximally  supersymmetric
gauge theory defined on ${\mathbb R}^{3,1} \times {\cal S}$. We follow conventions as in \cite{BHVI} (see also \cite{DWI}).
The four-dimensional effective theory is obtained by treating the eight-dimensional fields as a collection of four-dimensional
fields parameterized by points of ${\cal S}$. In \cite{BHVI}, the component form of the
Lagrangian was presented in which the eight-dimensional gauge symmetry was manifest. For brevity, here
we present a formulation of the higher-dimensional gauge theory which is manifestly $\mathcal{N} = 1$ supersymmetric
\cite{SiegelTEND,WackerGregoire}. Organize the eight-dimensional fields according to
four-dimensional $\mathcal{N}=1$ supermultiplets, the gauge sector of the seven-brane theory then
consists of: an adjoint-valued vector multiplet $V$, which represents an ordinary function on
${\cal S}$, a collection of  adjoint-valued chiral superfields $A$, which transform as a
$(0,1)$-form along the internal directions, and a collection of adjoint-valued chiral
superfields $\Phi$ which combine into a $(2,0)$-form on ${\cal S}.$

The eight-dimensional Lagrangian ${\cal L}$ can then be written in four-dimensional superspace notation as
\bea \label{DTERMS}
{\cal L} \, = \, \int \! d^{4}\theta \; {\cal K}  \; + \; \,  \int \! d^{2}\theta\;  {\cal W} \; +\; h.c.\; + \rm{WZW}\, ,
\eea
where the WZW term is a non-local term which vanishes in WZ gauge \cite{SiegelTEND}.
The gauge sector K\"ahler potential decomposes into a contribution from $\Phi$ and $A$:
\begin{equation}
{\cal K}_{\rm gauge} = {\cal K}_{\Phi} + {\cal K}_{A}
\end{equation}
where the K\"ahler potential is fixed by the inner product $(\cdot , \cdot)$ induced by the open string
Hermitian metric associated with the gauge bundles for $\Phi$ and $A$:
\bea
{\cal K}_{\Phi} \is \spc\bigl(\spc\Phi{}^{\dag} \spc e^{-V }\! ,\spc \Phi \, e^{ V} \spc\bigr) \\[3mm]
{\cal K}_{A} \is \spc\bigl(\spc (\ddd +  A)^{\dag}e^{- V}\! , \spc (\ddd + A)\spc e^{V}\spc \bigr)
- (\spc \ddd ^{\dag}e^{- V}\! , \spc \ddd \spc e^{V}\spc \bigr).
\eea
For example, a canonical K\"ahler potential for $\Phi$ is given by the pairing
$(\Phi^{\dag} , \Phi)_{\rm{canonical}}~=~\Phi^{\dag} \Phi$.\footnote{To emphasize the connection with the
usual K\"ahler potential, we deviate from the standard inner product notation, and
include the explicit Hermitian conjugation.} The F-terms~are:
\bea \label{FTERMS}
{\cal W}_{\rm gauge} 
 \is \; 
 {\rm Tr} \spc\bigl(\spc W_\alpha^{2} \spc\bigr)\, 
 \; + \;
{\rm Tr}\spc\bigl(
\spc \Phi \wedge(\ddd A + A\wedge A)\spc \bigr).
\eea
The fields $V$, $A$ and $\Phi$ all have non-trivial profiles
on the internal manifold ${\cal S}$, and thus decompose into a full KK tower of massive four dimensional fields.
The bulk zero mode content will in general depend on the choice of background gauge field flux.
The zero mode spectrum will consist of the constant modes of the vector multiplet, and possibly
additional chiral superfields descending from $\Phi$ and $A$. In realistic applications, one
typically requires the flux to be chosen so as to exclude these additional zero modes.

\subsection{Matter Localization}\label{ssec:LOCALIZE}

The singularity type of an elliptic fibration enhances along the intersection
of seven-branes. Consider two stacks of seven-branes, one wrapping the four-cycle ${\cal S}$ and the other one wrapping a transverse
four-cycle ${\cal S}_\alpha$, with respective gauge groups $G$ and $G_\alpha$. The physics near the intersection curve $\Sigma_\alpha = {\cal S} \cap {\cal S}_\alpha$ can be modelled in terms of a
breaking pattern of $G_{\Sigma}\supset G \times G_{\alpha}$, where
$G_{\Sigma}$ is the gauge group that specifies the enhanced singularity
type along the matter curve $\Sigma_\alpha$ \cite{KatzVafa}. In a local patch on
the GUT seven-brane ${\cal S}$ near the matter
curve we can view both stacks of seven-branes as part of one stack,  and
adopt a $G_{\Sigma}$ gauge theory perspective.
The breaking pattern can then be viewed as induced via a vev $\Phi_0$ of the Higgs scalar field valued in the adjoint of the enhanced gauge group
$G_\Sigma$.  The Higgs field $\Phi_0$ represents the geometric distance between the two seven-brane stacks on
${\cal S}$ and ${\cal S}_\alpha$. Hence it is non-zero everywhere except at the location of the matter curve.

The $G_\Sigma$ gauge theory has the same matter content $(V,\Phi,A)$ as the $G$ gauge theory on the GUT seven-brane,
except that all fields transform as adjoints of the enlarged gauge group $G_\Sigma$.
$\Phi_0$ is a vev of the (2,0) field $\Phi$, that takes the  form of a vortex configuration centered at the matter curve $\Sigma_\alpha$.
In the vicinity of the matter curve, localized fluctuation modes $\Psi_\alpha$ of $\Phi$ and $A_\alpha$ of $A$ satisfy the equations
\cite{BHVI}:
\begin{align}
\label{oldcommutator}
\overline{\partial}_{A}\Psi_\alpha \, +\, \bigl[\spc  \Phi_0,\spc  A_\alpha \bigr]    & =0 \\[2mm] 
k \wedge\partial_{A} A_\alpha \, +\, \bigl[ \spc  \Phi_0^{\dag},\spc \Psi_\alpha \bigr]    & =0 
\end{align}
where $k$ is the K\"ahler form in the patch. The
commutator with the Higgs field gives a mass to $A_\alpha$ and
$\Psi_\alpha$, and modes localize where this commutator vanishes.
These modes organize as six-dimensional fields which represent the
open strings that stretch between the two stacks of seven-branes.

The thickness of the vortex is governed by the slope of the Higgs field, or in geometric terms,
the angle at which the two seven-brane worldvolumes ${\cal S}$ and ${\cal S}_\alpha$ intersect each other.
When this angle is large, open strings that stretch between
the two sets of seven-branes acquire a large mass as soon as they separate from the matter curve.
In the large Higgs field limit, the derivative terms drop out of the equation of motion (\ref{oldcommutator}) of the
fluctuation fields %
\bea
\label{newcommutator}
\bigl[\spc  \Phi_0,\spc A_\alpha \spc \bigr]    \is 0, \qquad \qquad
\bigl[\spc   \Phi^{\dag}_0,\spc \Psi_\alpha \spc \bigr]  \,  = \, 0.
\eea
The commutator equation is equivalent to the algebraic data defining the
location of the holomorphic matter curves.

Working in a neighborhood of the compact curve $\Sigma_\alpha$,  it is enough to consider
a configuration where the Higgs field $\Phi_0$
takes values in the Cartan subalgebra of $G_{\Sigma}$.
Introduce Cartan generators $\mathfrak{h}_{i}$  and generators $\mathfrak{t}_{\alpha}$ parameterized by simple roots
of $G_{\Sigma}$.
For simplicity, let us now assume that the symmetry enhancement at the matter curve ${\cal S}_\alpha$ amounts to intersecting with
just a single seven-brane, so that $G_\alpha = U(1)$.
The matter curve $\Sigma_\alpha$ is then associated with some specified simple root $\alpha$, and
the corresponding localized matter fields can be decomposed as $\Psi = \Psi_\alpha\, {\mathfrak t}_\alpha$ and $A = A_\alpha  {\mathfrak t}_\alpha$.
Using that $[\, \mathfrak{h}_i ,\spc \mathfrak{t}_\alpha\spc ]\spc = \spc \langle \alpha,\mathfrak{h_i}\rangle \spc  \mathfrak{t}_\alpha$, the equation of motion (\ref{newcommutator}) then assumes the form
\bea
\quad \langle \spc \alpha,\spc \Phi_0\rangle\spc A_{\alpha}\is 0\, , \qquad \qquad \langle \spc \alpha,\spc \Phi^\dag_0\rangle\spc \Psi_{\alpha}\, =\, 0\,.
\eea
The fluctuation fields must vanish everywhere except at the location of the matter curve
\bea
\qquad \Sigma_{\alpha}\, =\, \bigl(\, \langle \spc \alpha,  \spc \Phi_0\spc  \rangle =0\, \bigr).
\eea
The large Higgs field limit thus amounts to a delta function approximation: the matter fields $\Psi_\alpha$ and $A_\alpha$ turn into distribution valued fields on ${\cal S}$, transforming in irreducible representations $R_\alpha$ of the unbroken gauge group $G$, specified by the simple root $\alpha$ of the enlarged gauge group associated with the matter curve $\Sigma_\alpha$.\footnote{The representation $R_\alpha$ is specified via the decomposition of the adjoint representation of $G_\Sigma$ into irreducible representations of $G \times G_{\alpha} \subset G_{\Sigma}$: $\rm{ad}(G_{\Sigma}) = \rm{ad}(G) \oplus \rm{ad}(G_{\alpha}) \oplus \oplus_{\tau} (R_{\tau} , R^{\prime}_{\tau})$, so that $\tau = \alpha$ corresponds to one or more of the summands. When $G_{\alpha} = U(1)$, $R_{\alpha}$ is an irreducible representation of $G$.} It is important to note that here we are describing the representation content of a
\textit{six-dimensional} field. Dimensionally reducing, four-dimensional matter fields localized on the curve
will transform in the representation $R_{\alpha}$, as well as the conjugate representation which we denote by $R_{- \alpha}$.
To emphasize this point, we shall sometimes write such fields as $A_{\alpha}$ and $\Psi_{- \alpha}$.

The above description of the matter localization in terms of a Higgs bundle of a gauge theory with an extended gauge group is in general
only valid in a local patch around each matter curve. In principle, one could consider the special situation in which all matter curves admit
a unified description as some appropriate Higgs bundle of some maximal
rank gauge theory defined globally over ${\cal S}$ \cite{DWIII}. The prescribed choice for this
maximal rank  gauge group would be $E_8$, and the local geometry would then take the form of a partially unfolded $E_8$ singularity.
However, the assumption that such a unified perspective exists is very restrictive, and the required global conditions are not easily
satisfied. Instead, we will view the collection of the matter curves and associated matter representations as extra data, external to
the seven-brane gauge theory on ${\cal S}$, and specified by the local geometry of the Calabi-Yau fourfold.

Along the matter curve $\Sigma_\alpha$, the equation of motion (\ref{oldcommutator}) for the localized fields reduces to
\be
\overline{\partial}_{A + A^{\prime}}\Psi_\alpha  = 0, \, \qquad \qquad
k \wedge \partial_{A + A^{\prime}} A_\alpha = 0\, .
\ee
Here $\ddd_{A + A^{\prime}}$ denotes the Dolbault operator associated with the difference of the background gauge fields defined on the two seven-brane stacks, restricted to $\Sigma_\alpha$. The independent solutions to these equations are
the charged massless chiral matter fields of the four-dimensional gauge theory. The number of
zero modes depends on the choice of gauge bundle. This bundle data is dictated by the
breaking patterns of the fibration singularities of the compactification, and
also by the choice of background fluxes. The number of chiral zero modes on a matter curve can be
adjusted by turning on an appropriate background $U(1)$ flux on the transverse seven-brane worldvolume ${\cal S}_\alpha$,
as well as by activating fluxes of the GUT brane.

The intersection of matter curves gives rise to Yukawa couplings in the four-dimensional theory. At the intersection, the singularity type of the ADE fibration enhances further to $G_{p}$. This situation can locally be modelled in terms of Higgsing of a $G_{p}$ gauge theory. Such
intersections can either involve just two matter curves meeting or three intersecting matter curves.
Though the latter possibility may look fine-tuned, both possibilities are generic from the perspective of the $G_p$ gauge theory, and in the geometry this
naturally occurs once the data of the fibration is included.

Putting this all together, the matter sector is described by an ${\cal N}=1$ supersymmetric Lagrangian defined on
${\mathbb R}^{3,1} \times {\cal S}$, specified by a K\"ahler and superpotential of the form \cite{BHVI}
\bea \label{matteraction}
{\cal K}_{\rm matter}\;
 \is  \left( \spc\Psi{}^{\; \dag}_{- \alpha} ,  e^{-V -V^{\prime}}\! \; \Psi_{-\alpha} \right)%
 \, + \,    \left( A_{\alpha}{}^{\!\!\dag}\! , e^{V + V^{\prime}} \! A_{\alpha} \right) %
\\[4mm]
{\cal W}_{\rm matter} \, \is \Psi_{- \alpha} \wedge \ddd_{A + A^{\prime}} A_{\alpha} + f_{\alpha\beta\gamma}\,
  \Psi_\alpha \wedge A_\beta \wedge A_\gamma.
\eea
In the above we have included possible couplings to the gauge fields of the
flavor seven-branes, with their associated vector multiplet $V^{\prime}$ and
internal gauge field $A^{\prime}$. In our conventions, matter fields $A_{\alpha}$ and $\Psi_{- \alpha}$
on a curve $\Sigma_{\alpha}$ respectively transform in the representation $R_{\alpha}$,
and its conjugate $R_{- \alpha}$. Here $f_{\alpha\beta\gamma}$ is only non-zero if the corresponding three matter curves intersect, and if the tensor product
of the three representations $R_\alpha \otimes R_\beta \otimes R_\gamma$ contains the trivial representation; in this case,
$f_{\alpha\beta\gamma}$ denotes the appropriate Clebsch-Gordon coefficient.
The matter fields are all viewed as distributions on ${\cal S}$ that are localized on the associated
matter curves.  Note that the Yukawa couplings involve the overlap of three such distributions. This triple overlap needs
to be regularized, by locally resolving the matter curves as Higgs vortices, as described above. In section \ref{sec:FUZZTHEORY}
we will see that in the non-commutative theory, this triple overlap is automatically well-defined, and the Yukawa couplings
are immediately finite.

\subsection{$U(1)$ Fluxes}

In perturbative IIB string theory, it is convenient to combine the $B$-field
and overall field strength in the direction of the
trace $U(1)$ of a $U(n)$ gauge theory into the quantity $\BB$. At first this may seem less straightforward in F-theory,
because the gauge group on a seven-brane is really $SU(n)$, and not $U(n)$. Let us first note that the
compactification geometry supports non-trivial three-form fluxes, and
so generically non-trivial $\BB$-periods will be present. As explained in the
Introduction, in the decoupling limit, the $\BB$-periods take quantized values.

In gauge theory terms there is also a notion of an overall $U(1)$ which is shared between intersecting seven-branes.
For example, $SU(6)$ contains a $5 \times 5$ $U(5)$ block and a $1 \times 1$ $U(1)$ block which are
subject to the condition that $\det_{U(5)} \times \det_{U(1)} = 1$, which is often denoted as the subgroup $S(U(5) \times U(1)) \subset SU(6)$.
Thus, switching on a flux on the $U(5)$ factor is compensated by a flux on the $U(1)$ factor. In F-theory, intersecting seven-brane configurations can be described by such breaking patterns, so we see that there is still a notion of $\BB$ on a seven-brane.

Activating the flux $\BB$ induces a net D3-brane and five-brane charge on the seven-brane worldvolume. In
an F-theory compactification, the D3-brane tadpole can be cancelled by the Euler character of the Calabi-Yau
fourfold \cite{Sethi:1996es}. Returning to the gauge theory description of $\BB$ fields on
seven-branes, we see that the five-brane tadpole induced by one stack of seven-branes can be
cancelled by a compensating contribution to the net five-brane charge induced by other intersecting seven-branes.

A convenient and by now standard way to break the $SU(5)$ gauge group down to the Standard Model gauge group
is to activate a background flux on the GUT brane, in the direction of the hypercharge $U(1)$ generator \cite{BHVII, DWII}
\be
\label{hyperdec}
{F}_Y = \sum_{a=1}^3 \, {F}_a\,  \mathds{1}_{a},
\ee
where $a=1,2,3$ labels the three SM gauge group factors.
The gauge and matter fields then organize into Standard Model representations. The hypercharge flux also
affects the line bundle data on the matter curves. This opens up the possibility of distinguishing the
Higgs fields of the MSSM from the other chiral matter content. The Higgs fields localize on matter
curves where the $U(1)$ hypercharge flux is non-vanishing, and the other chiral matter of the
MSSM localizes on curves where the net hyperflux vanishes. This gives a natural geometric
solution to the doublet-triplet splitting problem \cite{BHVII}.

The GUT symmetry breaking via hypercharge flux will
in general perturb  exact coupling constant unification. The geometric volume
one would associate with each gauge coupling constant will in general be different, because
the hyperflux contributes to the effective open string metric seen by each gauge group factor.

In the decoupling limit where the closed string volume is taken to zero, the geometric formula for the
three Standard Model gauge couplings is then:
\be
\label{threecouplings}
\frac{4 \pi}{g_{\rm a}^2} \spc =\, \frac{1}{(2\pi)^2g_s }\int_{{\cal S}} \! \BB_a \wedge \BB_a\;
\ee
with $\BB_a = {F}_a +  \BB$ where $\BB$ denotes the ambient two-form flux turned on to support the open string volume of the
cycle. To preserve coupling constant unification at this geometric level, one would need to choose the hyperflux
such that the right-hand side of (\ref{threecouplings}) takes approximately the same value for all three gauge groups. In
models with a realistic spectrum, this condition is already met.

\section{Fuzzy Toric Geometry\label{Fuzztor}}

Our aim in the following sections will be to develop the four-dimensional
field theory of a seven-brane wrapping a non-commutative four-cycle. To
this end, we now develop some tools for dealing with such geometries.

The standard prescription for defining a non-commutative space is to start with the algebra
of functions of a commutative space, and to then deform this structure to a more general algebra.
Though this provides a general way to work with non-commutative geometries, it can also be somewhat unwieldy. In this section we present a practical quantization prescription for working with
non-commutative or ``fuzzy'' geometries which exploits the additional
structure present in toric geometries. In subsequent sections we shall use this prescription
to study the low energy theory of seven-branes wrapping a fuzzy four-cycle.

In the commutative setting, one convenient way to realize a wide
variety of possible compact spaces is via toric constructions.
The main idea is that the compact space of
interest, such as the ones that are relevant to F-theory, can be viewed as the
intersection of hypersurfaces inside of a larger toric space.  Our strategy in
this section will be to first develop a suitable non-commutative analogue of toric spaces. This prescription
can be applied in situations where all toric coordinates are non-commutative, and also in situations where some
subset are non-commutative, while the others remain commutative.
Subspaces of these fuzzy toric geometries will then provide us with a
prescription for realizing non-commutative K\"{a}hler surfaces.
The construction we present bears a formal resemblance to that given in \cite{Iqbal:2003ds}, though as far as we are
aware, the geometric interpretation presented here has not appeared in the
literature before.\footnote{For another approach to defining non-commutative toric geometries, see
\cite{Saemann:2006gf}.} We shall therefore proceed somewhat systematically.

Let us first review classical toric geometry. In physics terms, toric geometries
are characterized by the gauged linear sigma model (GLSM). Recall that in a GLSM, there are $r$ chiral
superfields $z_{1},...,z_{r}$, and a gauge group $U(1)^{s}=U(1)_{(1)}%
\times...\times U(1)_{(s)}$ under which these chiral superfields are charged.
Hence each coordinate $z_{i}$ is labeled by an $s$-component charge vector $\overrightarrow
{q}_{i}$. In addition, the $z_{i}$ obey $s$
D-term constraints:%
\bea
\underset{i=1}{\overset{r}{%
{\displaystyle\sum}
}}\, \overrightarrow{\spc q}_{i}\left\vert z_{i}\right\vert ^{2}=\, \overrightarrow
{\spc \zeta}. \label{Dtermcon}%
\eea
The classical vacua of the $1+1$ dimensional theory then correspond to gauge
equivalent orbits of the $z_{i}$ subject to the D-term constraints of line
(\ref{Dtermcon}). These vacua then define the classical geometry of the
symplectic quotient
\be
X (\overrightarrow{\spc \zeta}\spc) =%
\mathbb{C}
^{r}/\! /\spc U(1)^{s}
\ee which is an $r-s$ complex dimensional manifold. More general spaces are obtained by
considering hypersurfaces in $X$, specified as the vanishing locus of one or more
weighted homogeneous polynomials $P_{\overrightarrow{w}}(z_{1},...,z_{r})$.
The weights $\vec{w}$ are constrained by the condition
that $P_{\vec w}$ has a well-defined GLSM\ charge.

This toric characterization can be straightforwardly
quantized and used to construct a general class of non-commutative geometries.  In principle one
would need to show that this particular quantization prescription is the correct one for
describing the zero slope limit of an open string theory on a wrapped seven-brane in a $\BB$-field.
We will not try to give such a derivation here, but do note that the GLSM formulation given here
appears to be a particularly suitable starting point for making such an attempt.

\def\FF{{\cal F}}
\subsection{Quantization Procedure}

We now provide a quantization prescription for this toric setup. In general terms, the data necessary to quantize our system
consists of a symplectic form $\omega$ defined on the ambient space $\mathbb{C}^{r}$, which we use to define commutation relations.
The D-term constraint of the classical toric geometry is then promoted to the
Hamiltonian constraint of the quantized system. Imposing the Hamiltonian constraint on the state space associated with
the quantized ambient space produces the non-commutative toric space $X$.

Let us now describe each step in more detail. We begin by quantizing $\mathbb{C}^{r}$, by viewing it as a symplectic manifold
with a closed symplectic form $\omega$. To keep our treatment as simple as possible, we choose $\omega$ to be a (1,1)-form.
In the F-theory context, this means that we assume that the $\BB$-flux is a $(1,1)$-form
on the seven-brane worldvolume.\footnote{For a discussion of local F-theory models with non-commutativity induced by a
holomorphic bi-vector, see \cite{FGUTSNC,Marchesano:2009rz}.}
Assuming $\omega$ does not vanish, Darboux's theorem ensures that there exists
a general coordinate redefinition on $\mathbb{C}^r$ such that $\omega$ takes the form
\be
\label{symp}
\omega \, =\,   \sum_{i=1}^{r} i dz_i \wedge d\overline{z}_i
\ee
This defines a Poisson bracket for the $z$'s and $\overline{z}$'s: $\{z_i, \overline{z}_j\}_{PB} = \delta_{ij}$. This means that the metric on $\mathbb{C}^r$ takes the
general form
\be
ds^2 = \sum_{i,j=1}^{r} G_{ij} dz_i d\overline{z}_j
\ee
with $G_{ij}$ some general matrix. In this section, we will mostly focus on the symplectic geometry. In particular, we shall  define the toric $U(1)^s$ action
on the $z_i$'s and the GSLM D-term constraint with respect to this symplectic coordinate basis.

We now take the quantization step and replace each coordinate $z_i$ and $\overline{z}_j$ by operators $Z_{i}$ and $Z^\dag_j$ obeying the commutation relations:%
\bea
\label{nco}
\qquad \bigl[  Z_{i},Z^{\dag}_{j}\bigr]    \is  \hbar_{NC} \spc \delta_{ij}\, , 
\\[3mm]
\bigl[  Z_{i},Z_{j}\bigr] \is \bigl[  Z^\dag_{i},Z^\dag_{j}\bigr]\, = \, 0\, , 
\eea
for $i,j=1,...,r$. The parameter $\hbar_{NC}$ is Planck's constant, which here has the dimension of an area: it sets the size of the Planck cell on the fuzzy space. For convenience, we shall mostly work in units where $\hbar_{NC}=1$.

To build up the quantized geometry, we introduce a vacuum state $\left\vert 0\right\rangle $
annihilated by the $Z_{i}$. Acting with creation operators on the vacuum realizes a Fock space%
\bea
\FF\spc(\mathbb{C}^r) =\text{span}\left\{ \; \underset{i=1}{\overset{r}{%
{\displaystyle\prod}\,
}}\frac{\bigl(  Z_{i}^{\dag}\bigr)  ^{n_{i}}}{\sqrt{n_{i}!}}\left\vert\,
0\,\right\rangle \; \right\}  .
\eea
The basis elements of $\FF(\mathbb{C}^r)$ can be thought of as the quantized points
of $%
\mathbb{C}
^{r},$ and the Fock space is obtained by treating
these points as basis elements of a linear vector space. For brevity,
we shall often refer to these vector spaces as defining the non-commutative space
itself, since the distinction will be largely unimportant in what follows.

We now want to take the symplectic quotient, and define the non-commutative version of the
toric space $X(\vec{\zeta})$. As we will see, the FI-parameters $\vec{\zeta}$ that determine the overall size and
shape of  $X(\vec{\zeta})$ will be quantized (in the sense that they can only take integer values)
via the requirement that the various compact two-cycles inside  $X$ support an integer number of Planck cells. Treating the $Z_{i}$ as the quantized analogues of the GLSM\ fields, there is
an associated set of $U(1)^{s}$ charges for each $Z_{i}$. Concretely,
the vector of D-term operators
\bea
\overrightarrow{\spc D}=\underset{i=1}{\overset{r}{%
{\displaystyle\sum}
}}\; \overrightarrow{\spc q}_{i}\spc Z_{i}^{\dag}Z_{i},
\eea
now assumes the natural role of the vector of conserved charges that generate the $U(1)^s$ rotations.
These conserved charges will act like Hamiltonians on the large phase space ${\mathbb C}^r$. The toric space is obtained by treating
these Hamiltonians as operator constraints.

The Fock space $\FF(\mathbb{C}^r)$ admits a grading in terms of the
spectrum of the $\overrightarrow{\spc D}$ operator%
\begin{equation}
\FF\spc(\mathbb{C}^r)=\underset{\overrightarrow{\spc \zeta}}{\mbox{\LARGE $\oplus$}}\, \FF_X%
(\overrightarrow{\spc \zeta}\spc )
\end{equation}
so that for all states $|\spc \Psi\spc \rangle$ in the subspace $\FF_X(\spc\overrightarrow{\zeta}\, ),$ the  vector
$\overrightarrow{D}$ takes the fixed value~$\overrightarrow{\spc \zeta}$:
\be
\label{dterm}
\overrightarrow{\spc D} \spc | \spc \Psi \spc \rangle \spc = \spc \overrightarrow{\spc \zeta} \, |\spc \Psi \spc \rangle\,. 
\ee
This defines the quantized analogue of the D-term equation (\ref{Dtermcon}) in the commuting case. Geometrically,
the $\overrightarrow{\spc \zeta}$'s define quantized versions of the
FI\ parameters of the GLSM, and thus specify the  K\"ahler data of the non-commutative
toric space $X(\overrightarrow{\spc \zeta}\spc)$. Roughly speaking, we
view $\mathcal{F}_{X}(\overrightarrow{\spc \zeta})$ as the vector space of fuzzy points on $X(\overrightarrow{\spc \zeta}\spc)$.\footnote{Later
when we discuss differential forms we will extend this definition to also include ``fermionic  points''. Since the context will hopefully be clear, we shall use similar notation to refer to both the bosonic space, and its fermionic extension.}
We recover the commutative geometry in the limit in which we simultaneously
rescale all of the components of the vector $\overrightarrow{\zeta}$ to
infinity, while also requiring that the quantization parameter $\hbar_{NC}$ tends
to zero.

\begin{figure}[t]\label{fuzzP1P1}
\begin{center}
 \epsfig{figure=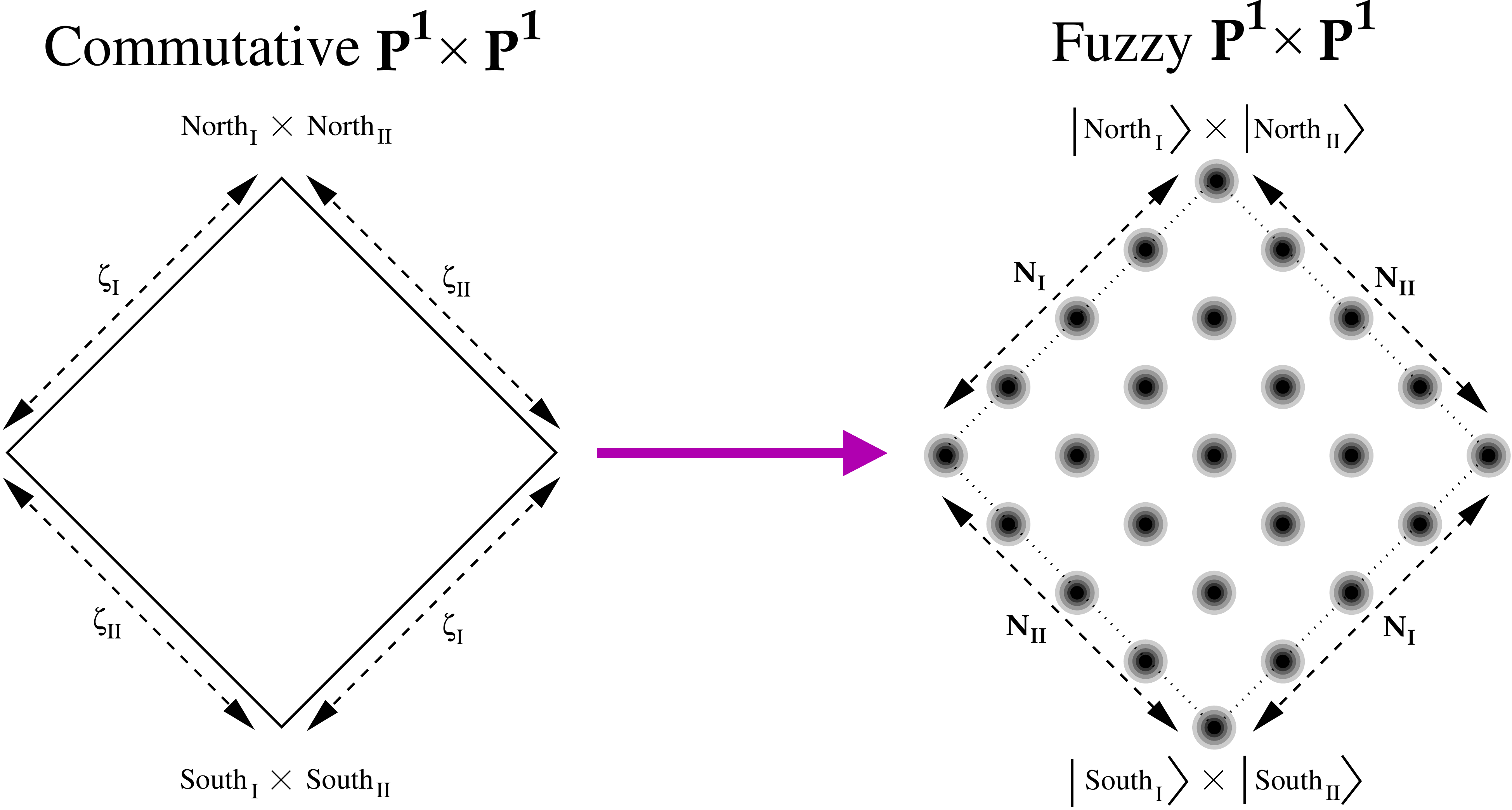, scale=.4}%
\caption{Toric diagram for commutative $\mathbb{P}^{1} \times \mathbb{P}^{1}$ and its fuzzy analogue. The commutative geometry is described
  by the classical vacua of a $U(1) \times U(1)$ GLSM with four chiral superfields $u_1$, $u_2$, $v_1$, $v_2$ with GLSM charges $(+1,0)$ for the $u_i$'s and $(0,+1)$ for the $v_i$'s. Viewing the norms $|u_{i}|^{2}$ and $|v_{i}|^{2}$ as coordinates of $\mathbb{R}^{4}_{\geq 0}$, the D-term constraints $|u_1|^2 + |u_2|^2 = \zeta_I$ and $|v_1|^2 + |v_2|^2 = \zeta_{II}$ define two hypersurfaces which intersect over a two-dimensional compact subspace. Here we have projected this space onto a two-dimensional plane, where it corresponds to a rectangle with sides of lengths $\zeta_I$ and $\zeta_{II}$. Each side corresponds to a $\mathbb{P}^{1}$ factor of the geometry. In the non-commutative theory, the D-term becomes a quantized Hamiltonian constraint, and the state space of points is finite-dimensional. The North and South poles of the commutative $\mathbb{P}^{1}$'s then become highest and lowest $su(2)$ angular momentum
  states of the fuzzy theory. See Appendix A for further discussion of fuzzy $\mathbb{P}^{1}$.}%
\end{center}
\end{figure}

The sign and magnitude of the FI parameters determine the compact and non-compact directions of the geometry. Indeed, the values of the FI parameters determine the K\"{a}hler volumes for the toric space. Viewing the norms $|z_i|^2$ as coordinates of $\mathbb{R}^{r}_{\geq 0}$, each D-term constraint defines a hyperplane in this space, and the toric space lies at the mutual intersection of all the hyperplanes. In the non-commutative setting, the intersection of these hyperplanes is replaced by discretized data. In particular, this means that finite area intersections of hyperplanes translate to subspaces with an upper bound on the oscillator number. Non-compact intersections of hyperplanes correspond to subspaces with no such bound. Changing the values of the FI parameters can cause the geometry to undergo a flop transition. Such transitions correspond to changing the oscillator content of the Fock space. See section \ref{sec:FORMAL} for further discussion of a fuzzy flop transition.

For compact toric manifolds, the subspaces $\FF_X(\overrightarrow{\spc \zeta}\spc )$ are finite dimensional.
The basis elements of $\FF_X(\overrightarrow{\spc \zeta}\spc )$ correspond to the
points of the fuzzy toric space $X(\overrightarrow{\spc \zeta}\spc )$, and hence the total number of
points equals the dimension of this vector space %
\bea
\# \; \text{points in} \, X(\overrightarrow{\spc \zeta}\spc) \equiv \dim\FF_X(\overrightarrow{\spc \zeta}\spc ) \, .
\eea
From a semi-classical viewpoint, this equals the number of Planck cells in $X$.

\subsection{Functions and Line Bundles}

Having specified the formal spaces of interest, we now introduce functions
and line bundles over a non-commutative toric variety $X(\vec{\zeta}\spc )$.

\bigskip

\noindent
{\it Fuzzy Functions }

In commutative geometry, a complex function on a space $X$ is defined as a map from
$X$ to $\mathbb{C}$. In the non-commutative setting, one can similarly consider linear maps from
the state space ${\mathcal F}_X(\vec{\zeta})$ to $\mathbb{C}$. The collection of all such maps is
a linear space with the same dimension as the state space
${\mathcal F}_X(\vec{\zeta})$ itself. There is, however, a more substantive
notion of functions on a fuzzy space, namely as the space of functions of the non-commutative coordinates
on $X$.

Since the fuzzy coordinates do not commute among each other, we cannot view them as having
some precise  classical value, but rather as operators that act non-trivially on the other coordinates.
Taking this into account, it is natural to define the space of non-commutative functions as the space of linear maps
from the state space ${\cal F}_X(\vec\zeta)$ to itself. Indeed, every linear operator from the state
space ${\cal F}_X(\vec\zeta)$ to itself can be presented in terms of polynomials built from
the creation and annihilation operators $Z^{\dag}$ and $Z$. More precisely, they are given by
polynomials $F(Z^\dag\! , Z)$, which have zero GLSM\ charge:%
\bea
\Omega^{0}(X)\equiv \left\{  \;
F(Z^{\dag},Z)\,  : \
 \bigl[ \, \overrightarrow
{D}\spc ,\,F \, \bigr]  =0 \; \, \right\}  .
\eea
A matrix representative for an element of $\mathcal{F}_X%
(\overrightarrow{\zeta}\spc)\otimes\mathcal{F}_X^{\ast}(\overrightarrow{\zeta}\spc)$ is
then given by evaluating the operator in a specified basis of states.

On the function space we can define a natural non-commutative star product,
given by operator multiplication
\bea
\label{starproduct}
(F * G)(Z^{\dag},Z) = F(Z^\dag,Z) \cdot G(Z^\dag,Z)
\eea
Since functions are defined as normal ordered expressions,
the right-hand side of (\ref{starproduct}) must be decomposed into a sum of normal ordered terms via Wick contractions.
In the matrix representation of functions, the star product is simply given by matrix multiplication. In the following we will not explicitly write the
star product $*$, as it will be automatically understood that functions are normal ordered expressions in the oscillators.

An important feature of our present discussion is that when all of the
GLSM charges are positive, it is not possible to write down any functions
which are purely holomorphic in the $Z$'s. This reflects the fact that on a
space of positive curvature, a holomorphic function will have a pole somewhere
on the surface. Holomorphic sections of non-trivial line bundles correspond
to holomorphic polynomials in the $Z$'s.

\bigskip

\noindent
{\it Fuzzy Line Bundles. }

We now present a prescription for defining line bundles on a fuzzy toric space.
Roughly speaking, non-trivial line bundles can be thought of as non-square
matrices.\footnote{Line bundles on fuzzy $\mathbb{P}^{1}$ and the generalization to fuzzy $\mathbb{P}^{n}$ have
been discussed in \cite{Grosse:1995jt} and \cite{Dolan:2006tx}.}
In terms of GLSM data, we can characterize the data of a line bundle in terms of operators built from the $z_i$'s. In particular, we can organize all operators
according to their GLSM charges. The space of operators with the same GLSM charge $\overrightarrow{Q}$ then define sections of a line bundle of degree $\overrightarrow{Q}$, which we call $L^{(\overrightarrow{Q})}$.

In the non-commutative setting these sections are represented by polynomials in the $Z_i$ and $Z_i^\dag$ oscillators with
GLSM\ charge $\overrightarrow{Q}$. In other words, the space of sections valued in
the line bundle $L^{(\overrightarrow{Q})}$ is defined as the space of polynomials $G(Z^{\dag
},Z)$ that satisfy\footnote{Our sign conventions are chosen to maximally adhere to the
sign conventions in the commutative theory. For example, a positive degree line bundle
on $\mathbb{P}^{1}$ has holomorphic sections.}
\bea
L^{(\overrightarrow{Q})} = \,   \left\{  \;  G(Z^\dag , Z) \, : \
\bigl[\,  \overrightarrow{D}\spc ,\, G
\, \bigr]  = - \overrightarrow{Q}\, G \; \right\}
\, .
\eea
Expressions given purely in terms of the $Z$'s with no $Z^{\dag}$'s correspond to
holomorphic sections of non-trivial line bundles.

Fixing the K\"ahler class of the fuzzy space, we can present sections
of the line bundle $L^{(\overrightarrow{Q})}$ in terms
of a collection of rectangular matrices valued in
$\mathcal{F}_X(\overrightarrow{\zeta} - \overrightarrow{Q})\otimes\mathcal{F}_X^{\ast}(\overrightarrow{\zeta})$.

\def\oC{\overline{C}}

\subsection{Fuzzy p-Forms}

We now construct differential forms on a non-commutative toric space $X$. Our strategy will be to first
construct the requisite geometry on $\mathbb{C}^{r}$, and to then perform the symplectic quotient to construct
differential forms on $X$.

Let us begin with differential forms on $\mathbb{C}^{r}$. To this end, introduce a set of fermionic oscillators
obeying the anti-commutation relations:%
\bea
\quad \bigl\{  C_{i},C_{j}^{\dag}\bigr\}    \is  \hbar_{NC}\spc \delta_{ij}\, , 
\\[3mm]
\bigl\{  C_{i},C_{j}\bigr\} \is \bigl\{  C^\dag_{i},C^\dag_{j}\bigr\}\, = \, 0\, , 
\eea
for $i,j=1,...,r$. In what follows we continue to set $\hbar_{NC}=1$. The $C_i$'s are the quantized versions of the differentials $dz_i$,
and the $C^\dag_i$'s are the quantized analogues of the $d\overline{z}_i$'s. Note that with this prescription the non-commutative analogue
of the identity $dz \wedge d\overline{z} = - d\overline{z} \wedge dz$ is violated due to Wick contraction between $C$ and $C^{\dag}$. This
additional non-commutative structure provides a covariant extension of functions to differentials. We can view the
oscillators $(Z_i, Z_j^\dag,C_i,C_j^\dag)$ as coordinates on the quantized cotangent bundle $T^{\ast}\mathbb{C}^{r}$.

Introducing a vacuum state $\left\vert 0 \right\rangle$ annihilated by the $Z_{i}$ and the $C_{i}$, the Fock
space of states spanned by acting with $Z_{i}^{\dag}$ and $C_{i}^{\dag}$
then provides a natural extension of the fuzzy points of $\mathbb{C}^{r}$:
\bea
\FF(T^{\ast}_{\rm{hol}}{\mathbb C}^{r}) =\text{span}\left\{ \; \underset{i=1}{\overset{r}{%
{\displaystyle\prod}\,
}}\frac{\bigl(  Z_{i}^{\dag}\bigr)^{n_{i}}}{\sqrt{n_{i}!}} \bigl(C^\dag_i\bigr)^{\sigma_i}\left\vert\,
0\,\right\rangle \; \right\}  .
\eea
Geometrically, the subspace of states with fermion number zero corresponds to the space of fuzzy points on $\mathbb{C}^{r}$, and
the subspace of states with fermion number zero or one corresponds to the fuzzy points of the holomorphic cotangent bundle
$T^{\ast}_{\rm{hol}}\mathbb{C}^{r}$. Here, we are viewing the cotangent bundle as a phase space in which the bosonic points correspond
to the ``position coordinates'' and the fermion number one points are the ``momentum coordinates''.
Similarly, the $n$-fermion subspace is the $n$th exterior power
$\Lambda^{n} T^{\ast}_{\rm{hol}}\mathbb{C}^{r}$. The Fock space corresponds
to exterior powers of the holomorphic cotangent bundle. We view the Hermitian
conjugate space of states as the fuzzy points
of the anti-holomorphic cotangent bundle of $\mathbb{C}^{r}$ and its exterior powers:
\begin{equation}
\FF(T^{\ast}_{\overline{\rm{hol}}}{\mathbb C}^{r})
= \FF(T^{\ast}_{\rm{hol}}{\mathbb C}^{r})^{\ast}.
\end{equation}

Differential forms of ${\mathbb C}^{r}$ correspond to maps which send one fuzzy point of
$\FF(T^{\ast}_{\rm{hol}}{\mathbb C}^{r})$ to another. The most general operator of this type can be expanded as a power series
in the oscillators which is normal ordered in the $Z$'s, and anti-normal ordered in the $C$'s.
An operator with $p$ $C$ oscillators and $q$ $C^{\dag}$ oscillators is then given by the $(p,q)$-form:
\begin{equation}
\omega_{(p,q)}=\underset{I,J}{\sum}f_{I,J}(Z^{\dag},Z)\; C_{i_{1}}%
\wedge...\wedge C_{i_{p}}\wedge C^{\dag}_{j_{1}}\wedge...\wedge
C^{\dag}_{j_{q}} \label{fuzzform}%
\end{equation}
where $I = (i_1,...,i_p)$ and $J = (j_1,...,j_q)$ are multi-indices running over $i$ and $j$ holomorphic and
anti-holomorphic indices respectively.

The space of all $(p,q)$-forms admits a $\mathbb{Z}_{2}$ grading according to whether there are an even or odd number of fermionic oscillators.
The wedge product of differential forms is given by operator multiplication
\bea
\bigl(\omega_1 \wedge_{\ast} \omega_2\bigr)(Z^\dag,Z, C,C^\dag)\; =\; \omega_1(Z^\dag,Z, C,C^\dag) \cdot \omega_2(Z^\dag,Z, C,C^\dag)
\eea
This non-commutative wedge product preserves the $\mathbb{Z}_2$ grading.
Note, however, that the $\wedge_{\ast}$-product involves non-trivial Wick contractions between the fermionic
oscillators. This means that, as opposed to the commutative theory,  the wedge product of a $(p,q)$- and $(p^{\prime} , q^{\prime})$-form
will generally be of mixed $(p,q)$-type. This feature should be viewed as additional
non-commutative structure along the cotangent space direction of the fuzzy space.

\subsubsection*{Dolbeault Operators}
\vspace{-2mm}

On the space of differential forms on $\mathbb{C}^{r}$, we can define the operators $\overline{\partial}$ and $\partial$ which
act on a form $\omega$ as:
\bea \label{diffops}
\ddd \omega \is \, \sum_{i=1}^r  \bigl[ \spc Z_i, \spc \omega\spc \bigr] \spc C^\dag_i\, , \qquad \quad
\partial \omega \, = \,  \sum_{i=1}^r C_i \bigl[ \spc Z^\dag_i, \spc \omega\spc \bigr] \, .
\eea
By virtue of the non-trivial commutator between $Z$ and $Z^\dag$, this acts as a differential operator.
As opposed to the commutative theory, $\partial$ and $\ddd$ do not commute.  Nevertheless, we can still define Dolbeault cohomology.
Just as in the commutative theory, the operators $\overline{\partial}$ and $\partial$
map $(p,q)$ forms to $(p,q+1)$ and $(p+1,q)$-forms, and are both nilpotent:
\bea
\overline{\partial}^{2} = 0 \qquad \text{and}\qquad \partial^{2} = 0.
\eea
Note that when $\overline{\partial}$ acts on $(0,q)$-forms, the definition given in equation
(\ref{diffops}) can also be written as a super-commutator with the operator $
\overline{\partial} = \sum_{i} Z_{i} C_{i}^{\dag},$ which manifestly satisfies $\ddd ^{2} = 0$.
Thus, we can also speak of a Dolbeault complex in this case as well. This latter version
of $\ddd$ can then be interpreted as a BRST operator. As a matter of notation, we will sometimes write
\bea
\overline{\partial} \omega_{(0,q)} \spc= \spc\bigl[\spc \overline{\partial},\spc \omega_{(0,q)}\bigr\}, \qquad \quad
\;\; \partial \omega_{(p,0)} \spc = \spc\bigl[\spc \partial, \omega_{(p,0) }\bigr\}.
\eea

\subsubsection*{p-Forms on Toric Spaces}
\vspace{-2mm}

We now turn to differential forms on the symplectic quotient $X =
\mathbb{C}^r/\!/U(1)^s$. To frame our discussion, first
recall how differentials are constructed in the symplectic quotient of the
commutative geometry. There, the restriction to the cotangent
bundle $T^*X$ requires us to impose the projection onto the subspace of the
holomorphic cotangent bundle orthogonal to  the Hamiltonian vector fields
associated with the
$U(1)^s$ action, and to the gradients $d{\vec D}$ of the D-term constraint.
Concretely, this means that we need to project the forms onto those
components that are orthogonal to the holomorphic and anti-holomorphic
gradients $\partial \overrightarrow{\spc D}$ and $\ddd \overrightarrow{\spc
D}$ of the D-terms.

We now extend our Hamiltonian constraints to the
extended phase space. In the GLSM, we can consider operators made from
bosons, but also operators involving fermions. These bosons and fermions
combine into supermultiplets of the two-dimensional gauge theory. Thus, we
see that the fermions also possess a GLSM charge, and our Hamiltonian
constraint must be modified to reflect this fact. The natural extension of
the original Hamiltonian constraint is then given by extending the
definition of the D-term constraint operators $\overrightarrow{\spc D}$ to
act on the fermionic coordinates:
\bea
\label{newd}
\overrightarrow{\spc D}\; =\; \underset{i=1}{\overset{r}{%
{\displaystyle\sum}
}}\; \, \bigl(\, \overrightarrow{\spc q}_{i}\,
 Z_{i}^{\dag}Z_{i} \spc + \spc \overrightarrow{\spc q}_i C^\dag_i C_i  \,
\bigr)
\eea
This operator $\overrightarrow{\spc D}$ acts  on functions and forms via the
commutator
${\rm ad} (\overrightarrow{\spc D}) \, \omega \equiv [ \overrightarrow{\spc
D} , \omega]$.
The constraint on the rest of the ``phase space'' for the holomorphic
cotangent bundle of $X$ is enforced by the operator:
\bea
\partial {\vec D} = \sum_{i} {\vec q}_{i} Z_{i}^{\dag} C_{i}.
\eea
The ket states $\left\vert \Psi \right\rangle$  that define the holomorphic
cotangent bundle and its exterior powers
$\Lambda^{\bullet}T^{\ast}_{\rm{hol}}X$ are then given by the subspace of
$\mathcal{F}(T^{\ast}_{\rm{hol}} \mathbb{C}^{r})$ which
satisfy the D-term constraint and are annihilated by $\partial {\vec D}$:
\bea
\label{ketcon}
{\vec D} \left\vert \Psi \right\rangle = {\vec \zeta} \left\vert \Psi
\right\rangle \, , \qquad\ \text{and} \qquad \partial {\vec D} \left\vert
\Psi \right\rangle = 0.
\eea

Importantly, all of the original bosonic fuzzy points are retained by this
extension. Indeed, what has been added is the additional structure of the
holomorphic cotangent bundle.
Further note that precisely because this is a holomorphic object,
it is appropriate to only enforce a vanishing condition from
$\partial {\vec D}$, and \textit{not} its Hermitian conjugate:
\bea
\overline{\partial} {\vec D} = \sum_{i} {\vec q}_{i} Z_{i} C_{i}^{\dag}.
\eea
To avoid cluttering notation, we shall refer to the entire state space
defined in this way as $\mathcal{F}_{X}({\vec \zeta})$.

The bra states $\langle \Psi |$ which are elements of the anti-holomorphic cotangent
bundle and its exterior powers $\Lambda^{\bullet}T^{\ast}_{\overline{\rm{hol}}}X$ similarly satisfy
\bea
\label{bracon}
\langle \Psi | \spc {\vec D}\spc  = \spc \langle \spc
\Psi\spc| \spc\vec{\zeta}  \, ,  \qquad\ \text{and} \qquad \langle\spc \Psi\spc |\spc \ddd
{\vec D} \spc  =\, 0\, .
\eea

Differential forms on $X$ are given by operators which send a state of
$\mathcal{F}_{X}(\vec{\zeta}\spc )$ to another state of
$\mathcal{F}_{X}(\vec{\zeta}\spc )$. Compatibility with the
Hamiltonian constraints (\ref{ketcon}) imposed by ${\vec D}$ and
$\partial{\vec D}$ require that a $(p,q)$-form $\omega_{(p,q)}$ on $X$
satisfies:
\bea
\label{dcomcon}
\bigl[\, \overrightarrow{D},\omega_{(p,q)}\bigr]\, =\, 0 \quad\ \text{and}
\qquad \bigl[\, \partial\overrightarrow{D},\omega_{(p,q)}\bigr]\, =\, 0.
\eea
In other words, $\omega_{(p,q)}$ must have zero GLSM charge and preserve the
structure of $T_{\rm{hol}}^{\ast}X$.

Note that in (\ref{dcomcon}) we do not impose that $\omega_{(p,q)}$ commutes
with  $\ddd{\vec D}$.
So when acting to the left, the form does not need to preserve the hermitian
conjugate GLSM constraints (\ref{bracon}). Our definition of differential
forms on $X$
thus looks  asymmetric under complex conjugation on $X$. However, the
following observation restores the symmetry.  Physical quantities are given
by the matrix element of (products of) differential forms
between bra and ket states, that satisfy (\ref{ketcon}) and
(\ref{bracon}). Consider such a matrix element
\bea
\label{ommatrix}
\langle \spc \mu \spc | \, \omega_{(p,q)} |\spc \nu \spc \rangle.
\eea
The bra states span the dual space to ${\cal F}_X(\vec{\zeta})$, in the
sense that {\it any} bra state in the big Fock space ${\cal
F}(T^{\ast}_{\rm{\overline{hol}}} \mathbb{C}^{r})$ can be decomposed into an
element of ${\cal F}^*_X(\vec{\zeta})$ -- that is, a state that satisfies
the conditions
(\ref{bracon}) --  plus a state in the orthocomplement of
${\cal F}^{\ast}_X(\vec{\zeta})$. So in equation (\ref{ommatrix}), the inner product with the
ket state $|\nu\rangle$
automatically projects the state $\langle \spc \mu \spc | \, \omega_{(p,q)}$
along the component in ${\cal F}^*_X(\vec{\zeta})$, that satisfies
(\ref{bracon}).
By explicitly implementing this projection, we can map a fuzzy differential
form $\omega_{(p,q)}$ on $X$ in the `holomorphic representation'
(\ref{dcomcon}) into an equivalent differential form
$\overline{\omega}_{(p,q)}$ on $X$ in the `anti-holomorphic representation',
satisfying
\bea
\bigl[\, \overrightarrow{D},\overline{\omega}_{(p,q)}\bigr]\, =\, 0 \quad\
\text{and}
\qquad \bigl[\, \ddd\overrightarrow{D},\overline{\omega}_{(p,q)}\bigr]\, =\,
0.
\eea

Next consider the Dolbeault operator on $X$. The adjoint action of the
D-term
constraint $\overrightarrow{\spc D}$ and the Dolbeault operators on
functions and forms are compatible, in the sense that
\bea
\label{compat}
\bigl[ \spc {\rm ad}\bigl(\vec D\bigr), \ddd\, \bigr] \, \is \, \bigl[\spc
{\rm ad} \bigl(\vec D\bigr), \partial\spc \bigr]\, = \, 0\,
\eea
This property guarantees that the conditions of line (\ref{dcomcon})
are preserved by the action of the Dolbeault operators.
In other words, $\partial$ and $\ddd$ map differential forms on $X$ to
differential forms on $X$, and thus properly project
to Dolbeault operators on $X$. Just as for the commutative theory, the
explicit form of the
Dolbeault operator on $X$ will in general not be as simple as its
presentation on $\mathbb{C}^{r}$.

This provides us with a prescription for constructing differential forms
on~$X$.
In general, given a differential form of the commutative geometry, there is
a corresponding
differential form on $X$, given as a polynomial expression in the $Z$ and
$C$ oscillators.
Since Dolbeault cohomology is also retained in the fuzzy theory, this means
that in practice,
there is a direct translation of all of the necessary algebraic data of the
commutative geometry.

At a practical level, all of the differential forms on $X$ can be viewed as
descending from $(p,q)$-forms on $\mathbb{C}^{r}$. Viewing $X$ as embedded in $\mathbb{C}^{r}$, there is a
corresponding projection operator given by summing over an orthonormal basis
of
states for $\mathcal{F}_{X}(\vec{\zeta})$:
\bea\label{projop}
 \pi_{X}(\vec{\zeta}) =\underset{\left\vert \mu\right\rangle
\in\mathcal{F}_{X}%
(\overrightarrow{\zeta})}{\sum}\left\vert \mu\right\rangle
\left\langle\mu\right\vert .
\eea
Given a $(p,q)$-form $\omega_{(p,q)}$ on $\mathbb{C}^{r}$, the restriction
to $X(\vec{\zeta})$ is given
by sandwiching between projection operators:
\bea
\omega_{(p,q)}|_{X(\vec{\zeta})} = \pi_{X}(\vec{\zeta}) \cdot \omega_{(p,q)} \cdot
 \pi_X(\vec{\zeta}).
\eea

For actual computations, the explicit basis of differential forms for a
toric space will
depend on details of the geometry. Indeed, there is a certain amount of
redundancy built into our description because  there are more $Z$ and
$Z^\dag$ coordinates than the dimension of the
fuzzy toric space. For explicit computations, it is often convenient to
eliminate this redundancy. To do this,
we choose a collection of holomorphic and anti-holomorphic monomials in
the $Z$'s and ${Z}^\dag$'s which we refer to as $E_{I}$ and
$\overline{E}_{J}$. The $E_{k}$'s should correspond to monomials in the
$Z$'s such that for each monomial, it is not possible for all of the $Z$'s
to
simultaneously vanish (while satisfying the D-term constraint). The
collection of such $E$'s determines the Stanley-Reisner
ideal of the toric variety $X$ \cite{FultonToric}.\footnote{See \cite{Blumenhagen:2010pv}
for a recent discussion of Stanley-Reisner ideals
in the physics literature.} We then introduce a collection of differentials:
$dE_{I} = [\partial, E_{I}]$ and $d\overline{E}_{J} = [\ddd,
\overline{E}_J]\, .$
A priori there could be additional non-vanishing differentials on $X$, which
we can add to the
collection of $dE$'s and $d\overline{E}$'s. This is really an issue in the
commutative geometry,
so we shall not dwell on this subtlety here. A general differential form on
$X$ can then be expanded in
terms of this basis as
\begin{equation}\label{pqSR}
\omega_{(p,q)}=\underset{I,J}{\sum} : f_{I,J}(Z^{\dag},Z)\otimes\wedge_{I}dE_{I}%
\wedge_{J}d\overline{E}_{J} : .%
\end{equation}

In the following, we will also consider $(p,q)$-forms on $X$ that take
values in a non-trivial line bundle $L^{(\overrightarrow{Q})}$.
These are characterized as forms with non-zero $U(1)^s$ GLSM charge:
\begin{equation}
[{\vec D} , \omega_{(p,q)}] = - {\vec Q} \omega_{(p,q)}.
\end{equation}
Note that varying with respect to the $Z$'s, the phase space constraint on
differential forms is the same as before
\begin{equation}
[\partial {\vec D}, \omega_{(p,q)}] = 0.
\end{equation}
Indeed, a $(p,q)$-form valued in the line bundle $L^{(\overrightarrow{Q})}$
can be viewed as a linear map from $\mathcal{F}_{X}(\vec{\zeta})$ to
$\mathcal{F}_{X}(\vec{\zeta} - \vec{Q})$.
The Dolbeault operator $\ddd_A$ acting on such line-bundle valued $(p,q)$
forms requires a non-trivial connection, which is defined using the projection operators
introduced in equation (\ref{projop}) as
\be
\ddd_A \omega_{(p,q)}\,  =\,  \pi_{X}(\vec{\zeta} - \vec{Q}) \cdot \ddd
\omega_{(p,q)} \cdot  \pi_X(\vec{\zeta})
\ee

\subsection{Subspaces}\label{ssec:SUBSPACE}

Consider a subspace $P$ of the ambient toric space $X$ defined as the vanishing locus of a
weighted homogeneous polynomial which we denote (by abuse of notation) by
$P_{\overrightarrow{w}}(z_{1},...,z_{r})$. To specify the
embedding of $(P = 0)$ in $X$, we can also introduce a local basis of differential forms
corresponding to directions normal to the divisor labelled as $dP$, and directions parallel
to the divisor, which we label as $dz_{\parallel}$.

In the quantized version of this construction, the polynomial $P$ becomes an operator
$P(Z_{1},...,Z_{r})$ with fixed GLSM\ charges:%
\begin{equation}
\bigl[ \, \overrightarrow{\spc D}, \, P \spc \bigr]  =\overrightarrow{\spc w}\, P .
\end{equation}
Moreover, the formal differential $dP$ is now given by
\begin{equation}
dP = \bigl[\spc \partial ,\spc  P\spc \bigr],
\end{equation}
with $\partial = \sum_i C_i Z^\dag_i$. The formal differential $dP$ can be
viewed as the local holomorphic direction normal to the fuzzy subspace.

The subspace of states in $\mathcal{F}_{X}({\vec \zeta})$ annihilated by both
$P$ and $dP$ then defines the vector space of fuzzy points associated with $P$:
\begin{equation}
\mathcal{F}_{P}(\overrightarrow{\zeta}\spc) = \ker P \cap \ker dP.
\end{equation}
Provided the polynomial $P$ has positive GLSM charge with respect to each $U(1)_{i}$ factor,
the space of states annihilated by $P$ will always be
non-trivial.\footnote{To prove this, we
view $P$ as a linear map from $\mathcal{F}_{X}(\overrightarrow{\zeta})$ to
$\mathcal{F}_{X}(\overrightarrow{\zeta}-\overrightarrow{w})$, with
$\overrightarrow{w}$ the weight of the polynomial $P$. It follows from
the rank nullity theorem that $
\dim\ker P+\dim\text{im }P=\dim\mathcal{F}_{X}(\overrightarrow{\zeta}).$
Since all components in the vector $\overrightarrow{w}$ are positive this implies:
$\dim\ker P=\dim\mathcal{F}_{X}(\overrightarrow{\zeta})-\dim\text{im }%
P\geq\dim\mathcal{F}_{X}(\overrightarrow{\zeta})-\mathcal{F}_{X}(\overrightarrow{\zeta}-\overrightarrow{w})>0
$, where the final inequality follows from the fact that the degrees of the
polynomial in $P$ are all positive.} Hence, there is always a sense in which we
can speak of the collection of fuzzy points associated with this divisor. We can also consider more general intersections of
hypersurfaces $(P_{1}=0)\cap...\cap(P_{M}=0)$. The associated vector space generated by the
collection of points of this subspace are then given by the overlap: %
$\mathcal{F}_{P_{1}\cap...\cap P_{M}}(\overrightarrow{\zeta})=
\mathcal{F}_{P_{1}}(\overrightarrow{\zeta})\cap...\cap\mathcal{F}_{P_{M}}(\overrightarrow{\zeta}) $.

Functions and differential forms on a subspace $P$ are defined
by a similar procedure to that used for $X$. Viewing $\FF_{P}(\vec{\zeta})$ as
the exterior powers of the holomorphic cotangent bundle of $P$,
a differential form corresponds to a map from $\FF_{P}(\vec{\zeta})$ back
to itself. Let us now describe this procedure in practical terms.

One way to obtain the function space on $P$ is to introduce projection operators $\pi_{P}$ given
in terms of an orthonormal basis $\{|\mu \rangle \}$  spanning $\mathcal{F}_{P}(\overrightarrow{\zeta})$ as:%
\bea \pi_{P}=\underset{\left\vert \mu\right\rangle \in\mathcal{F}_{P}%
(\overrightarrow{\zeta})}{\sum}\left\vert \mu\right\rangle \left\langle\mu\right\vert .\eea
The projection operator $\pi_P$ can be viewed as the characteristic function localized on
the subspace $\mathcal{F}_P(\overrightarrow\zeta)$.
A general function $G_{P}$ on $P$ is then given by sandwiching a polynomial $G(Z^{\dag
},Z)$ in the oscillators $Z^{\dag}$ and $Z$ between two projection factors%
\bea
G_{P}
=\pi_{P}\cdot G
\cdot\pi_{P}.
\eea
Again by evaluating specific matrix elements in a basis for
$\mathcal{F}_{P}(\overrightarrow{\zeta})\otimes\mathcal{F}_{P}^{\ast
}(\overrightarrow{\zeta})$, we can present such zero forms as explicit
matrices. Similar considerations hold for further restrictions onto
intersections of subspaces $P_{1},...,P_{n}$.

Differential forms of a subspace $P$ of $X$ can be obtained in the same way, by restriction of
$(p,q)$ forms on $X$ to $P$. In the commutative theory, we can decompose a general $(p,q)$ form as
$\omega_{(p,q)} = \omega_{\parallel} + \omega_{\bot}$ where $\omega_{\bot}$ has formal differentials which include $dP$ or $\overline{dP}$, and
$\omega_{\parallel}$ is defined as the remainder. Formally setting to zero $\omega_{\bot}$ then defines the restriction of the form. In the fuzzy theory,
the analogous prescription is given by projecting out $dP$, and performing an appropriate restriction on the remaining components. All of this is captured by sandwiching
a general fuzzy $(p,q)$ form between two projection operators:
\begin{equation}
\omega_{(p,q)}{}_{| P} \, = \, \pi_{P}\cdot \omega_{(p,q)}\cdot\pi_{P}.
\end{equation}
In the commutative theory, it is often convenient to adopt
differentials intrinsic to a subspace such as $P$. In most of our applications, it will be enough to work in
terms of the ambient space differentials defined on the larger toric space $X$.

\subsection{Integration} \label{ssec:INTEGRATION}

Finally, we need to provide a suitable prescription for integrating functions and forms.

Consider some non-commutative subspace $P$ of dimension $d$ inside of the
toric space $X = {\mathbb C}^r/\!/ U(1)^s$. We wish to define integration over $P$. To motivate our definition,
we first characterize a natural prescription for integrating forms over
the underlying commutative space $P$. This space $P$ is a symplectic manifold with a symplectic form $\omega_P$ that  gets
inherited from the standard symplectic form $\omega$ (given in (\ref{symp})) defined on ${\mathbb C}^r$, by applying the appropriate projection operator.
As is clear from the construction of $P$, the symplectic form $\omega_P$ is a (1,1)-form on $P$.
Hence we can use it define the integration of general $(p,p)$ forms $W_{(p,p)}$ over $P$, by
taking the wedge product with the required number of $\omega_P$'s
to produce a volume form which is then integrated over $P$
\bea
\int_{P} W = \int_{P}  \omega_{P}\wedge  \ldots\wedge \omega_{P} \wedge {W}\, .
\eea
Equivalently, we can first use the inverse of the symplectic form $\omega_P$
to contract all indices of $W_{(p,p)}$, and define a function $W^{(0)}$ via
\be
\label{contraction}
W^{(0)}\,  =\, W_{i_1 \ldots i_p \bar{j_1} \ldots \bar{j_p}}\, \omega_P^{i_1 \bar{j_1}} \ldots \omega_P^{i_p \bar{j_p}}
\ee
We can then multiply this function with the standard symplectic volume element and integrate
\bea
\label{sympint}
\int_{P} W = \int_{P}  \omega_{P}\wedge  \ldots\wedge \omega_{P}\cdot W^{(0)}\, .
\eea
The latter prescription is most directly generalized to the non-commutative context.

Given a function $W^{(0)}(Z^\dag , Z)$ defined on the space $P$,
the analogue of the integral (\ref{sympint}) is to take the trace over the bosonic state space ${\cal F}_P(\vec{\zeta})$
\bea
\label{fuzzint}
\underset{\left\vert \mu
\right\rangle \in\mathcal{F}_{P}(\vec{\zeta})}{%
{\displaystyle\sum}
}\left\langle \mu \right\vert \; W^{(0)}(Z^\dag , Z) \; \left\vert \mu \right\rangle
\eea
Indeed, the states in ${\cal F}_P(\vec{\zeta})$ represent the points on $P$, and the trace amounts to summing the values of the function $W^{(0)}(Z^\dag , Z)$
evaluated at all the points of the geometry.

The contraction (\ref{contraction}) from a differential form $W$ to a function $W^{(0)}$ also has a direct non-commutative realization. The fuzzy $(p,p)$-form
represents an operator $W(Z^\dag , Z,C,C^\dag)$ on the extended state space, that includes the anti-commuting oscillators. Let $|0\rangle_{{}_C}$ denote
the fermionic vacuum state annihilated by the $C$ oscillators. We then define
\bea
\label{ncontract}
W^{(0)}(Z^\dag , Z) = {}_{{}_C}\! \langle \, 0\, | W(Z^\dag, Z ,C,C^\dag) | \, 0 \, \rangle_{{}_{\! C}}
\eea
Note that we defined the $(p,q)$-differentials $W(Z^\dag , Z ,C,C^\dag)$ to be anti-normal ordered in the fermionic oscillators. The above
vacuum expectation value thus generates Wick contractions between the $C^\dag$ and $C$ oscillators. These Wick contractions
contract the form indices with the inverse symplectic form. So the two steps (\ref{fuzzint}) and (\ref{ncontract})
provide a direct fuzzy realization of the commutative integration prescription as given above.

\section{F(uzz) Theory}\label{sec:FUZZTHEORY}

After collecting the necessary geometrical tools,
we now return
to our original task of describing the worldvolume gauge theory of a stack of seven-branes
wrapping a fuzzy internal del Pezzo four-cycle ${\cal S}$, supported by a non-zero $\BB$-flux, within an F-theory compactification.
Our approach will be as follows.

We can locally specify the elliptically fibered Calabi-Yau four-fold in the
neighborhood of the seven-branes via a Weierstrass equation
\begin{equation}
\label{fuzzweierstrass}
y^{2}=x^{3}+f(\spc z \spc ;\spc  Z_{j})x+g(\spc z\spc  ;  Z_{j}).
\end{equation}
Here $z$ denotes an ordinary commuting local coordinate, chosen such that
the stack of seven-branes is located at $z=0$. The other coordinates $Z_{j}$
will be used to represent non-commutative coordinates that
parametrize the fuzzy four-cycle ${\cal S}$.

Following the general treatment presented in section \ref{Fuzztor},
we start by defining the $Z_i$ as coordinates on some ambient toric variety $X = {\mathbb C}^r/\! / U(1)^s$,
which we can view as the classical vacuum of a GLSM.  The four-cycle
${\cal S}$ wrapped by the GUT seven-brane is defined as an intersection of hypersurfaces within $X$.
Concretely, $X$ has dimension $r-s$ so the K\"ahler surface ${\cal S}$ is defined as the intersection of $r-s-2$
hypersurfaces $P_{1},...,P_{r-s-2}$, each of complex codimension one, inside of $X$.

We can now apply the quantization procedure for a general toric space $X$ given
in section \ref{Fuzztor}, and promote the coordinates $Z_i$ and $Z_i^\dag$ to annihilation and creation operators. This provides us with a non-commutative deformation of the toric space $X$, and an associated finite-dimensional state space ${\cal F}_{X}(\vec{\zeta})$, specified by a set of discrete FI-parameters $\vec{\zeta}$. From this ambient toric space, we obtain the fuzzy four-cycle ${\cal S}$ and the associated state space ${\cal F}_{\cal S}(\vec{\zeta})$
by applying the quantized version of the projections onto the hypersurfaces  $P_{1},...,P_{r-s-2}$, following the procedure outlined in subsection \ref{ssec:SUBSPACE}.

Our main interest will be in formulating the gauge theory living on the seven-brane worldvolume ${\mathbb R}^{3,1} \times {\cal S}$, and the
spectrum and interactions of
chiral matter localized on fuzzy divisors inside of this space.  Applying our discussion of non-commutative spaces provided above, we introduce a Fock space $\mathcal{F}_{\cal S}(\vec{\zeta})$ associated with the
vector space of points of ${\cal S}$. The vector space of points of the matter curves then correspond to
subspaces of this larger Fock space.
We can expand the functions $f(z; Z_i)$ and $g(z;Z_j)$ that appear in the
Weierstrass equation (\ref{fuzzweierstrass}) in powers  of $z$. The coefficients are then defined
as sections of the appropriate line bundles on the ambient toric space $X$, restricted to the hypersurface
${\cal S}$. The discriminant $\Delta = 4f^3 + 27 g^2$ then locally looks as
\begin{equation}\label{fuzzydisc}
\Delta=z^{5}\underset{\Delta_{\alpha}\neq\Delta_{GUT}}{%
{\displaystyle\prod}
}\Delta_{\alpha}(Z_{j},z) \, .%
\end{equation}
The discriminant locus $\Delta=0$ decomposes into the set of fuzzy divisors $\Delta_\alpha(Z_j,z) = 0$.
Along the locus $z = 0$, the component $\Delta_{\alpha}(Z_{j},z = 0)$ defines the equation of a matter curve in $\cal{S}$. These divisors need
to be treated as operator equations that restrict the state and function space of the matter curves, on which the chiral matter lives.

\subsection{Fuzzy Seven-Branes}

In this section we will describe how to construct the
eight-dimensional Lagrangian $\cal{L}$ of a fuzzy seven-brane. This Lagrangian defines
an operator ${\cal L}(Z^\dag,Z,C,C^\dag)$ built from the bosonic and fermionic oscillators. The four-dimensional
Lagrangian $L$ is given by integrating $\cal{L}$ over the non-commutative internal space
${\cal S}$ using the prescription outlined in subsection \ref{ssec:INTEGRATION}:
\bea \label{LagInt}
L \, \is\! \underset{\left\vert \mu \right\rangle \in\mathcal{F}_{\cal S}(\vec{\zeta})}{%
{\displaystyle\sum}
}\!  \langle\, \mu\,\vert \, {\cal L}^{(0)}(Z^\dag , Z) \, \vert \, \mu \, \rangle
\eea
where in the above, the sum runs over an orthonormal basis of bosonic states of
$\mathcal{F}_{\cal S}(\vec{\zeta})$. Here, ${\cal{L}}^{(0)}$ is the bosonic projection of $\cal{L}$:
\bea
{\cal L}^{(0)}(Z^\dag , Z) = {}_{{}_C}\! \langle \, 0\, | \, {\cal L}(Z^\dag , Z ,C,C^\dag)\, | \, 0 \, \rangle_{{}_{\! C}}
\eea
Since $\cal{S}$ is compact and its state space is finite-dimensional, the sum in equation (\ref{LagInt}) truncates to a finite number of
terms. The effective action then involves only a finite number of four-dimensional fields.

\subsubsection*{Gauge Sector}
\vspace{-2mm}

The worldvolume theory of the seven-brane  describes an eight-dimensional
gauge theory of $\mathbb{R}^{3,1} \times {\cal S}$ with gauge group $G$, as well as
chiral modes which localize on matter curves $\Sigma_\alpha$.

As before, we shall find it convenient
to organize the fields in terms of
four-dimensional $\mathcal{N}=1$ supersymmetry.
The field content of the gauge sector includes an adjoint-valued vector
multiplet $V$, a collection of adjoint-valued
chiral superfields $A$ and $\Phi$, which transform respectively as an ordinary function, a $(0,1)$-form and a $(2,0)$-form on ${\cal S}.$
So the dependence on the internal fuzzy coordinates of these fields is of the form
\bea
V \is V(Z^\dag , Z)\nonumber \\[2mm]
A \is A_i(Z^\dag , Z) \, C^\dag_i\\[2mm]  
 \Phi \is \Phi_{ij} (Z^\dag , Z) \, C_i \wedge C_J \nonumber
\eea
The presentation of the Lagrangian is similar to the commutative case, though the interpretation of the fields is quite different.
Returning to our discussion around equations (\ref{DTERMS})-(\ref{FTERMS}), the Lagrangian density $\cal{L}$ takes the form:
\be
{\cal L} \, = \, \int \! d^{4}\theta \; {\cal K}  \; + \; \,  \int \! d^{2}\theta\;  {\cal W} \; +\; h.c.\; + \rm{WZW}\,.
\ee
where the WZW term is a non-local term which vanishes in WZ gauge \cite{SiegelTEND}.
Repeating the equations already presented in the commutative case, first consider the eight-dimensional gauge sector
of the seven-brane action. In terms of four-dimensional $\mathcal{N} = 1$ superspace, the D-terms are:
\begin{equation}
{\cal K}_{\rm gauge} = {\cal K}_{\Phi} + {\cal K}_{A}
\end{equation}
where the K\"ahler terms for $\Phi$ and $A$ are:
\begin{equation}
{\cal K}_{\Phi} = \spc\bigl(\spc\Phi{}^{\dag}\spc e^{-V }\! ,\spc \Phi \, e^{ V} \spc\bigr)%
\end{equation}
\begin{equation}
{\cal K}_{A} = \spc\bigl(\spc (\ddd +  A)^{\dag}e^{- V}\! , \spc (\ddd + A)\spc e^{V}\spc \bigr)
- (\spc \ddd ^{\dag}e^{- V}\! , \spc \ddd \spc e^{V}\spc \bigr).
\end{equation}
The F-terms are:
\bea
{\cal W}_{\rm gauge} 
 \is \; 
 {\rm Tr} \spc\bigl(\spc W_\alpha^{2} \spc\bigr)\, 
 \; + \;
{\rm Tr}\spc\bigl(
\spc \Phi \wedge(\ddd A + A\wedge A)\spc \bigr).
\eea
Here we chose the gauge kinetic term to be quadratic. Since all fields are now
normal ordered functions of the oscillators, the multiplication between fields
proceeds via a star product.
In addition the pairing $(\cdot , \cdot)$ present in the above definition represents the fact that
in general, the pairing of a field with its conjugate need not be the same as
the inner product associated with the Fock space.
Each field
consists of a zero mode component, as well as
an infinite sum of Kaluza-Klein excitations. Note, however, that the finite
size of the fuzzy space causes this infinite sum to truncate to a finite
number of dynamical degrees of freedom.

\subsubsection*{Fuzzy Matter}
\vspace{-2mm}

In the non-commutative theory, matter localization proceeds via the same mechanism as
in the commutative case. A matter curve
\bea
\Sigma_\alpha = \bigl(\, \Delta_\alpha(Z_{j},z) = 0 \, \bigr) \cap (z = 0)
\eea
is the intersection of the GUT seven-brane at $z = 0$ with another
component of the discriminant locus of equation (\ref{fuzzydisc}),
and indicates a local enhancement of the singularity type of
the elliptic fibration. We denote by $\Delta_{\alpha}(Z_{j})$ the restriction of
$\Delta_{\alpha}(Z_{j} , z)$ to $(z = 0)$.

The degrees of freedom in a local patch near $\Sigma_\alpha$  can then be
captured by means of a gauge theory with an enlarged gauge
group $G_\Sigma$, in the presence of a vortex shaped Higgs profile $\Phi_0$
that induces a breaking pattern
$G_{\Sigma}\supset G \times G_{\alpha}$.
The matter curve $\Sigma_\alpha$ then corresponds to the locus
where the component of the Higgs field along the simple root vanishes
$\langle \alpha, \Phi_0\rangle = 0$. All these data
can be straightforwardly reproduced within the
non-commutative setting, by using the technology developed in section \ref{Fuzztor}.

Here we will follow a more practical approach, and directly take the limit
in which the Higgs field generating the breaking pattern
is very large. This limit is particularly natural in the decoupling limit,
in which the closed string volume of the four-cycle wrapped by the seven-brane is taken to zero.
Then following the discussion in subsection \ref{ssec:LOCALIZE}, we can introduce the chiral
matter content by associating to $\Sigma_\alpha$ two
chiral superfields
\bea
A_\alpha \is A_{\alpha, i}(Z^\dag , Z) \, C^\dag_i\\[2mm]
 \Psi_\alpha \is \Psi_{\alpha, ij} (Z^\dag , Z) \, C_i \wedge C_j \nonumber
\eea
transforming in the representation $R_\alpha$ of the unbroken gauge group
$G$. These fields are localized on the matter curve. Just as in
the commutative theory, the six-dimensional mode
content is chiral, and therefore the theory of an isolated matter curve is anomalous.
These anomalies are cancelled through the presence of Chern-Simons terms of the seven-brane,
which induce anomaly inflow (and outflow) between the curves and closed string modes.
In practice, the cancellation of all such anomalies is guaranteed by compactifying
on an elliptic Calabi-Yau fourfold.

To describe the localization of matter fields in the non-commutative geometry, we first construct the
state space ${\cal F}_{\Sigma_\alpha}(\vec{\zeta})$ associated to the matter curve.
Following the prescription outlined in section \ref{ssec:SUBSPACE}, we introduce the operator
$\Delta_\alpha(Z)$. We then define  ${\cal F}_{\Sigma_\alpha}(\vec{\zeta})$  as the subspace
of the total state space ${\cal F}_{\cal S}(\vec{\zeta})$ annihilated by
$\Delta_\alpha(Z)$
\bea
{\cal F}_{\Sigma_\alpha}(\vec{\zeta}) = {\rm ker}(\Delta_\alpha).
\eea
The fact that $A_\alpha$ and $\Psi_{-\alpha}$ localize on
$\Sigma_\alpha$ means that they are given by differential forms
on ${\cal S}$, though the bosonic content of this form is localized on a subspace.
Introduce a bosonic projection operator $\pi_{\Sigma}^{(0)}$
onto the subspace ${\cal F}_{\Sigma_\alpha}$ which acts as the identity on the fermionic oscillators.
Given a bulk form $\Psi$ on $\cal{S}$, one way to localize its bosonic content is to sandwich it
in between such bosonic projections via $\pi_{\Sigma}^{(0)} \cdot \Psi \cdot \pi_{\Sigma}^{(0)}$.
Let us stress that we view these matter fields as forms on ${\cal S}$.\footnote{This is essentially the same
procedure one follows in the spectral cover construction \cite{Hayashi:2009ge,DWIII}
in which an auxiliary local Calabi-Yau threefold $O(K_{\cal{S}}) \rightarrow \mathcal{S}$
is erected over the GUT seven-brane. Zero modes are identified with elements of
appropriate $Ext$ groups on the threefold as $Ext^{p}(\mathcal{V} , \mathcal{V}^{\prime})$
for sheaves $\mathcal{V}$ and $\mathcal{V}^{\prime}$ with support on subspaces.}
This is distinct from the notion of a differential form defined exclusively on a subspace, as
in subsection \ref{ssec:SUBSPACE}. In addition, in the commutative theory,
there is the phenomenon of seven-brane monodromy \cite{Hayashi:2009ge,BHSV,DWIII,EPOINT,Marsano:2009gv}. Again, the main point here
is that algebraic expressions of the commutative theory have direct analogues in the fuzzy theory, so the explicit presentation
of the higher-dimensional modes in such cases is really an issue in the commutative theory. Once we have this commutative
data, the non-commutative translation follows. Accordingly, we can write an eight-dimensional matter sector Lagrangian,
specified by a K\"ahler potential and superpotential of the same form as equation (\ref{matteraction}):
\bea
{\cal K}_{\rm matter}\;
 \is  \left( \spc\Psi{}^{\; \dag}_{- \alpha} ,  e^{V}\! \; \Psi_{-\alpha} \right)%
 \, + \,    \left( A_{\alpha}{}^{\!\!\dag}\! , e^{V} \! A_{\alpha} \right) %
\\[4mm]
{\cal W}_{\rm matter} \, \is \Psi_{- \alpha} \wedge \ddd_A A_{\alpha} + f_{\alpha\beta\gamma}\,
  \Psi_\alpha \wedge A_\beta \wedge A_\gamma.
\eea

\subsubsection*{Mode Decomposition \label{ssec:fuzzCOHOM}}

In the commutative setting the matter fields of the higher-dimensional
theory transform as forms of appropriate line bundles. The zero mode content
is computed by the Dolbeault cohomology, and the KK spectrum is given by the higher harmonics
of the Laplacian operator.

All of these statements have analogues in the non-commutative setting. Consider first the spectrum of zero modes.
Because we still have a notion of a Dolbeault operator, we can define Dolbeault cohomology for fuzzy line bundles,
following the same algebraic prescription in the commmutative theory.
Similar considerations hold for the cohomology theory of line bundles on subspaces.
Our discussion also holds for non-toric spaces as well. Indeed, in the
commutative theory, there is a well-known prescription for determining
cohomology groups based on the Koszul complex. The main idea in Koszul complex
computations is that once we have specified a divisor $D$ of the toric space
$X$, the cohomology theory of a line bundle defined over $D$ can be computed
in terms of the cohomology theory on $X$ \cite{Griffiths}.\footnote{See for example \cite{Cvetic:2010rq}
for a recent discussion in the physics literature for how this algorithm is
implemented in practice.} Roughly speaking, the elements of the cohomology groups should be viewed as operators which are
``holomorphic'' in the sense that they are annihilated by the Dolbeault operator of the fuzzy subspace.

Sections and forms of a line bundle not annihilated by the Dolbeault operator
then determine the spectrum of KK modes. To illustrate
the general treatment of chiral matter fields, consider
the modes associated with a higher-dimensional field
$\Psi^{(\overrightarrow{Q})}$ corresponding to a section of a degree $\overrightarrow{Q}$ fuzzy
line bundle $L^{(\overrightarrow{Q})}$. This field is given
by a formal sum of normal ordered polynomials in the oscillators $Z$
and $Z^{\dag}$ of weighted degree $\overrightarrow{Q}$:%
\bea \label{psiexpand}
\quad \Psi^{(\overrightarrow{Q})}=\sum_{I , \; \vec{v} - \vec{u} = \vec{Q}}\psi_{I(\overrightarrow{u},\overrightarrow
{v})}
\overline\sigma_{I(\overrightarrow{u})}(Z^\dag)\, \sigma_{I(\overrightarrow{v})}(Z) \label{psiQdef}%
\eea
where in the above, the $\sigma$'s are a basis of polynomials of a given weight
\bea
\bigl[ \overrightarrow{\spc D},  \overline{\sigma}_{I(\overrightarrow{u})} \bigr] \, =\, + \overrightarrow{\spc u} \, \overline{\sigma}_{I(\overrightarrow{u})}
\;\;\;\;
\bigl[ \overrightarrow{\spc D},  \sigma_{I(\overrightarrow{v})} \bigr] \, =\, - \overrightarrow{\spc v} \, \sigma_{I(\overrightarrow{v})} \, .
\eea
The overall weight of $\Psi^{(\overrightarrow
{Q})}$ is fixed to be equal to
$\overrightarrow{v}-\overrightarrow{\spc u}\, =\, \overrightarrow{Q}\text{.}$
The indices $I$ label the linearly independent weighted homogeneous
polynomials, and the coefficients $\psi_{I(\overrightarrow{u},\overrightarrow
{v})}$ correspond to four-dimensional chiral superfields. Note that the sum in
$\Psi^{(\overrightarrow{Q})}$ runs over an infinite set of terms. The terms
with $\overrightarrow{u}=0$ are to be viewed as the massless modes
(corresponding to holomorphic sections), and the terms with $\overrightarrow
{u}\neq0$ define  massive Kaluza-Klein excitations.

The presence of an infinite number of terms might at first appear contrary to the
intuition that making the geometry non-commutative truncates the Kaluza-Klein spectrum.
Here this occurs once we fix an overall size for the fuzzy space
on which $\Psi^{(\overrightarrow{Q})}$ localizes.
Indeed, tracing over the state space of bosonic points,
the kinetic term is
\begin{equation}
\sum_{\left\vert \mu \right\rangle \in\mathcal{F}_{\cal S}({\vec \zeta})}
\int d^{4} \theta \left \langle \mu \right \vert \left( \Psi^{(\overrightarrow{Q}) \dag} , \Psi^{(\overrightarrow{Q})} \right) \left \vert \mu \right \rangle.
\end{equation}
Writing the explicit bundle indices in the
open string metric, the pairing is given by inserting the appropriate Hermitian metric in
between the two factors, so that
\begin{equation}
\left( \Psi^{(\overrightarrow{Q}) \dag} , \Psi^{(\overrightarrow{Q})} \right)
= \Psi^{(\overrightarrow{Q}) \dag} \cdot G_{\rm{open}} \cdot \Psi^{(\overrightarrow{Q})}.
\end{equation}
Thus, once the magnitude of the vector $\overrightarrow{v}$ given in equation (\ref{psiexpand})
becomes sufficiently large, all states of $\mathcal{F}_{\cal S}({\vec \zeta})$ will be annihilated.
Hence the number of dynamical four-dimensional fields is finite.
Similar considerations hold for the massive modes descending from
differential-forms valued in the line bundle $L^{(\overrightarrow{Q})}$.

The field content of the seven-brane theory is then given by sums over
operators defined as in equation (\ref{psiQdef}). These fields correspond to
chiral superfields in the case of the modes $\Phi$ and $A$, which propagate in
the bulk, and their analogues $\Psi_{\alpha}$ and $A_{\alpha}$ which localize on a curve. To
treat all such modes on a uniform footing, we shall simply write all of the
modes as $\Psi_{i}$ and $A_{i}$, where $i = \cal{S}$ denotes the modes propagating in
the bulk, and $i=\Sigma$ denotes a mode localized on the subspace $\Sigma$.

One of the virtues of the above description is that it provides us with an
explicit list of the Kaluza-Klein modes. The actual mass matrix for the KK
modes will depend on the details of the open string metric. Nevertheless, the fact
that we can now catalogue this spectrum means that we can treat this theory
in four-dimensional terms.

GUT breaking by a $U(1)_{Y}$ hypercharge flux works just as in the commutative theory \cite{BHVII, DWII}.
In four dimensions, the unbroken gauge group is the commutant of $U(1)_{Y}$ inside of $SU(5)_{GUT}$, and
the fuzzy modes organize into representations of $SU(3)_{C} \times SU(2)_{L} \times U(1)_{Y}$.
The hypercharge flux (\ref{hyperdec}) can either be viewed as a component of the non-commutativity
parameter of each MSSM gauge group factor, or as a background flux of the $SU(5)$  gauge theory,
with non-commutativity parameter set by the overall  $U(1)$ flux $\BB$. From our discussion above
we see that these two ways of representing the hypercharge flux are completely equivalent.

Fluxes from the GUT seven-brane as well as flavor seven-branes affect the line bundle data of the gauge
and matter fields. After activating a hyperflux, the massless mode content in the
gauge sector corresponds to the Standard Model gauge group, as well
as massive vector multiplets which also transform in the adjoint
representation of the Standard Model gauge group. In addition, there will also
be Kaluza-Klein modes corresponding to the off-diagonal components of the GUT group.

\subsubsection*{Matter Delocalization}

Fuzz theory retains much of the structure of the commutative local F-theory model, though there
are also differences. An example of a new phenomenon is that matter localization on curves
may break down in the non-commutative setting. To illustrate the general issue, consider the weighted
homogeneous polynomial $P(Z_{1},...,Z_{r})$ in the $Z$'s. Our definition
of matter localized along the vanishing locus of $P$ is given by states
$\left\vert \psi\right\rangle $ in $\mathcal{F}_{X}(\vec{\zeta})$ subject
to the condition:%
\begin{equation}
P(Z_{1},...,Z_{r})\left\vert \psi\right\rangle =0.
\end{equation}
Note, however, that when the FI parameter $\vec{\zeta}$ is sufficiently
small and the weighted degree of the operator $P(Z_{1},...,Z_{r})$ is sufficiently high, all of the
states of $\mathcal{F}_{X}(\vec{\zeta})$ will be annihilated. In other words, in this
case, there is no sense in which the matter remains localized. Rather, the matter field
spreads out over the entire space. In the commutative theory, the degree
of $P$ determines the genus of the curve. Here, we see that high genus
matter curves do not really correspond to curves at all.

This in turn raises the question as to how to count the number of such
delocalized matter fields. Ordinarily in the commutative theory, the main task
is to specify a line bundle over an ambient toric space. Once this is
accomplished, we can perform an appropriate restriction of this line bundle to
subspaces, and count the number of associated zero modes. Here we face the
issue that when the space is sufficiently small, the
notion of a subspace itself breaks down, and thus the restriction of line
bundles onto a given subspace also becomes ill-defined.

Our definition of holomorphic sections of a line bundle, however, remains
well-defined at arbitrary values of the radii because it is determined by purely algebraic data.
Indeed, at an abstract level, we have specified the zero modes localized on a matter
curve simply as appropriate polynomials in the $Z$ and $C$ oscillators sandwiched between
projection factors $\pi_{P}$. When the projection $\pi_{P}$
acts as the identity, the matter field is no longer localized.

One consequence of this delocalization is that the number of points sampled by the
Yukawa coupling will now increase. This then causes a jump in the rank of the Yukawa
coupling matrix. Assuming a suitable hierarchy in the presentation of the
original holomorphic sections, matter delocalization then induces
novel textures for Yukawas.

\subsection{A Toy Example: Yukawa Couplings on Fuzzy $\mathbb{P}^{1}\times\mathbb{P}^{1}$}

In this section we illustrate the general ideas developed earlier, and compute the
Yukawa couplings in a toy model based on fuzzy $\mathbb{P}^{1}\times
\mathbb{P}^{1}$. Even though the general discussion we have presented applies
equally well to cases where the GUT\ surface $\mathcal{S}$ is non-toric, note that because
$\mathbb{P}^{1}\times\mathbb{P}^{1}$ is itself toric, we can bypass some of
this general discussion and simply treat the GLSM\ for $\mathbb{P}^{1}%
\times\mathbb{P}^{1}$. For simplicity, we focus on the case where all matter fields
remain localized on well-defined subspaces. We find that just
as in the commutative theory, only one overlap of wave
functions is non-zero. However, background fluxes can
distort this structure, producing flavor hierarchies \cite{HVCKM,FGUTSNC}.

Let us first describe the geometry of fuzzy $\mathbb{P}^{1} \times \mathbb{P}^{1}$.
The classical GLSM for $\mathbb{P}^{1}\times\mathbb{P}^{1}$ is a
$U(1)^{2}=U(1)_{I}\times U(1)_{II}$ gauge theory with chiral superfields
$u_{i}$ and $v_{i}$ for $i=1,2$. Under the two $U(1)$ factors, the charges of
$u_{i}$ and $v_{i}$ are $(+1,0)$ and $(0,+1)$. The fields are subject to the
D-term constraints:%
\begin{align}
\left\vert u_{1}\right\vert ^{2}+\left\vert u_{2}\right\vert ^{2}  &
=\zeta_{I} \nonumber\\[-2.5mm] & \\[-2.5mm]
\left\vert v_{1}\right\vert ^{2}\spc +\spc \left\vert v_{2}\right\vert ^{2}  &
=\zeta_{II}.\nonumber
\end{align}
The symplectic quotient $%
\mathbb{C}
^{4}//U(1)^{2}$ yields $\mathbb{P}^{1}\times\mathbb{P}^{1}$. The $u_{i}$ and
$v_{i}$ correspond to projective coordinates for the two $\mathbb{P}^{1}$ factors.

We now turn to the non-commutative geometry. Here we focus on the
bosonic content, treating all differential forms more implicitly.
Introduce bosonic oscillators for each $\mathbb{P}^{1}$ factor, $U_{i}$ for $\mathbb{P}^{1}_{I}$ and $V_{i}$ for
$\mathbb{P}^{1}_{II}$. These are subject to the commutation relations:
\begin{equation}
\bigl[  U_{i},U_{j}^{\dag}\bigr]  = \bigl[  V_{i},V_{j}^{\dag}\bigr] = \delta_{ij}%
\end{equation}
with all other commutators vanishing.

Next, introduce a vacuum $\left\vert 0\right\rangle $ annihilated by the $U$'s and $V$'s.
The Fock space ${\cal F}({\mathbb{C}^{4}})$ is given
by acting on the vacuum state with the creation operators. The basis elements
of ${\cal F}({\mathbb{C}^{4}})$ are
\bea
| n_1,n_2 ; m_1,m_2 \rangle \, = \,  \frac{\left(  U_{1}^{n_{1}}U_{2}^{n_{2}}%
V_{1}^{m_{1}}V_{2}^{m_{2}}\right)  ^{\dag}}{\sqrt{n_{1}!n_{2}!m_{1}!m_{2}!}%
} \left\vert 0\right\rangle.
\eea
We introduce $\mathcal{F}_{X}(N_{I},N_{II})$ as the subspace of $\mathcal{F}(\mathbb{C}^{4})$ subject to the
oscillator number constraint
\bea
n_1\spc +\spc n_2 \, = \spc N_I\; , \nonumber \\[-2.5mm]\\[-2.5mm] \nonumber
m_1 + m_2 = N_{II}\, .
\eea  This corresponds to the linear space of bosonic points for
a $\mathbb{P}_{I}^{1}\times\mathbb{P}_{II}^{1}$ with FI parameters $N_{I}$ and $N_{II}$.

We now compute the Yukawa coupling associated with a triple intersection of matter curves.
In the commutative geometry, we consider a configuration of matter fields in the local patch
where $u_{2}$ and $v_{2}$ are both non-zero. Local coordinates for this patch are
$u = u_{1} / u_{2}$ and $v = v_{1} / v_{2}$. The configuration of matter curves we
consider are described by Higgsing of $U(3)$ gauge theory down to $U(1)^{3}$ according to
the background vev:%
\begin{equation}
\Phi_{0}=\left[
\begin{array}
[c]{ccc}%
v & 0 & 0\\
0 & 0 & 0\\
0 & 0 & u%
\end{array}
\right]  .
\end{equation}
In this patch, the three matter curves of this configuration correspond
to enhancements in the singularity type at $u=0$, $v=0$ and
$u = v$. In terms of the original $u_i$ and $v_i$ variables, the three matter curves are:%
\begin{align}
\mathbb{P}_{I}^{1}  &  =\mathbb{P}_{(12)}^{1}=(v_{1}=0)\nonumber \\
\mathbb{P}_{II}^{1}  &  =\mathbb{P}_{(23)}^{1}=(u_{1}=0)\\
\mathbb{P}_{\text{diag}}^{1}  &  =\mathbb{P}_{(13)}^{1}=(u_{1}v_{2}=v_{1}%
u_{2}),\nonumber
\end{align}
where we have indicated the color space assignment associated with each curve.

We now discuss matter localization. We are interested in
states of $\mathcal{F}_{X}(N_{I},N_{II})$ which
are respectively annihilated by $V_{1}$, $U_{1}$ or $U_{1}V_{2}-U_{2}V_{1}$.
The set of states annihilated by $V_{1}$ and $U_{1}$ are:%
\bea
\ker V_{1}  \is \text{span}\bigl\{\;  |\spc n_1,n_2; \spc 0, N_{II}\spc \rangle\; \, :\, n_{1}\spc +n_{2}\spc =\spc N_I\spc \, \bigr\}
\nonumber \\[-2mm]\\[-2mm] \nonumber
\ker U_{1}  \is \text{span}\bigl\{  \; |\spc 0,N_I\spc ; m_1,m_2\spc \rangle\; :m_{1}+m_{2}=N_{II}\, \bigr\}
.
\eea
Geometrically, $\ker V_{1}$ corresponds to
$\mathbb{P}_{I}^{1}\times\vert$South$\rangle_{II}$ and $\ker U_{1}$
corresponds to $\vert\text{South}\rangle_{I}%
\times\mathbb{P}_{II}^{1}$. See Appendix A for further discussion of fuzzy $\mathbb{P}^{1}$.

Next consider the set of states annihilated by $U_{1}V_{2}-U_{2}V_{1}$. One
state which is annihilated by this operator is the South pole state of both
$\mathbb{P}^{1}$ factors:%
\bea
\label{doublesouth}
\vert\text{South} \rangle_{I}\otimes\vert\text{South}\rangle
_{II}\, =\, |0,N_I; 0,N_{II} \rangle
\eea
To generate the remaining states of the diagonal $\mathbb{P}^{1}$, note
that $U_{1}V_{2}-U_{2}V_{1}$ commutes with the diagonal raising operator%
\begin{equation}
J_{+}^{\text{diag}}=J_{+}^{(I)}+J_{+}^{(II)}=U_{1}^{\dag}U_{2}+V_{1}^{\dag
}V_{2}%
\end{equation}
In other words, the remaining states of the diagonal $\mathbb{P}^{1}$ are
obtained by acting by successive powers of $J_{+}^{\text{diag}}$:%
\begin{equation}
\ker(U_{1}V_{2}-U_{2}V_{1})=\text{span}\left\{  \bigl(  \, J_{+}^{\text{diag}%
}\bigr)  ^{m}|\spc 0\spc ,N_I\spc; 0,N_{II}\rangle\, :\, m\geq0\, \right\}  .
\end{equation}

We now present a computation of Yukawa couplings for this
configuration of matter curves. To keep the computation as symmetric as possible,
we assume that the modes on $\mathbb{P}_{I}^{1}$ and $\mathbb{P}_{II}^{1}$
correspond to $(0,1)$-forms with localized bosonic content, while the modes of
$\mathbb{P}_{\text{diag}}^{1}$ correspond to $(2,0)$-forms with localized bosonic content.

The number of zero modes is controlled by the line bundle assignments
on each matter curve. For simplicity, we view the line bundles as
descending from bulk line bundles on $\mathbb{P}^{1} \times \mathbb{P}^{1}$. The localized matter fields
are then given by a projection on the bosonic content. Actually, the line bundle assignments need only be
defined over the matter curves, so we merely view this as a useful device for deducing the operator content of the
localized modes. Stated in terms of the commutative geometry, the
configuration of line bundles we take for the three matter curves are:
\begin{align}
\qquad \qquad L_{I}  &  =O_{\mathbb{P}_{I}^{1}\times\mathbb{P}_{II}^{1}}( \spc d\, \sigma_{2}\spc ), \nonumber \\
L_{II}  &  =O_{\mathbb{P}_{I}^{1}\times\mathbb{P}_{II}^{1}}(\spc d \, \sigma_{1}\spc ), \qquad \qquad \ \ d = g-1,\\
L_{\text{diag}}  &  =O_{\mathbb{P}_{I}^{1}\times\mathbb{P}_{II}^{1}%
}(-d \, ( \sigma_{1}+\sigma_{2}) ),\nonumber
\end{align}
where $\sigma_{1}$ and $\sigma_{2}$ denote the divisor classes for
$\mathbb{P}_{I}^{1}$ and $\mathbb{P}_{II}^{1}$. In the fuzzy setting, differential forms
of these bundles correspond to operators of fixed degree. Note that the tensor product
$L_{I}\otimes L_{II}\otimes L_{\text{diag}}$ is indeed trivial, so the
product $A_I \wedge_{\ast} A_{II} \wedge_{\ast} \Psi_{\rm{diag}}$ can indeed
form a Yukawa. Further, the restriction of each line bundle onto the appropriate matter curve yields a
degree $d$ line bundle on both $\mathbb{P}_{I}^{1}$ and $\mathbb{P}_{II}%
^{1}$, while on $\mathbb{P}_{\rm{diag}}^{1}$ we find a degree
$-2d$ line bundle.

To compute the Yukawa couplings for this configuration, we first specify the profile of the zero mode wave-functions.
First consider the zero modes localized on $\mathbb{P}_{I}^{1}$. Such modes correspond to zero-forms on $\mathbb{P}^{1}_{I}$,
and have the form content of a one-form on $\mathbb{P}_{II}^{1}$. Since line bundles on $\mathbb{P}^{1}$ are fully classified by
their degree, holomorphic sections of this line bundle can be presented as holomorphic polynomials
in $U_{1}$ and $U_{2}$ of homogeneous degree $d$. The $g = d + 1$ zero mode wave functions are then:
\begin{equation}
A_{(12)}  = \pi_{\mathbb{P}_{(12)}^{1}}\cdot\underset{i=0}{\overset{d}{\sum}}a_{i}^{(12)}U_{1}^{i}U_{2}^{d-i}\cdot\pi_{\mathbb{P}_{(12)}^{1}}
\end{equation}
here we are suppressing the one-form content to avoid cluttering the presentation.
The $\pi$'s denote bosonic projection factors to the matter curves and
the $a$'s correspond to four-dimensional chiral superfields, and define the coefficients of the degree $d$ polynomials.
Similar considerations hold for the $g$ zero modes localized on $\mathbb{P}^{1}_{II}$ so that:
\begin{equation}
A_{(23)}  = \pi_{\mathbb{P}_{(23)}^{1}}\cdot\underset{i=0}{\overset{d}{\sum}}a_{i}^{(23)}V_{1}^{i}V_{2}^{d-i}\cdot\pi_{\mathbb{P}_{(23)}^{1}}.
\end{equation}

Next consider the zero modes associated with $\Psi_{\rm{diag}}$. Since we have already demanded
the bulk line bundle assignments tensor to the trivial bundle, we deduce that the GLSM
charge is $(-d,-d) + (2,2)$ for the zero modes of $\Psi_{\rm{diag}}$. Here we have split
up the contributions to the GLSM charge from the $(2,0)$ form, and its localized coefficient.
Suppressing the form content, such zero modes can all be written as:
\begin{equation}
\Psi_{(31)}=\pi_{\mathbb{P}_{(31)}^{1}}\cdot \sum_{i} a_{i}^{(31)}O_{i}(U^{\dag} , V^{\dag})\cdot\pi_{\mathbb{P}_{(31)}^{1}}.
\end{equation}
To be a section of the appropriate line bundle, an operator $O_{i}$ must
commute with $U_1 V_2 - U_2 V_1$. In other words, acting on the vacuum by $O_{i}$ can be viewed as generating a
state of a $\mathbb{P}^{1}_{\rm{diag}}$ with $2d + 1$ points:
\begin{equation}
\mathcal{F}_{\mathbb{P}^{1}_{\rm{diag}}}(2d+1) = \text{span}\left\{  \bigl(
J_{+}^{\text{diag}}\bigr)  ^{m}\bigl(  U_{2}^{\dag}\bigr)  ^{d}\bigl(
V_{2}^{\dag}\bigr)  ^{d}\bigl\vert \spc 0\spc \bigr\rangle :0\leq m\leq 2d \right\}  .
\end{equation}
There are $2d+1$ linearly independent $O_{i}$'s of this form. We order
the modes in accord with the basis of states given above so that for
example $O_{0} = \bigl(  U_{2}^{\dag}\bigr)  ^{d}\bigl(V_{2}^{\dag}\bigr)  ^{d}$ and
$O_{2d} = \bigl(  U_{1}^{\dag}\bigr)  ^{d}\bigl(V_{1}^{\dag}\bigr)  ^{d}$.
The zero modes localized on the diagonal $\mathbb{P}_{\text{diag}}^{1}$ are then given as:%
\begin{equation}
\Psi_{(31)}=\pi_{\mathbb{P}_{(31)}^{1}}\cdot\underset{i=0}{\overset
{2d}{\sum}}a_{i}^{(31)}O_{i}(U^{\dag} , V^{\dag})\cdot\pi_{\mathbb{P}_{(31)}^{1}}
\end{equation}
where in the above expression we have suppressed the form content of the zero mode.

Having specified the zero mode content of the theory, we now compute the
Yukawa couplings for this toy model.
We introduce a convenient basis of states for the zero mode wave functions:%
\begin{align}
\Psi_{(31)}^{(i)}  &  =\pi_{\mathbb{P}_{(31)}^{1}}\cdot a_{i}^{(31)}%
O_{i}(U^{\dag},V^{\dag})\cdot\pi_{\mathbb{P}_{(31)}^{1}}\text{.}\nonumber \\
A_{(12)}^{(j)}  &  =\pi_{\mathbb{P}_{(12)}^{1}}\cdot a_{j}^{(12)}U_{1}%
^{j}U_{2}^{d-j}\cdot\pi_{\mathbb{P}_{(12)}^{1}}\\
A_{(23)}^{(k)}  &  =\pi_{\mathbb{P}_{(23)}^{1}}\cdot a_{k}^{(23)}V_{1}%
^{k}V_{2}^{d-k}\cdot\pi_{\mathbb{P}_{(23)}^{1}}\nonumber
\end{align}
Working at fixed discretized FI parameters $(N_{I},N_{II})$ for the two $\mathbb{P}^{1}$
factors, we must evaluate the superpotential:%
\begin{equation}
W_{ijk}=\underset{\left\vert \mu \right\rangle \in\mathcal{F}_{X}(N_{I},N_{II})}{%
{\displaystyle\sum}
}\left\langle \mu \right\vert \Psi_{i}\wedge_{\ast} A_{j}\wedge_{\ast} A_{k}\left\vert \mu \right\rangle .
\end{equation}

By inspection, the presence of the projection factors causes most of the
Yukawas to vanish. Indeed, working at fixed K\"ahler volumes, the only state common to
all three fuzzy divisors is the mutual South pole (\ref{doublesouth}).
Thus, we can replace the general projection operators presented by the
projection onto the South pole state, which we denote by $\pi_{\text{South}}$. As a geometric operation,
the projection makes sense regardless of the FI parameters for our toric space.
The relevant projection is then given by a direct sum over the projection operators we would obtain
by working at a fixed value of the FI parameters:
\begin{equation}
\pi_{\text{South}}\spc=\, \sum_{M_{I} , M_{II}} \bigl| 0, M_I ; 0, M_{II}\bigr\rangle\,\bigl\langle 0, M_I;0,M_{II}\bigr| \, .
\end{equation}
The superpotential can therefore be written as:%
\bea
W_{ijk}=\bigl\langle 0, N_{I}; 0 , N_{II}\bigr | \,
a_{i}^{(31)}O_{i}(U^{\dag}, V^{\dag})\! \cdot \pi_{{}_{\! \text{South}}}\cdot a_{j}^{(12)}%
U_{1}^{j}U_{2}^{d-j}\! \cdot\pi_{{}_{\text{South}}}\cdot a_{k}^{(23)}V_{1}^{k}%
V_{2}^{d-k}\, \bigl|0, N_{I}; 0 , N_{II}
\bigr\rangle .\nonumber
\eea
Since we have fixed the overall K\"ahler class of $X$, we now substitute the projection
operator $\pi_{\text{South}}$:
\begin{align}
W_{ijk}  &  =a_{i}^{(31)}a_{j}^{(12)}a_{k}^{(23)} \times \langle \, 0,N_I; 0, N_{II}| \, O_{i}(U^{\dag} , V^{\dag})\,
|0, N_{I}\! -d; 0 , N_{II}\! -d\,
\rangle\; \times\nonumber\\[2mm]
& \qquad\quad  \times \; \bigl\langle 0,N_I\! -d; 0, N_{II}\! -d\bigr| \, U_{1}^{j}U_{2}^{d-j}\, \bigl|0, N_{I}; 0 , N_{II}\! -d\,
\bigr\rangle \;
\times\nonumber\\[2mm]
&  \qquad \qquad \ \ \times \ \bigl\langle 0,N_I; 0, N_{II}\! -d\bigr|\,V_{1}^{k}V_{2}^{d-k}\, \bigl|0, N_{I}; 0 , N_{II}
\bigr\rangle.
\end{align}
To obtain a non-zero result, it is necessary to exclude all
dependence on the oscillators $U_{1}$, $V_{1}$, $U_{1}^{\dag}$ and $V_{1}^{\dag}$.
In other words, the only non-zero Yukawa coupling occurs for $i=j=k=0$. A  straightforward calculation then gives
\begin{equation}
W_{ijk}=\frac{N_{I}!N_{II}!}{(N_{I}-d)!(N_{II}-d)!}\times\delta
_{i0}\delta_{j0}\delta_{k0}\times a_{i}^{(31)}a_{j}^{(12)}a_{k}^{(23)}.
\end{equation}
Thus, as expected, we obtain only a single non-vanishing Yukawa coupling.

For commutative seven-branes, small changes to the rank of the superpotential can be phrased in terms of
non-commutative deformations of the holomorphic structure in a patch of the Yukawa point \cite{FGUTSNC}. This
corresponds to a holomorphic deformation of the oscillator algebra.

\section{Geometric Applications}\label{sec:FORMAL}

In this section we study some other potential applications of the formulation of non-commutative toric geometry developed in section \ref{Fuzztor}.

\subsection{A Fuzzy Flop}

A well known phenomenon in commutative toric geometry are flop transitions. These correspond to transitions in the large volume regime where changing the values of the FI parameters of the GLSM causes some two-cycles to collapse, and new ones to take their place. In this section we study the corresponding fuzzy flop.

To keep our discussion concrete, we focus on the flop transition associated with the small resolution of the conifold geometry $O(-1) \oplus O(-1) \rightarrow \mathbb{P}^{1}$. The GLSM data of the conifold is given in terms of a single $U(1)$ gauge theory and four chiral superfields $u_1$, $u_2$, $v_1$ and $v_2$ so that the $u$'s have GLSM charge $+1$ and the $v$'s have GLSM charge $-1$. The D-term constraint is:
\begin{equation}
|u_1|^2 + |u_2|^2 - |v_1|^2 - |v_2|^2 = \zeta.
\end{equation}
When $\zeta > 0$, the $v$'s should be viewed as normal coordinates to a $\mathbb{P}^{1}$ with K\"ahler class proportional to $\zeta + |v_1|^2 + |v_2|^2$, so that the $u$'s define coordinates for the
$\mathbb{P}^{1}$. When $\zeta < 0$, the geometry undergoes a flop transition and the $u$ and $v$ coordinates exchange roles.

We now describe the fuzzy flop transition.
Introduce operators $U_i$ and $V_i$ subject to the commutation relations:
\begin{equation}
[U_{i} , U_j^{\dag}] = [V_{i} , V_{j}^{\dag}] = \delta_{ij}
\end{equation}
with all other commutators zero. Introducing the D-term operator:
\begin{equation}
D = U^{\dag}_{1} U_{1} + U^{\dag}_{2} U_{2} - V^{\dag}_{1} V_{1} - V^{\dag}_{2} V_{2},
\end{equation}
the Fock space $\mathcal{F}_{X}(\zeta)$ of bosonic states for the conifold with resolution parameter $\zeta$ is spanned by
\be
|\spc  n_1,n_2; , m_1,m_2 \spc \rangle = \underset{i,j=1}{\overset{2}{%
{\displaystyle\prod}\,
}}\frac{\bigl(  U_{i}^{\dag}\bigr)  ^{n_{i}}  \bigl(  V_{i}^{\dag}\bigr)  ^{m_{j}} }{\sqrt{n_{i}!} \sqrt{m_{j}!}}\left\vert\,
0\,\right\rangle
\quad \text{with} \quad n_1 \!+ n_2\! - m_1 \! - m_2 = \zeta . 
\ee

The $\mathbb{P}^{1}$ of minimal size is the subspace of states in $\mathcal{F}_{X}(\zeta)$ with either trivial $U$ oscillator number, or trivial $V$ oscillator number. In particular, we see that for $\zeta > 0$, there is a $\mathbb{P}^{1}$ generated by just the $U$ creation operators
\be
\mathcal{F}_{\mathbb{P}^{1}}(\zeta >0)=\bigl\{ \,
|n_1,n_2; 0, 0 \rangle
\; ; \; n_{1}+n_{2}=\zeta\, \bigr\}\, ,
\ee
while for $\zeta < 0$, there is a $\mathbb{P}^{1}$ generated by just the $V$ oscillators:
\be
\mathcal{F}_{\mathbb{P}^{1}}(\zeta <0)=
\bigl\{ \, |0,0; m_1,m_2\rangle
\; ; \; m_{1}+m_{2}=\zeta\, \bigr\}\, .
\ee

\subsection{Fuzzy Landau-Ginzburg}

One of the remarkable features of the GLSM description of a target space is
that it provides a common language for describing both the geometric and Landau-Ginzburg phases of a
string compactification \cite{WittenPhases}. In this paper we have been primarily focussed on the
toric phase of the GLSM. It is tempting, therefore, to see whether we can also extend this discussion
to provide a fuzzy analogue of Landau-Ginzburg vacua.

Rather than be overly general, we follow the main example presented in \cite{WittenPhases} and
focus on the GLSM defined by a single $U(1)$ with $r$ chiral superfields
$z_{i}$ of charges $q_{i} > 0$, and another chiral superfield $p$ with charge $n < 0$. A compact hypersurface
is specified by including a superpotential:
\begin{equation}
W = p \cdot G(z_{1},...,z_{r}).
\end{equation}
Anomaly freedom of the GLSM is equivalent to the Calabi-Yau condition that
$q_1 + ... + q_r = n$. Further note that the superpotential $W$ is neutral with respect to the
$U(1)$ gauge symmetry. The D-term and F-term equations of motion are respectively:
\bea
\sum_{i} q_{i} |z_{i}|^{2} - n|p|^{2} = \zeta\nonumber\\[-2.5mm]\\[-2.5mm]
G = 0 \; \; \; \text{and} \; \; \; p \partial_{i}G = 0.\nonumber
\eea
In the geometric phase, $\zeta \gg 0$ and $G = p = 0$. In the Landau-Ginzburg phase, we instead
have $\zeta \ll 0$, $p = \sqrt{- \zeta / n}$, and the vacua
are described in terms of the Landau-Ginzburg superpotential
$\widetilde{W} = \sqrt{- \zeta / n} G(z_{1},...,z_{r})$ with vacua
described by the critical points of $\widetilde{W}$.

Again, the data of the GLSM provides a canonical quantization prescription for the fuzzy target. We restrict our
attention to the bosonic states. Introduce operators $Z_{i}$ and $P$ subject to the commutation relations:
\bea
[Z_{i} , Z_{j}^{\dag}] \is \delta_{ij}, \qquad \quad
 [P,P^{\dag} ]\, = \, 1,\nonumber
\eea
with all other commutators vanishing. Next introduce a vacuum $\vert 0 \rangle$ annihilated by the $Z_{i}$'s and $P$. We then construct
a Fock space $\mathcal{F}(\mathbb{C}^{r + 1})$ by acting on the vacuum state with the
$Z_{i}^{\dag}$'s and $P^{\dag}$.

The geometric phase (G) and Landau-Ginzburg phase (LG) of the target
are then given in terms of the Hamiltonian constraint operators:
\bea
D_{G} \is \sum_{i} q_{i} Z_{i}^{\dag} Z_{i} - n P^{\dag} P\\[3mm]
&& \hspace{-5mm} D_{LG} = -n P^{\dag} P
\eea
The hypersurface constraint in the geometric phase is that $G$ and $P$ both annihilate a state.
The hypersurface constraint in the Landau-Ginzburg phase is that $\partial_{i} G$ annihilates a state for all
$i = 1,...,r$.

Let us consider now the vector space of points for each phase. In the geometric phase, since $P$ must annihilate
a state, we see that the space of states is entirely described by acting on the vacuum with the $Z_{i}^{\dag}$'s. This then reproduces
the toric prescription we have been using throughout this paper.

The Landau-Ginzburg phase is qualitatively different. At radius $N = - k n$, the vector space of points is given by:
\bea
\mathcal{F}_{LG} 
= \Bigl\{  \; \bigl| \psi \, \rangle \in \mathcal{F}(\mathbb{C}^{r+1})\;  : \; \partial
_{i}G\left\vert \psi\right\rangle =0\, \Bigr\}
\eea
where $|\psi\rangle$ is an element of the Fock space ${\cal F}(\mathbb{C}^{r+1})$ spanned by states of the form
\bea
 \left\vert n_i;  k\right\rangle =\underset
{i=1}{\overset{r}{%
{\displaystyle\prod}
}}\frac{\bigl(  Z_{i}^{\dag}\bigr)  ^{n_{i}}}{\sqrt{n_{i}!}}\frac{\left(
P^{\dag}\right)  ^{k}}{\sqrt{k!}}\left\vert 0\right\rangle\;
\eea
A candidate state is annihilated by $r$ operators. This significantly
limits the number of points, at any choice of the fuzzy
radius. For example, for the quintic $G = Z_{1}^{5} + ... + Z_{5}^{5}$, we have
that for each $i$, $\partial_{i}G = 5 Z_{i}^{4}$ annihilates the set of bosonic states. This means:
\bea
\mathcal{F}_{LG_{quintic}}
= \text{span}\Bigl\{
|\spc n_i ; \spc k\spc  \rangle\, : \, 0\leq n_{i}%
\leq3\, \Bigr\}
\eea
Similar considerations hold for more generic choices of $G$.

One aspect of the commutative description which is missing from this discussion is the fuzzy
analogue of renormalization group flow to the Landau-Ginzburg phase. It would be interesting to develop
this discussion further. Let us also note that the approach to local mirror symmetry in \cite{HoriVafa} (see also \cite{Morrison:1995yh})
relies heavily on the interplay between the toric phase of one geometry, and the Landau-Ginzburg phase
of its mirror.

More generally, mirror symmetry constitutes a deep feature of stringy geometry.
It is quite tempting to speculate that since toric geometry provides a natural setting for
setting up mirror pairs \cite{Roan}, the Batyrev construction \cite{Batyrev}
may have a non-commutative analogue.

\section{Physical Applications}\label{sec:Physical}

In this section we consider some other aspects and possible applications of Fuzz theory.

\subsection{Discretuum of Gauge Couplings}

One application already mentioned in the Introduction pertains to moduli stabilization.
There is a natural limit where the volumes of four-cycles collapse to zero size. In this
paper we have focussed on a limit where we expect these volumes to be at zero size, but
in which we nevertheless retain a perturbative gauge coupling constant,
which is effectively quantized in units of $1 / g_{s}$. The basic relation
we have found is that the number of points $N$ determines the Yang-Mills coupling via:
\begin{equation}
\frac{g^{2}_{YM}}{4 \pi} = \frac{g_{s}}{N}
\end{equation}
where $N$ is in turn fixed by the amount of $\mathcal{B}$-field flux passing through the seven-brane.

Different values of the discrete parameter $N$ define different vacua of the theory. These vacua are connected by domain walls in $\mathbb{R}^{3,1}$
which change the amount of $\BB$-flux threading the seven-brane. A domain wall which can change the amount of $\BB$-flux must
be charged under the $B$-field, so we expect (in a perturbative IIB frame) the domain wall is given by
an NS5-brane wrapping a three-chain $\Gamma_{3}$ with boundary on the two-cycle Poincar\'e dual to the cohomology class of $\BB$.

To motivate this answer, consider again the case of the conifold transition with $N_c$ D5-branes wrapping the $S^2$ of the
resolved conifold. After passing through the geometric transition, the amount of $\BB$-flux through the $S^{2}$ becomes the amount of NS three-form flux
threading the non-compact three-cycle of the deformed conifold:
\begin{equation}
\int_{S^{2}} B_{NS} = \int_{B_{3}} H_{NS}.
\end{equation}
In a non-compact setting, $H_{NS}$ becomes quantized once we
impose a suitable boundary condition for the non-compact directions
of the geometry. In this case, we write $H_{NS} = k \cdot H^{min}_{NS}$ where
$H^{min}_{NS}$ integrates over $B_3$ to one unit of three-form flux. Changing the amount of
three-form flux threading $B_{3}$ amounts to wrapping an NS5-brane on the three-cycle Poincar\'e dual to $H^{min}_{NS}$. In the
present case, this is just the three-cycle $B_{3}$. Thus, the tension of the domain wall connecting vacua is $|\Delta k| \Vol(B_{3}) / g^{2}_{s}$, where
$|\Delta k|$ denotes the change in $H_{NS}$ across the domain wall. Passing
back to the original small resolution, this NS5-brane wraps a non-compact three-chain with boundary on the $S^2$.

One expects something similar to happen in the case of the fuzzy seven-brane theory. Here, the Poincar\'e dual of the
flux $\mathcal{B}$ defines a two-cycle $[\mathcal{B}]$ inside of the
surface wrapped by $\cal{S}$. An NS5-brane wrapping a non-compact three-chain with boundary on $[\mathcal{B}]$ connects
vacua with different values of the $\BB$-flux. Similar considerations hold in other duality frames.

Each value of $N$ defines an effective tension for the seven-brane. Coupling this system to gravity and assuming
all other contributions to the vacuum energy density remain fixed,
this suggests that vacua with different values of $N$ have different energy densities,
which are connected by domain walls with tension $\Vol(\Gamma_{3})$, where $\Gamma_{3}$
is the three-chain wrapped by the five-brane. Here we drop factors of $g_s$ since it is an order one quantity
in F-theory. In the thin-wall approximation, the tunnelling rate is then given by:
\begin{equation}
\Gamma_{N \rightarrow N - k} \sim \exp \left( - \frac{27 \pi^{2}}{2} \cdot k \Vol(\Gamma_{3})^{4}    \right).
\end{equation}

\subsection{Threshold Corrections}

One of the limitations of higher-dimensional gauge theories is that at energy scales above the Kaluza-Klein scale, an entire
tower of KK modes enter the spectrum limiting the range of validity for the effective field theory. It is then
necessary to embed the theory in an ultraviolet completion of the field theory, such as string theory.
In the case of fuzzy extra dimensions, the number of dynamical KK modes truncates at finite order, regulating the higher-dimensional
theory.

At a practical level, making the internal directions non-commutative still retains much of the
geometric structure of the original commutative space. Truncating the KK spectrum has another
benefit, which is that it makes threshold corrections
both easier to catalog and to compute. One loop perturbation theory
in a gauge theory with $N_{\deg}$ degrees of freedom is valid provided:
\begin{equation}  \label{TrustPert}
\frac{g_{YM}^{2}N_{\deg}}{16\pi^{2}}\ll 1.
\end{equation}
For a fuzzy seven-brane theory, we have the basic
relation $\alpha_{GUT} = g_{s} / N$. In tandem with the expectation that $N_{\deg} \sim N^{2}$,
this suggests $ g_{YM}^{2} N_{deg} / 16 \pi^{2} = g_{s} N / 4 \pi$. For
$N \sim O(30)$, this might at first suggest that one loop perturbation theory breaks down, once all of the KK modes are
included.

This counting is a bit naive, however, because supersymmetry will induce
some additional suppression in the size of threshold corrections. Consider the contribution of the bulk KK modes
of the seven-brane to the running of the four-dimensional gauge coupling. Up to subtleties connected with line bundle
data (which will make some KK modes fill out rectangular matrices),
the KK modes transform as $N \times N$ matrices, where $N$ is the number of points. The mode content can be organized as
$N^{2}$ $\mathcal{N} = 1$ adjoint-valued massive vector multiplets, and $N^{2}$ $\mathcal{N} = 2$ adjoint-valued hypermultiplets.
Since each massive vector multiplet can be viewed as a massless vector multiplet and an adjoint-valued chiral
superfield, we see that the beta function vanishes at leading order in $N^{2}$. One does expect subleading corrections of order $N$, for example,
from matter fields localized on curves. In this case, we obtain the more reliable estimate that the effective number of degrees of freedom
entering a threshold correction is more on the order of $N_{deg} \sim N$. Thus, we obtain
\begin{equation}
\frac{g_{YM}^{2} N_{deg}}{16 \pi^{2}} \sim \frac{g_{s}}{4 \pi} \ll 1
\end{equation}
so that perturbation theory remains meaningful. Detailed studies of threshold effects
in commutative seven-branes in F-theory GUTs have
appeared in \cite{Blumenhagen:2008aw,DWII,Conlon:2009qa}. It would be interesting
to revisit these analyses in the present context.

\subsection{Low Scale Fuzz}

In theories with a low scale for quantum gravity, there is a potential worry that processes otherwise
protected by making the Planck scale very high could now be violated more significantly.
In some sense, this is just a reflection of the fact that a typical string
compactification can look quite complicated at the string or Planck scale.

As we have emphasized in this paper, there are (at least)
two Kaluza-Klein scales associated with an extra-dimensional theory in string theory:
One for open string KK modes $M^{KK}_{open}$, and one for the closed string KK modes $M^{KK}_{closed}$.
Taking this distinction seriously, we can in principle lower the scale of open string
KK excitations all the way to the TeV scale:
\begin{equation}
M^{KK}_{open}\sim\text{TeV}
\end{equation}
while still keeping the closed string modes and quantum gravity effects decoupled.\footnote{In a scenario with
TeV scale non-commutativity, but with $M^{KK}_{closed} = M_{pl}$ \textit{finite}, the string scale $M_{string}$
is equal to the geometric mean $M_{string} \sim \sqrt{M^{KK}_{open} \cdot M^{KK}_{closed}} \sim 10^{11}$ GeV.}

This type of scenario certainly shares features similar to those of TeV scale extra dimensions \cite{Antoniadis:1990ew},
those of large extra dimensions \cite{ADD,AADD} and those with warped extra dimensions
\cite{RS1,RS2}.\footnote{See for example \cite{Carlson:2001bk} for a collider study
of some aspects of non-commutative extra dimensions.}
Let us stress, however, that our limit is qualitatively different from these scenarios,
because we have explicitly decoupled the closed string modes from the description. Indeed,
since $M^{KK}_{closed} \gg M^{KK}_{open}$, non-commutative extra dimensions allow KK\ gluons
and other KK\ open string excitations with no accompanying KK\ gravitons.
This leads to some differences in the low energy phenomenology. For example,
a common collider signature of extra-dimensional scenarios involves production of KK gravitons.
With fuzzy extra dimensions, this signature would be absent.

There are clearly many possible variants of extra dimensional ideas and many
mechanisms previously discussed in such frameworks likely possess fuzzy
analogues. Though it is in principle possible to still maintain some aspects of GUTs in TeV to intermediate scale
extra-dimensional models, (see for example \cite{DienesI, DienesII, HallNomura, ArkaniHamed:2001vr, Goldberger:2002pc}),
putting $M^{KK}_{open}$ near the TeV scale instead suggests widening our scope beyond the GUT paradigm.

Indeed, one might contemplate models in which some or all of the Standard Model fields have non-commutative counterparts.
In extra-dimensional models, it is common to regulate the extra dimensions with a
lattice formulation. At the very least, working with a fuzzy prescription would provide a way to maintain more of
the smooth structure of the internal geometry, and may also open up new avenues for model building with
extra dimensions.

\section{Towards a Holographic Dual}\label{sec:Holo}

The seven-brane on a non-commutative four-cycle with $N$ points can be viewed as tiled by $N$ D3-branes that
each occupy one Planck cell. In the limit where the number of fuzzy points becomes large,
the D3-branes form a dense mesh, and the non-commutative geometry
converges to the commutative description. This structure hints at a possible
holographic dual description.\footnote{In practical
applications, $\alpha_{GUT} = g_{s}/ N $ so that $N \sim O(30)$
when $g_{s}$ is an order one parameter (as in F-theory). Insofar as this is a large number,
this expansion is also potentially relevant in more phenomenological applications.} Here we make
some qualitative remarks about the form of this holographic dual. Though the application is
different, the appearance of fuzzy geometry in Klebanov-Strassler type solutions has been studied
for example in \cite{Maldacena:2009mw}. Our situation also shares some similarities with the description given in \cite{Gaiotto:2004pc} of
extremal black holes in IIA string theory in terms of D0-branes tiling the horizon (see also \cite{Douglas:2008es}).

Consider a seven-brane with gauge group $G = SU(N_{c})$ wrapped on a
non-commutative four-cycle with a large number of fuzzy points, $N$. The qualitative behavior
of this theory depends on the energy scale. At energies
below the scale of Kaluza-Klein excitations, the low energy theory consists
of a single four-dimensional $SU(N_{c})$ gauge theory. The number of seven-branes
$N_{c}$ cannot be too large without overclosing the compactification \cite{VafaFTHEORY}.
As we proceed up in energy, additional light modes begin to enter the spectrum. For instance, the vector multiplet
of the non-commutative theory is given by a $(0,0)$-form which can be viewed as an $N \times N$ matrix:
\begin{equation}
V^{i \overline{i}}\in Mat(N \times N).
\end{equation}
The other bulk KK modes roughly fill out $N \times N$ matrices with shifts $N \rightarrow N + q_{i}$ set by line bundle data.
Matter on curves crudely correspond to $N^{1/2} \times N^{1/2}$ matrices, though the precise exponent
of the $N$ scaling depends on details of the matter curve.

Once we pass the threshold for the open string KK states, we find additional degrees of freedom. For example,
the $N^{2}$ massive vector multiplets can be viewed as $N^{2}$ massless vector multiplets transforming in $N^{2}$ copies of
$SU(N_{c})$, and $N^{2}$ adjoint-valued chiral superfields which give mass to these
vector multiplets through their vevs.

The appearance of so many additional vector multiplets suggests that we have tiled the seven-brane with a collection of D3-branes. Indeed,
turning on a non-trivial $\mathcal{B}$-flux through the worldvolume of the seven-brane induces a net D3-brane charge equal to the number of
points $N$. These D3-branes have finite extent because they experience a collective Myers effect induced by the background fluxes \cite{Myers:1999ps}, and
so rather than sitting at commutative points, the D3-branes puff up and sit at fuzzy points of the non-commutative geometry.
Indeed, though it is tempting to speak of the classical position of $N$ D3-branes located at various positions
on the seven-brane, in the non-commutative setting we cannot simultaneously diagonalize these positions. The
configuration of D3-branes is therefore better viewed as filling out an $N \times N$ matrix.
The gauge group of this configuration exactly matches what we expect in the fuzzy theory. The
bifundamentals correspond to $3 - 3^{\prime}$ strings stretched between the various
D3-brane factors, and set the mass scale of open string Kaluza-Klein excitations.

At even higher energies, we pass back to an unHiggsed theory in which the $3 - 3^{\prime}$ strings no longer condense. In this limit,
the $N$ D3-branes sit on top of each other at the tip of the singularity,
and organize according to an $SU(N_{c} \times N)$ gauge theory, with some appropriate
matter content and scalar potential that reflects the presence of the local del Pezzo geometry.
Viewed as a collection of $N_{c} \times N$ D3-branes sitting at a single point of the geometry,
the $SU(N_{c} \times N)$ gauge theory has Yang Mills coupling $g^{2}_{\rm{HI}} = 4 \pi g_{s}$. Taking the large
$N$ limit suggests organizing our perturbative expansion according to the 't Hooft parameter:
\begin{equation}
\lambda_{\rm{HI}} = g^{2}_{\rm{HI}} N_{c} \times N = 4 \pi g_{s} N_{c} \times N.
\end{equation}
At large $\lambda_{\rm{HI}}$, we expect to obtain a holographic dual to the stack of $N_c \times N$ D3-branes, supported by
the five-form flux $F_{5}$ \cite{juanAdS}. The closed string modes in this dual geometry correspond to glueball like excitations
of the large $N$ gauge theory. Since the number of chiral matter fields is assumed to be much smaller than $N_c \times N$,
the flavor seven-branes can to first order be treated in a probe approximation.

We could ask how one would recover the low energy fuzzy description from this holographic dual perspective.
Descending from the high energy theory, one enters a regime where the wave functions
of the D3-branes spread out over the seven-brane.
Viewed from a classical perspective, we are performing a breaking pattern
of the form $SU(N_{c} \times N) \supset SU(N_{c})^{N}$, where we rearrange the three-brane configuration into
$N$ stacks of $N_{c}$ D3-branes.
The gauge coupling of the parent gauge theory descends directly to each stack of D3-branes so that $g^{2}_{\rm{HI}} = g^{2}_{\rm{MID}}$.
Note, however, that in the vicinity of each single stack, the gauge group is
only $SU(N_{c})$. The radius of curvature of a putative holographic dual to this smaller stack
would then be set by the 't Hooft coupling $\lambda_{\rm{MID}} = 4 \pi g_{s} N_{c}$.
Insofar as $g_{s}$ is at most order one, and since we cannot arbitrarily increase
the rank of a seven-brane gauge group, we see that the holographic dual of this
configuration will have a string scale curvature.

Finally, we descend back down to the Higgs branch of the D3-brane configuration,
in which only a single $SU(N_{c})$ gauge group factor remains. This can be
viewed as embedded in the diagonal subgroup of $SU(N_{c})^{N}$. This breaking pattern affects the value of
the gauge coupling constant of the gauge theory on the Higgs branch so that:
\begin{equation}
g^{2}_{\rm{LO}} = \frac{g^{2}_{\rm{HI}}}{N} = \frac{4 \pi g_{s}}{N}.
\end{equation}
Thus we recover the by now familiar relation $ g^2_{YM} = g^2_{LO} = 4 \pi g_{s} / N$.

\begin{figure}[t]
\begin{flushright}
 \epsfig{figure=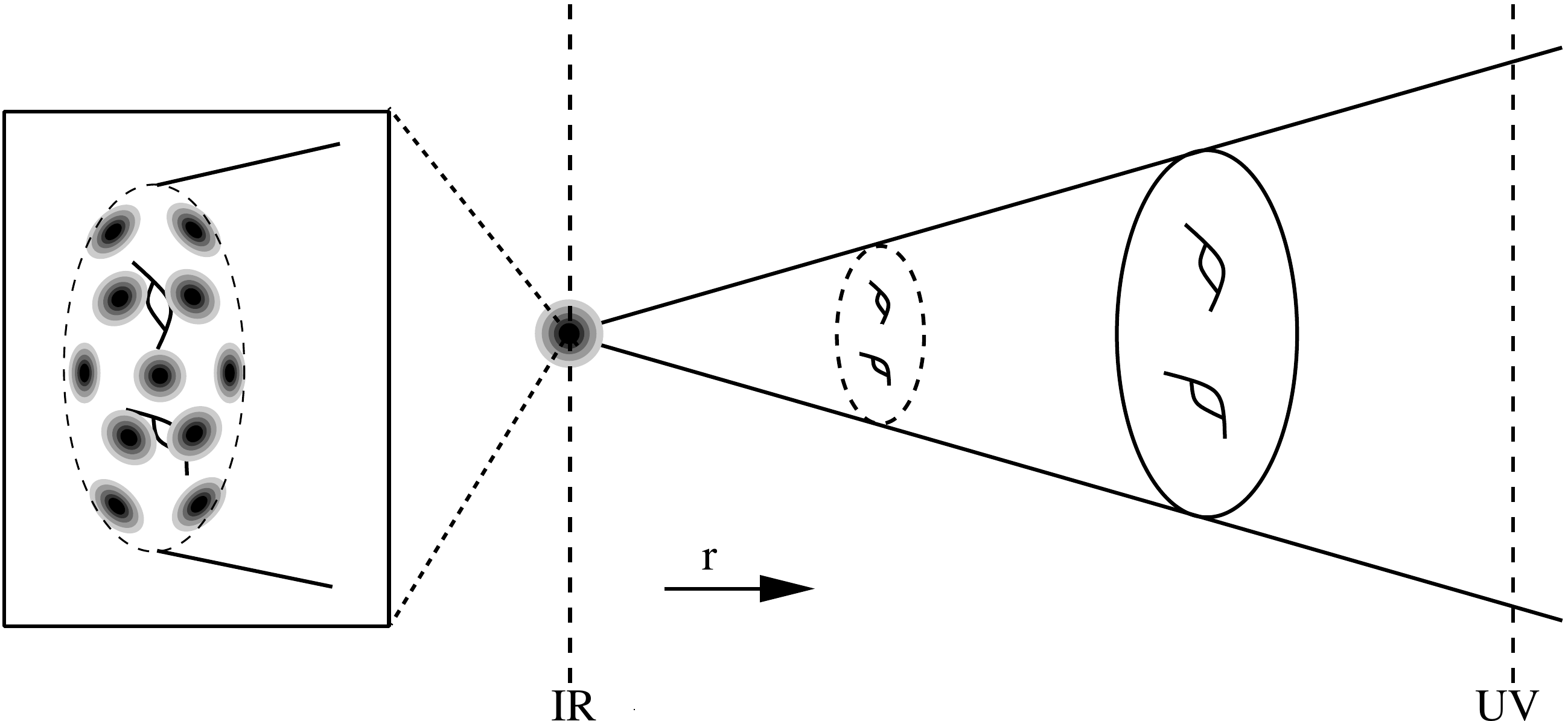,scale=.6}
\end{flushright}
  \caption{Depiction of the holographic dual to a seven-brane wrapping a non-commutative four-cycle.
  In the open string picture, this configuration can alternatively be viewed as
  a collection of $N$ D3-branes sitting at the fuzzy points of
  the geometry. At higher holographic energy scales $r$, the D3-branes can clump together,
  and we recover a corresponding weakly curved holographic dual.}
  \label{HoloFuzz}
\end{figure}

Even though in the IR, the gauge symmetry has been broken to $SU(N_c)$, the UV presence of
the large gauge group $SU(N \times N_c)$ is still reflected in the spectrum of KK excitations
of the low energy theory.
This allows one to organize the perturbation theory of the fuzzy seven-brane in terms
of a large $N$ expansion \cite{'tHooft:1973jz}.
For example, the vector multiplets organize as an
$N \times N$ matrix $V^{i \overline{i}}$. The Feynman rules
require that in interactions involving
the vector multiplets $V^{i \overline{i}}$, an $i$ index
must link up with an $\overline{i}$ index. This means that
we have an analogue of the double line notation. In this $1/N$ expansion,
the corresponding 't Hooft coupling would now be:
\begin{equation}
\lambda_{LO} \equiv g_{YM}^{2} N_{c} \times N = 4 \pi g_{s}N_{c}.
\end{equation}
The resulting large $N$ expansion for amplitudes can
then be interpreted as the loop expansion of a closed
string theory which lives at the highly curved
tip of the closed string geometry. Indeed, for small $N_{c}$
the 't Hooft coupling is order one indicating that the holographic dual geometry has
string scale curvature. See figure \ref{HoloFuzz} for a depiction.


\section{Conclusions}\label{sec:CONCLUDE}

Local model building provides a general strategy for embedding the Standard Model in string theory
by focussing first on the open string degrees of freedom, keeping the closed string degrees of freedom decoupled. For
seven-branes wrapping a compact four-cycle, fully decoupling the closed string sector requires working in
the zero slope limit, and simultaneously demanding that this four-cycle collapses to zero closed string volume. The open
string volume is still non-zero, and is supported by a two-form flux $\mathcal{B}$. This is a quite natural regime
for high scale stabilization of K\"ahler moduli, and further, leads to significant simplifications in the dynamics of the
higher-dimensional gauge theory defined by the seven-brane. In this limit,
the internal geometry becomes non-commutative, and the spectrum of KK excitations for the seven-brane theory truncates. We have developed a
general effective action for fuzzy seven-branes, and have explored some of the consequences of non-commutative extra dimensions.
We now turn to some further avenues of investigation.

Our approach to non-commutative toric spaces has been based on effectively quantizing the data of a classical gauged linear sigma model.
This appears to be a physically reasonable starting point in particular because $\BB$-fields threading the worldvolume of D-branes on
$\mathbb{C}^{r}$ are well-understood, and the symplectic quotient provides a natural formulation for treating more general toric spaces.
Even so, the quantization procedure is still at the level of a motivated ansatz. \textit{Deriving} this
structure from the gauged linear sigma model would likely provide further insight into stringy realizations of fuzzy toric spaces. Along these
lines, we have seen that there are analogues of many by now standard features of GLSM constructions such as flop transitions and Landau-Ginzburg phases.
It is far from clear that this exhausts the full list of possibilities, and it may be that there are
transitions which do not possess commutative analogues.

We have seen that the dynamics of seven-branes admits a $1/N$ expansion in the number of points $N$ of the non-commutative geometry. This
suggests an intuitive picture based on viewing the seven-brane as a collection of $N$ puffed up D3-branes
tiling the interior. Developing gravity duals of such D3-brane configurations would likely provide
further insight into strongly coupled gauge theories built from such seven-branes, and would open up the possibility of finding
a holographic dual of weakly coupled low rank gauge theories, like the Standard Model. This picture also
suggests an intriguing interpolation between a single unified $SU(5)$ GUT model on a fuzzy seven-brane,
and the collective dynamics of $5 \times N$ D3-branes. Though this is similar to what is expected in deconstructed extra-dimensional
theories \cite{ArkaniHamed:2001ca}, the construction we have presented retains additional geometric
structure which is typically absent in lattice regularization of higher-dimensional theories.

\section*{Acknowledgements}

We thank M. Bianchi, B. Jurke and C. Vafa for helpful discussions. Some of the work
of this paper was completed at the ``Strings at the LHC and Early Universe''
workshop at the KITP in Santa Barbara, and JJH and HV thank the KITP for
hospitality. The work of JJH is supported by the NSF under grant PHY-0503584.  The work of
HV is supported by the NSF under grant PHY-0243680.

\appendix

\section*{Appendix A: Fuzzy $\mathbb{P}^{1}$}

In this Appendix we present some additional features of fuzzy
$\mathbb{P}^{1}$. We begin by making contact
with an alternative formulation of this space, which is equivalent to what we have
discused.

We start with commutative coordinates $x_{i}$ of $\mathbb{R}^{3}$ subject to the condition:
\begin{equation}
x_{1}^{2} + x_{2}^{2} + x_{3}^{2} = R^{2}.%
\end{equation}
The ordinary non-commutative construction then
quantizes this information by promoting the coordinates $x_{i}$ to operators $X_{i}$ which
satisfy the commutation relations of the $su(2)$ algebra:
\begin{equation}
[X_{i},X_{j}] = i \theta \varepsilon_{ijk} X_{k}.
\end{equation}
Here $\theta$ is a formal quantization parameter.
The overall size of the sphere is then
fixed by specifying a spin $j$ representation of
the angular momentum algebra:
\begin{equation}
X_{1}^{2} + X_{2}^{2} + X_{3}^{2} = \theta^{2} j(j+1).
\end{equation}
Rescaling the $X$'s can be absorbed into a change of the quantization
parameter $\theta$. One common convention is $\theta^{2} = 1/j$, so that the volume
of the sphere scales as $j+1$.

To make contact between this description and the construction we have been using in this paper,
we first review the commutative toric description of $\mathbb{P}^{1}$.
Recall that in the commutative theory, $\mathbb{P}^{1}$ can be defined in terms of the GLSM of a single $U(1)$ with two
chiral superfields $u_{1}, u_{2}$ of charge $+1$. The volume of the $\mathbb{P}^{1}$ is
then fixed by the D-term constraint:
\begin{equation}
\left\vert u_{1}\right\vert ^{2}+\left\vert u_{2}\right\vert ^{2}  =\zeta.
\end{equation}

We now quantize this data. Denote the analogue of the $u_{i}$
by the quantized variables $U_{i}$. Let $\vert 0 \rangle$
denote the vacuum state annihilated by the $U$'s. The Fock space
$\mathcal{F}(\mathbb{C}^{2})$ is given by all states obtained by acting on $\vert 0 \rangle$ with the $U^{\dag}$'s.
The spectrum of the D-term operator $D = U_{1}^{\dag} U_{1} + U_{2}^{\dag}U_{2}$ then defines a grading of
$\mathcal{F}(\mathbb{C}^{2})$. We denote by $\mathcal{F}_{X}(N)$ the bosonic points of
$X = \mathbb{P}^{1} $ with quantized FI parameter $N$. The space $\mathcal{F}_{X}(N)$ is spanned by the states:
\begin{equation}
|\, n_1,\spc n_2\spc \rangle\,  =\, \frac{(U^{\dag}_{1})^{n_{1}}(U^{\dag}_{2})^{n_{2}}}%
{\sqrt{n_{1}!n_{2}!}} \vert0 \rangle
\end{equation}
with $n_1 + n_2 = N$.

Associated with the $U$'s there is a canonical $su(2)$ algebra given as:
\begin{equation}
J_{i} = \frac{1}{2} U^{\dag} \sigma_{i} U
\end{equation}
where $U^{\dag}$ is a two-component vector, and the $\sigma$'s are the Pauli
matrices. Introducing $J_{+} = J_{1} + i J_{2}$ and $J_{-} = J_{1} - i J_{2}$,
the angular momentum generators are:
\begin{align}
J_{+}  &  = U_{1}^{\dag}U_{2}\nonumber\\[1mm]
J_{-}  &  = U_{2}^{\dag}U_{1}\\
J_{3}  &  = \frac{1}{2}(U_{1}^{\dag}U_{1} - U_{2}^{\dag}U_{2}).\nonumber
\end{align}
The South pole state of $\mathbb{P}^{1}$ corresponds to:
\begin{equation}
\vert\text{South} \rangle = |\, 0, \spc N\, \rangle . 
\end{equation}
Indeed, acting by $J_{-}$ annihilates this state, and acting by $J_{+}$ $N
+ 1$ times yields, up to a constant of proportionality, the North pole state:
\begin{equation}
\vert\text{North} \rangle = | \, N, \spc 0 \, \rangle .
\end{equation}

\subsection*{Line Bundle Cohomology on Fuzzy $\mathbb{P}^{1}$}

We now discuss line bundle cohomology for fuzzy $\mathbb{P}^{1}$. Our general
strategy is to map the algebraic data of the commutative theory
to the non-commutative geometry.

A general zero form of the degree $Q$ line bundle $\mathcal{O}_{\mathbb{P}^{1}}(Q)$
can be presented as a polynomial in the $u$'s and $\overline{u}$'s with net
GLSM\ charge $Q$. Global sections of $\mathcal{O}_{\mathbb{P}^{1}}(Q)$
correspond to degree $Q$ polynomials in the $u$'s:%
\begin{equation}
H^{0}(\mathbb{P}^{1},\mathcal{O}(Q))=\bigl\{  G(u_{1},u_{2})|\deg G=Q\bigr\}
\end{equation}
This implies that there are global sections for $Q\geq0$, but for $Q<0$,
$\mathcal{O}_{\mathbb{P}^{1}}(Q)$ does not possess any global sections. In
particular, for $Q\geq0$, the number of global sections is $Q+1$.

The case of $(0,1)$-forms is similar. The canonical line
bundle of $\mathbb{P}^{1}$ is $\mathcal{O}_{\mathbb{P}^{1}}(-2)$, so a natural
basis element for holomorphic one-forms will be $de_{\mathbb{P}^{1}}$, where $e_{\mathbb{P}^{1}}$ has
GLSM charge $+2$. The condition that this monomial remains well-defined over
all of $\mathbb{P}^{1}$ effectively fixes $e_{\mathbb{P}^{1}}$ to be $u_{1}u_{2}$, since the
differentials $d(u_{1}^{2})$ and $d(u_{2}^{2})$ both vanish on $\mathbb{P}%
^{1}$. Note that $d(u_{1}u_{2})=u_{1}du_{2}+u_{2}du_{1}$, so that since
$u_{1}$ and $u_{2}$ cannot both simultaneously vanish, this differential is
the unique choice.\footnote{The set $\{u_{1}u_{2}\}$ for $\mathbb{P}^{1}$ generates the
Stanley-Reisner ideal for $\mathbb{P}^{1}$. More generally, for a toric variety $X$,
we can form a collection of minimal monomials such that for each monomial, not all of the factors can
simultaneously vanish. The main utility for our purposes is that such elements
provide a convenient basis of monomials to construct differential forms from.
Stanley-Reisner ideals figure prominently in the proposed algorithmic approach to line bundle cohomology in
\cite{Blumenhagen:2010pv}. It would be interesting to use this algorithmic approach to line bundle cohomology
to find explicit representatives of fuzzy cohomology groups.}

To fix actual representatives, it is actually enough to work in terms of the
Serre dual description of the cohomology group. Using the isomorphism:%
\begin{equation}
H^{1}(\mathbb{P}^{1},\mathcal{O}_{\mathbb{P}^{1}}(Q))\simeq H^{0}%
(\mathbb{P}^{1},\mathcal{O}_{\mathbb{P}^{1}}(-Q)\otimes\mathcal{O}%
_{\mathbb{P}^{1}}(-2))^{\ast},
\end{equation}
we conclude that the elements of $H^{1}(\mathbb{P}^{1},\mathcal{O}%
_{\mathbb{P}^{1}}(Q))$ can be viewed as living in the dual space to
$H^{0}(\mathbb{P}^{1},\mathcal{O}_{\mathbb{P}^{1}}(-Q-2)).$ In other words, a
convenient basis for this cohomology group can be presented as:%
\begin{equation}
H^{1}(\mathbb{P}^{1},\mathcal{O}_{\mathbb{P}^{1}}(Q))\simeq\left\{
G(u_{1}^{\ast},u_{2}^{\ast})\otimes d(\overline{u_{1}u_{2}}) : -\deg
G-2=Q\right\}
\end{equation}
where the $u^{\ast}$'s are to be viewed as dual coordinates such that
$u^{\ast}(u)=1$. This convention requires the dual coordinates $u^{\ast}$ to have GLSM charge $-1$.
Since the overall GLSM charge of the one-forms is fixed to
be $Q$, this implies $\deg G=-Q-2$, so that for $Q\geq-1$, the cohomology
group vanishes, and for $Q<-2$, the dimension of the cohomology group is
$-Q-1$.

We now turn to the fuzzy cohomology groups. The quantized analogue of $du_{i}$ is the fermionic
oscillator $C_{i}$. To distinguish from the cohomology of the commutative geometry, we
refer to our $\mathbb{P}^{1}$ as $X$. Fixing a degree $Q$ line bundle
$L^{(Q)}$, the space $H^{0}(X,L^{(Q)})$ is spanned by the operators:%
\begin{equation}
H^{0}(X,L^{(Q)})=\left\{  G(U_{1},U_{2}) : [D , G] = - Q G \right\}
\end{equation}
Indeed, acting on the bra vacuum state $\left\langle 0\right\vert $ by
elements of $H^{0}(X,L^{(Q)})$ generates a $Q+1$
dimensional vector space which is isomorphic to the dual vector space of
points of a fuzzy $\mathbb{P}^{1}$ of size $Q$:%
\begin{equation}
\mathcal{F}_{X}^{\ast}(Q)=\left\{  \left\langle 0\right\vert G(U_{1},U_{2}) : [D , G] = - Q G \right\}  .
\end{equation}
Note that having non-trivial $H^{0}(X,L^{(Q)})$ requires $Q \geq 0$.

Next consider the elements of the first cohomology group. The fuzzy analogue
of Serre duality is provided by acting with elements of
$H^{0}(X,L^{(Q)})$ on the bra vacuum state $\left\langle 0\right\vert $, and elements of
$H^{1}(X,L^{(Q)})$ on the ket vacuum state $\left\vert 0\right\rangle $. Indeed,
there is a canonical pairing between $U$ and $U^{\dag}$. Thus, elements of
$H^{1}(X,L^{(Q)})$ correspond to operators built exclusively from creation operators.
We restrict our discussion to cases where the commutative theory cohomology group is
non-trivial so that $Q < -1$. Exploiting the presence of this canonical pairing, we can make
the identification:
\begin{equation}
H^{1}(X,L^{(Q)})=\bigl\{  G(U_{1}^{\dag},U_{2}^{\dag}) \times
(U_{1}^{\dag} C_{2}^{\dag} + U_{2}^{\dag} C_{1}^{\dag}) : [D , G] = (Q + 2) G \bigr\}  .
\end{equation}
Indeed, acting on the ket vaccum state $\left\vert 0\right\rangle $ by degree
$-Q-2$ polynomials in the $U^{\dag}$'s, we see that each of the states of
$\mathcal{F}_{X}(-Q-2)$ naturally pairs with a state of $\mathcal{F}_{X}^{\ast}%
(-Q-2)$ associated with acting on the bra vacuum state $\left\langle 0\right\vert $ with operators of
the cohomology group $H^{0}(X,L^{(-Q-2)})$.

Let us note that in our discussion the $U^{\dag}$'s play two roles: first
as anti-holomorphic coordinates for functions, and second as Serre dual
coordinates for cohomology groups.

\section*{Appendix B: Intersection Theory on Fuzzy $\mathbb{P}^{1}\times\mathbb{P}^{1}$}

In this subsection we develop intersection theory on fuzzy $\mathbb{P}^{1}\times\mathbb{P}^{1}$.
To set notation, we introduce oscillators $U_{i}$ and $V_{i}$ for the respective factors
$\mathbb{P}_{I}^{1}$ and $\mathbb{P}_{II}^{1}$ of
$\mathbb{P}^{1}_{I} \times \mathbb{P}^{1}_{II}$. We denote the vector space of bosonic points
for $\mathbb{P}^{1}_{I} \times \mathbb{P}^{1}_{II}$ at K\"ahler classes $N_{I}$ and $N_{II}$ by
$\mathcal{F}_{X}(N_{I},N_{II})$.

In the commutative theory, divisors are defined as the vanishing locus of a bihomogeneous
polynomial in the $u$'s and $v$'s. In the non-commutative geometry, this locus is associated
with the linear subspace of $\mathcal{F}_{X}(N_{I},N_{II})$ annihilated by the corresponding
operator in $U$ and $V$.

In the commutative geometry, divisors will typically intersect at some number
of points. We now discuss the analogous intersection number for fuzzy
divisors. Given two operators $P$ and $Q$ of respective bidegrees
$(p_1,p_2)$ and $(q_1,q_2)$, the number of points in the intersection is:%
\begin{equation}
\# (P = 0) \cap (Q = 0) =\dim(\ker P\cap \ker Q). \label{PQintersection}%
\end{equation}

Intersection theory in the commutative theory is fully determined by the
bidegrees of the bihomogeneous polynomials $P$ and $Q$. Indeed, letting
$\sigma_{1}$ and $\sigma_{2}$ denote the divisors corresponding to the two
$\mathbb{P}^{1}$ factors of $\mathbb{P}^{1}\times\mathbb{P}^{1}$ such that
$\sigma_{i}\cap\sigma_{j}= 1 - \delta_{ij}$, the divisor classes for $P$ and $Q$ are:%
\begin{align}
\left[  P\right]   &  =p_{1}\sigma_{2}+p_{2}\sigma_{1}\nonumber\\[-3mm]\\[-3mm]\nonumber
\left[  Q\right]   &  =q_{1}\sigma_{2}+q_{2}\sigma_{1}.
\end{align}
The intersection number of $P$ and $Q$ in the commutative theory is then given
by:%
\begin{equation}
[P] \cap [Q] =\left(  p_{1}\sigma_{2}+p_{2}\sigma
_{1}\right)  \cap\left(  q_{1}\sigma_{2}+q_{2}\sigma_{1}\right)  =p_{1}%
q_{2}+p_{2}q_{1}\text{.}%
\end{equation}

We now establish a similar relation in the non-commutative geometry, with the intersection
number given by equation (\ref{PQintersection}). In order to keep our
discussion as \textquotedblleft generic\textquotedblright\ as possible, we
assume that each linear map induced by acting by either $P$ or $Q$ has trivial
cokernel.\footnote{Recall that given a linear map $T: A\rightarrow B$ between
vector spaces $A$ and $B$, the cokernel is defined as coker$(T)\equiv
B/$im$(T)$. Having trivial cokernel means that the map $T~$is onto, so that
for every vector $b$ in $B,$ there exists a vector $a$ in $A$ such that
$T(a)=b$.}

At the level of set theory, the mutual intersection of the two kernels is:
\begin{equation} \label{kernels}
\dim(\ker P \cap \ker Q ) = \dim \ker P + \dim \ker Q - \dim \ker PQ.
\end{equation}
Since we are assuming trivial cokernel for each operator, the rank nullity theorem implies:
\bea
\dim \ker P = N_{I} N_{II} - (N_{I} - p_1)(N_{II} - p_2)\nonumber\\[2mm]
\dim \ker Q = N_{I} N_{II} - (N_{I} - q_1)(N_{II} - q_2)\\[2mm]
\dim \ker PQ = N_{I} N_{II} - (N_{I} - p_{1} - q_{1})(N_{II} - p_{2} - q_{2}).\nonumber
\eea
Returning to equation (\ref{kernels}), we find:
\begin{equation}
\dim(\ker P \cap \ker Q ) = p_{1}q_{2} + p_{2}q_{1},
\end{equation}
which we recognize as the intersection number for divisors in the commutative
geometry. Let us note that in less generic situations where some of the maps
are not onto, or in cases where the action by an operator annihilates all
points, the effects of the non-commutativity will be more prominent. It
would be interesting to develop intersection theory for more general
fuzzy toric spaces.

\newpage
\
\bibliographystyle{ssg}
\bibliography{fuzztheory}

\begingroup\raggedright\begin{thebibliography}{10}

\bibitem{VerlindeWijnholtBottomUp}
H.~Verlinde and M.~Wijnholt, ``Building the standard model on a D3-brane,''
  {\em JHEP} {\bf 01} (2007) 106,
  \href{http://xxx.lanl.gov/abs/hep-th/0508089}{{\tt hep-th/0508089}}.

\bibitem{KiritsisBottomUp}
I.~Antoniadis, E.~Kiritsis, and T.~N. Tomaras, ``A D-brane alternative to
  unification,'' {\em Phys. Lett.} {\bf B486} (2000) 186--193,
  \href{http://xxx.lanl.gov/abs/hep-ph/0004214}{{\tt hep-ph/0004214}}.

\bibitem{UrangaBottomUp}
G.~Aldazabal, L.~E. Ib\'{a}\~{n}ez, F.~Quevedo, and A.~M. Uranga, ``D-branes at
  singularities: A bottom-up approach to the string embedding of the standard
  model,'' {\em JHEP} {\bf 08} (2000) 002,
  \href{http://xxx.lanl.gov/abs/hep-th/0005067}{{\tt hep-th/0005067}}.

\bibitem{BerensteinJejjalaLeigh}
D.~Berenstein, V.~Jejjala, and R.~G. Leigh, ``The Standard Model on a
  D-brane,'' {\em Phys. Rev. Lett.} {\bf 88} (2002) 071602,
  \href{http://xxx.lanl.gov/abs/hep-th/0105042}{{\tt hep-th/0105042}}.

\bibitem{DWI}
R.~Donagi and M.~Wijnholt, ``Model Building with F-Theory,''
  \href{http://xxx.lanl.gov/abs/arXiv:0802.2969 [hep-th]}{{\tt arXiv:0802.2969
  [hep-th]}}.

\bibitem{BHVI}
C.~Beasley, J.~J. Heckman, and C.~Vafa, ``GUTs and Exceptional Branes in
  F-theory - I,'' {\em JHEP} {\bf 01} (2009) 058,
  \href{http://xxx.lanl.gov/abs/arXiv:0802.3391 [hep-th]}{{\tt arXiv:0802.3391
  [hep-th]}}.

\bibitem{WatariTATARHETF}
H.~Hayashi, R.~Tatar, Y.~Toda, T.~Watari, and M.~Yamazaki, ``New Aspects of
  Heterotic--F Theory Duality,'' \href{http://xxx.lanl.gov/abs/arXiv:0805.1057
  [hep-th]}{{\tt arXiv:0805.1057 [hep-th]}}.

\bibitem{BHVII}
C.~Beasley, J.~J. Heckman, and C.~Vafa, ``GUTs and Exceptional Branes in
  F-theory - II: Experimental Predictions,'' {\em JHEP} {\bf 01} (2009) 059,
  \href{http://xxx.lanl.gov/abs/arXiv:0806.0102 [hep-th]}{{\tt arXiv:0806.0102
  [hep-th]}}.

\bibitem{DWII}
R.~Donagi and M.~Wijnholt, ``{Breaking GUT Groups in F-Theory},''
  \href{http://xxx.lanl.gov/abs/arXiv:0808.2223 [hep-th]}{{\tt arXiv:0808.2223
  [hep-th]}}.

\bibitem{Heckman:2010bq}
J.~J. Heckman, ``{Particle Physics Implications of F-theory},''
  \href{http://xxx.lanl.gov/abs/arXiv:1001.0577 [hep-th]}{{\tt arXiv:1001.0577
  [hep-th]}}.

\bibitem{DWIII}
R.~Donagi and M.~Wijnholt, ``Higgs Bundles and UV Completion in F-Theory,''
  \href{http://xxx.lanl.gov/abs/arXiv:0904.1218 [hep-th]}{{\tt arXiv:0904.1218
  [hep-th]}}.

\bibitem{Cordova:2009fg}
C.~C\'{o}rdova, ``{Decoupling Gravity in F-Theory},''
  \href{http://xxx.lanl.gov/abs/arXiv:0910.2955 [hep-th]}{{\tt arXiv:0910.2955
  [hep-th]}}.

\bibitem{Grimm:2009yu}
T.~W. Grimm, S.~Krause, and T.~Weigand, ``{F-Theory GUT Vacua on Compact
  Calabi-Yau Fourfolds},'' \href{http://xxx.lanl.gov/abs/arXiv:0912.3524
  [hep-th]}{{\tt arXiv:0912.3524 [hep-th]}}.

\bibitem{Dine:1985he}
M.~Dine and N.~Seiberg, ``{Is the Superstring Weakly Coupled?},'' {\em Phys.
  Lett.} {\bf B162} (1985) 299.

\bibitem{SeibergWitten}
N.~Seiberg and Witten, ``{String Theory and Non-Commutative Geometry},'' {\em
  JHEP} {\bf 09} (1999) 032, \href{http://xxx.lanl.gov/abs/hep-th/9908142}{{\tt
  hep-th/9908142}}.

\bibitem{Connes:1997cr}
A.~Connes, M.~R. Douglas, and A.~S. Schwarz, ``{Noncommutative Geometry and
  Matrix Theory: Compactification on Tori},'' {\em JHEP} {\bf 02} (1998) 003,
  \href{http://xxx.lanl.gov/abs/hep-th/9711162}{{\tt hep-th/9711162}}.

\bibitem{Alekseev:1999bs}
A.~Y. Alekseev, A.~Recknagel, and V.~Schomerus, ``{Non-commutative World-volume
  Geometries: Branes on SU(2) and Fuzzy Spheres},'' {\em JHEP} {\bf 09} (1999)
  023, \href{http://xxx.lanl.gov/abs/hep-th/9908040}{{\tt hep-th/9908040}}.

\bibitem{Alekseev:2000fd}
A.~Y. Alekseev, A.~Recknagel, and V.~Schomerus, ``{Brane Dynamics in Background
  Fluxes and Non-Commutative Geometry},'' {\em JHEP} {\bf 05} (2000) 010,
  \href{http://xxx.lanl.gov/abs/hep-th/0003187}{{\tt hep-th/0003187}}.

\bibitem{ConnesBook}
A.~Connes, {\em Noncommutative Geometry}.
\newblock Academic Press, 1994.

\bibitem{Douglas:2001ba}
M.~R. Douglas and N.~A. Nekrasov, ``{Noncommutative Field Theory},'' {\em Rev.
  Mod. Phys.} {\bf 73} (2001) 977--1029,
  \href{http://xxx.lanl.gov/abs/hep-th/0106048}{{\tt hep-th/0106048}}.

\bibitem{Balachandran:2005ew}
A.~P. Balachandran, S.~Kurkcuoglu, and S.~Vaidya, ``{Lectures on Fuzzy and
  Fuzzy SUSY Physics},'' \href{http://xxx.lanl.gov/abs/hep-th/0511114}{{\tt
  hep-th/0511114}}.

\bibitem{AntoniadisReview}
I.~Antoniadis, ``{Topics on String Phenomenology},''
  \href{http://xxx.lanl.gov/abs/arXiv:0710.4267 [hep-th]}{{\tt arXiv:0710.4267
  [hep-th]}}.

\bibitem{Aschieri:2006uw}
P.~Aschieri, T.~Grammatikopoulos, H.~Steinacker, and G.~Zoupanos, ``{Dynamical
  generation of fuzzy extra dimensions, dimensional reduction and symmetry
  breaking},'' {\em JHEP} {\bf 09} (2006) 026,
  \href{http://xxx.lanl.gov/abs/hep-th/0606021}{{\tt hep-th/0606021}}.

\bibitem{Steinacker:2007ay}
H.~Steinacker and G.~Zoupanos, ``{Fermions on spontaneously generated spherical
  extra dimensions},'' {\em JHEP} {\bf 09} (2007) 017,
  \href{http://xxx.lanl.gov/abs/arXiv:0706.0398 [hep-th]}{{\tt arXiv:0706.0398
  [hep-th]}}.

\bibitem{Chatzistavrakidis:2009ix}
A.~Chatzistavrakidis, H.~Steinacker, and G.~Zoupanos, ``{On the fermion
  spectrum of spontaneously generated fuzzy extra dimensions with fluxes},''
  {\em Fortsch. Phys.} {\bf 58} (2010) 537--552,
  \href{http://xxx.lanl.gov/abs/arXiv:0909.5559 [hep-th]}{{\tt arXiv:0909.5559
  [hep-th]}}.

\bibitem{HVinprog}
J.~J. Heckman and H.~Verlinde, ``{Work in Progress},''.

\bibitem{Bergshoeff:2006gs}
E.~A. Bergshoeff, M.~de~Roo, S.~F. Kerstan, T.~Ortin, and F.~Riccioni,
  ``{$SL(2,\mathbb{R})$-invariant IIB Brane Actions},'' {\em JHEP} {\bf 02}
  (2007) 007, \href{http://xxx.lanl.gov/abs/hep-th/0611036}{{\tt
  hep-th/0611036}}.

\bibitem{VafaFTHEORY}
C.~Vafa, ``Evidence for F-theory,'' {\em Nucl. Phys.} {\bf B469} (1996)
  403--415, \href{http://xxx.lanl.gov/abs/hep-th/9602022}{{\tt
  hep-th/9602022}}.

\bibitem{MorrisonVafaI}
D.~R. Morrison and C.~Vafa, ``Compactifications of F-theory on Calabi-Yau
  threefolds (I),'' {\em Nucl. Phys.} {\bf B473} (1996) 74--92,
  \href{http://xxx.lanl.gov/abs/hep-th/9602114}{{\tt hep-th/9602114}}.

\bibitem{MorrisonVafaII}
D.~R. Morrison and C.~Vafa, ``Compactifications of F-theory on Calabi-Yau
  threefolds (II),'' {\em Nucl. Phys.} {\bf B476} (1996) 437--469,
  \href{http://xxx.lanl.gov/abs/hep-th/9603161}{{\tt hep-th/9603161}}.

\bibitem{SiegelTEND}
N.~Marcus, A.~Sagnotti, and W.~Siegel, ``Ten-dimensional supersymmetric
  Yang-Mills theory in terms of four-dimensional superfields,'' {\em Nucl.
  Phys. B} {\bf 224} (1983) 159--179.

\bibitem{WackerGregoire}
N.~Arkani-Hamed, T.~Gregoire, and J.~Wacker, ``Higher dimensional supersymmetry
  in 4D superspace,'' {\em JHEP} {\bf 03} (2002) 055,
  \href{http://xxx.lanl.gov/abs/hep-th/0101233}{{\tt hep-th/0101233}}.

\bibitem{KatzVafa}
S.~H. Katz and C.~Vafa, ``{Matter from geometry},'' {\em Nucl. Phys.} {\bf
  B497} (1997) 146--154, \href{http://xxx.lanl.gov/abs/hep-th/9606086}{{\tt
  hep-th/9606086}}.

\bibitem{Sethi:1996es}
S.~Sethi, C.~Vafa, and E.~Witten, ``{Constraints on Low-Dimensional String
  Compactifications},'' {\em Nucl. Phys.} {\bf B480} (1996) 213--224,
  \href{http://xxx.lanl.gov/abs/hep-th/9606122}{{\tt hep-th/9606122}}.

\bibitem{Iqbal:2003ds}
A.~Iqbal, N.~Nekrasov, A.~Okounkov, and C.~Vafa, ``{Quantum Foam and
  Topological Strings},'' {\em JHEP} {\bf 04} (2008) 011,
  \href{http://xxx.lanl.gov/abs/hep-th/0312022}{{\tt hep-th/0312022}}.

\bibitem{Saemann:2006gf}
C.~Saemann, ``{Fuzzy Toric Geometries},'' {\em JHEP} {\bf 02} (2008) 111,
  \href{http://xxx.lanl.gov/abs/hep-th/0612173}{{\tt hep-th/0612173}}.

\bibitem{FGUTSNC}
S.~Cecotti, M.~C.~N. Cheng, J.~J. Heckman, and C.~Vafa, ``{Yukawa Couplings in
  F-theory and Non-Commutative Geometry},''
  \href{http://xxx.lanl.gov/abs/arXiv:0910.0477 [hep-th]}{{\tt arXiv:0910.0477
  [hep-th]}}.

\bibitem{Marchesano:2009rz}
F.~Marchesano and L.~Martucci, ``{Non-perturbative effects on seven-brane
  Yukawa couplings},'' \href{http://xxx.lanl.gov/abs/arXiv:0910.5496
  [hep-th]}{{\tt arXiv:0910.5496 [hep-th]}}.

\bibitem{Grosse:1995jt}
H.~Grosse, C.~Klim\v{c}\'{i}k, and P.~Pre\v{s}najder, ``{Topologically
  Nontrivial Field Configurations in Noncommutative Geometry},'' {\em Commun.
  Math. Phys.} {\bf 178} (1996) 507--526,
  \href{http://xxx.lanl.gov/abs/hep-th/9510083}{{\tt hep-th/9510083}}.

\bibitem{Dolan:2006tx}
B.~P. Dolan, I.~Huet, S.~Murray, and D.~O'Connor, ``{Noncommutative vector
  bundles over fuzzy $\mathbb{CP}^{N}$ and their covariant derivatives},'' {\em
  JHEP} {\bf 07} (2007) 007, \href{http://xxx.lanl.gov/abs/hep-th/0611209}{{\tt
  hep-th/0611209}}.

\bibitem{FultonToric}
W.~Fulton, {\em Introduction to Toric Varieties (AM-131)}.
\newblock Princeton University Press, 1993.

\bibitem{Blumenhagen:2010pv}
R.~Blumenhagen, B.~Jurke, T.~Rahn, and H.~Roschy, ``{Cohomology of Line
  Bundles: A Computational Algorithm},''
  \href{http://xxx.lanl.gov/abs/arXiv:1003.5217 [hep-th]}{{\tt arXiv:1003.5217
  [hep-th]}}.

\bibitem{Hayashi:2009ge}
H.~Hayashi, T.~Kawano, R.~Tatar, and T.~Watari, ``{Codimension-3 Singularities
  and Yukawa Couplings in F- theory},'' {\em Nucl. Phys.} {\bf B823} (2009)
  47--115, \href{http://xxx.lanl.gov/abs/arXiv:0901.4941 [hep-th]}{{\tt
  arXiv:0901.4941 [hep-th]}}.

\bibitem{BHSV}
V.~Bouchard, J.~J. Heckman, J.~Seo, and C.~Vafa, ``{F-theory and Neutrinos:
  Kaluza-Klein Dilution of Flavor Hierarchy},'' {\em JHEP} {\bf 01} (2010) 061,
  \href{http://xxx.lanl.gov/abs/arXiv:0904.1419 [hep-ph]}{{\tt arXiv:0904.1419
  [hep-ph]}}.

\bibitem{EPOINT}
J.~J. Heckman, A.~Tavanfar, and C.~Vafa, ``{The Point of $E_8$ in F-theory
  GUTs},'' {\em JHEP} {\bf 08} (2010) 040,
  \href{http://xxx.lanl.gov/abs/arXiv:0906.0581 [hep-th]}{{\tt arXiv:0906.0581
  [hep-th]}}.

\bibitem{Marsano:2009gv}
J.~Marsano, N.~Saulina, and S.~Schafer-Nameki, ``{Monodromies, Fluxes, and
  Compact Three-Generation F-theory GUTs},'' {\em JHEP} {\bf 08} (2009) 046,
  \href{http://xxx.lanl.gov/abs/arXiv:0906.4672 [hep-th]}{{\tt arXiv:0906.4672
  [hep-th]}}.

\bibitem{Griffiths}
P.~Griffiths and J.~Harris, {\em Principles of Algebraic Geometry}.
\newblock John Wiley \& Sons, Inc., New York, 1978.

\bibitem{Cvetic:2010rq}
M.~Cveti\v{c}, I.~Garcia-Etxebarria, and J.~Halverson, ``{Global F-theory
  Models: Instantons and Gauge Dynamics},''
  \href{http://xxx.lanl.gov/abs/arXiv:1003.5337 [hep-th]}{{\tt arXiv:1003.5337
  [hep-th]}}.

\bibitem{HVCKM}
J.~J. Heckman and C.~Vafa, ``{Flavor Hierarchy From F-theory},'' {\em Nucl.
  Phys.} {\bf B837} (2010) 137--151,
  \href{http://xxx.lanl.gov/abs/arXiv:0811.2417 [hep-th]}{{\tt arXiv:0811.2417
  [hep-th]}}.

\bibitem{WittenPhases}
E.~Witten, ``{Phases of $\mathcal{N} = 2$ Theories in Two Dimensions},'' {\em
  Nucl. Phys.} {\bf B403} (1993) 159--222,
  \href{http://xxx.lanl.gov/abs/hep-th/9301042}{{\tt hep-th/9301042}}.

\bibitem{HoriVafa}
K.~Hori and C.~Vafa, ``{Mirror Symmetry},''
  \href{http://xxx.lanl.gov/abs/hep-th/0002222}{{\tt hep-th/0002222}}.

\bibitem{Morrison:1995yh}
D.~R. Morrison and M.~R. Plesser, ``{Towards Mirror Symmetry as Duality for
  Two-Dimensional Abelian Gauge Theories},'' {\em Nucl. Phys. Proc. Suppl.}
  {\bf 46} (1996) 177--186, \href{http://xxx.lanl.gov/abs/hep-th/9508107}{{\tt
  hep-th/9508107}}.

\bibitem{Roan}
S.-S. Roan, ``{On Calabi-Yau orbifolds in weighted projective spaces},'' {\em
  Int. J. Math.} {\bf 1} (1990) 211--232.

\bibitem{Batyrev}
D.~R. Morrison and M.~R. Plesser, ``{Dual polyhedra and mirror symmetry for
  Calabi-Yau hypersurfaces in toric varieties},'' {\em J. Alg. Geom.} {\bf 3}
  (1994) 493--535.

\bibitem{Blumenhagen:2008aw}
R.~Blumenhagen, ``{Gauge Coupling Unification in F-Theory Grand Unified
  Theories},'' {\em Phys. Rev. Lett.} {\bf 102} (2009) 071601,
  \href{http://xxx.lanl.gov/abs/arXiv:0812.0248 [hep-th]}{{\tt arXiv:0812.0248
  [hep-th]}}.

\bibitem{Conlon:2009qa}
J.~P. Conlon and E.~Palti, ``{On Gauge Threshold Corrections for Local
  IIB/F-theory GUTs},'' {\em Phys. Rev.} {\bf D80} (2009) 106004,
  \href{http://xxx.lanl.gov/abs/arXiv:0907.1362 [hep-th]}{{\tt arXiv:0907.1362
  [hep-th]}}.

\bibitem{Antoniadis:1990ew}
I.~Antoniadis, ``{A Possible new dimension at a few TeV},'' {\em Phys. Lett.}
  {\bf B246} (1990) 377--384.

\bibitem{ADD}
N.~Arkani-Hamed, S.~Dimopoulos, and G.~R. Dvali, ``{The Hierarchy Problem and
  New Dimensions at a Millimeter},'' {\em Phys. Lett.} {\bf B429} (1998)
  263--272, \href{http://xxx.lanl.gov/abs/hep-ph/9803315}{{\tt
  hep-ph/9803315}}.

\bibitem{AADD}
I.~Antoniadis, N.~Arkani-Hamed, S.~Dimopoulos, and G.~R. Dvali, ``{New
  Dimensions at a Millimeter to a Fermi and Superstrings at a TeV},'' {\em
  Phys. Lett.} {\bf B436} (1998) 257--263,
  \href{http://xxx.lanl.gov/abs/hep-ph/9804398}{{\tt hep-ph/9804398}}.

\bibitem{RS1}
L.~Randall and R.~Sundrum, ``{A Large Mass Hierarchy from a Small Extra
  Dimension},'' {\em Phys. Rev. Lett.} {\bf 83} (1999) 3370--3373,
  \href{http://xxx.lanl.gov/abs/hep-ph/9905221}{{\tt hep-ph/9905221}}.

\bibitem{RS2}
L.~Randall and R.~Sundrum, ``{An Alternative to Compactification},'' {\em Phys.
  Rev. Lett.} {\bf 83} (1999) 4690--4693,
  \href{http://xxx.lanl.gov/abs/hep-th/9906064}{{\tt hep-th/9906064}}.

\bibitem{Carlson:2001bk}
C.~E. Carlson and C.~D. Carone, ``{Discerning Noncommutative Extra
  Dimensions},'' {\em Phys. Rev.} {\bf D65} (2002) 075007,
  \href{http://xxx.lanl.gov/abs/hep-ph/0112143}{{\tt hep-ph/0112143}}.

\bibitem{DienesI}
K.~R. Dienes, E.~Dudas, and T.~Gherghetta, ``{Extra Spacetime Dimensions and
  Unification},'' {\em Phys. Lett.} {\bf B436} (1998) 55--65,
  \href{http://xxx.lanl.gov/abs/hep-ph/9803466}{{\tt hep-ph/9803466}}.

\bibitem{DienesII}
K.~R. Dienes, E.~Dudas, and T.~Gherghetta, ``{Grand Unification at Intermediate
  Mass Scales Through Extra Dimensions},'' {\em Nucl. Phys.} {\bf B537} (1999)
  47--108, \href{http://xxx.lanl.gov/abs/hep-ph/9806292}{{\tt hep-ph/9806292}}.

\bibitem{HallNomura}
L.~J. Hall and Y.~Nomura, ``{Gauge Unification in Higher Dimensions},'' {\em
  Phys. Rev.} {\bf D64} (2001) 055003,
  \href{http://xxx.lanl.gov/abs/hep-ph/0103125}{{\tt hep-ph/0103125}}.

\bibitem{ArkaniHamed:2001vr}
N.~Arkani-Hamed, A.~G. Cohen, and H.~Georgi, ``{Accelerated Unification},''
  \href{http://xxx.lanl.gov/abs/hep-th/0108089}{{\tt hep-th/0108089}}.

\bibitem{Goldberger:2002pc}
W.~D. Goldberger, Y.~Nomura, and D.~Tucker-Smith, ``{Warped Supersymmetric
  Grand Unification},'' {\em Phys. Rev.} {\bf D67} (2003) 075021,
  \href{http://xxx.lanl.gov/abs/hep-ph/0209158}{{\tt hep-ph/0209158}}.

\bibitem{Maldacena:2009mw}
J.~Maldacena and D.~Martelli, ``{The unwarped, resolved, deformed conifold:
  fivebranes and the baryonic branch of the Klebanov-Strassler theory},'' {\em
  JHEP} {\bf 01} (2010) 104, \href{http://xxx.lanl.gov/abs/arXiv:0906.0591
  [hep-th]}{{\tt arXiv:0906.0591 [hep-th]}}.

\bibitem{Gaiotto:2004pc}
D.~Gaiotto, A.~Simons, A.~Strominger, and X.~Yin, ``{D0-branes in Black Hole
  Attractors},'' \href{http://xxx.lanl.gov/abs/hep-th/0412179}{{\tt
  hep-th/0412179}}.

\bibitem{Douglas:2008es}
M.~R. Douglas and S.~Klevtsov, ``{Black holes and balanced metrics},''
  \href{http://xxx.lanl.gov/abs/arXiv:0811.0367 [hep-th]}{{\tt arXiv:0811.0367
  [hep-th]}}.

\bibitem{Myers:1999ps}
R.~C. Myers, ``{Dielectric-Branes},'' {\em JHEP} {\bf 12} (1999) 022,
  \href{http://xxx.lanl.gov/abs/hep-th/9910053}{{\tt hep-th/9910053}}.

\bibitem{juanAdS}
J.~M. Maldacena, ``The Large N Limit of Superconformal field theories and
  supergravity,'' {\em Adv. Theor. Math. Phys.} {\bf 2} (1998) 231--252,
  \href{http://xxx.lanl.gov/abs/hep-th/9711200}{{\tt hep-th/9711200}}.

\bibitem{'tHooft:1973jz}
G.~'t~Hooft, ``{A PLANAR DIAGRAM THEORY FOR STRONG INTERACTIONS},'' {\em Nucl.
  Phys.} {\bf B72} (1974) 461.

\bibitem{ArkaniHamed:2001ca}
N.~Arkani-Hamed, A.~G. Cohen, and H.~Georgi, ``{(De)Constructing Dimensions},''
  {\em Phys. Rev. Lett.} {\bf 86} (2001) 4757--4761,
  \href{http://xxx.lanl.gov/abs/hep-th/0104005}{{\tt hep-th/0104005}}.

\end{thebibliography}\endgroup

\end{document}